%% file: main.tex
\documentclass[fleqn,usenatbib,useAMS]{mnras}

\usepackage[percent]{overpic}
\usepackage{newtxtext,newtxmath}

\usepackage[T1]{fontenc}
\usepackage{ae,aecompl}
\usepackage{multirow}

\usepackage{graphicx}	
\usepackage{amsmath}	

\newcommand{\DM}{{\rm DM}}

\newcommand{\DMISM}{{\rm DM}_{\rm ISM}}

\title[FRB $z$--DM distribution]{The $z$--DM distribution of fast radio bursts}

\author[C.W.~James et al.]{
C.W.~James,${^{1}}$\thanks{E-mail: clancy.james@curtin.edu.au}
J.X.~Prochaska${^{2,3}}$
J.-P.~Macquart,${{^1}}$\footnote{Deceased}
F.O.~North-Hickey${^{1}}$
K.~W.~Bannister${^{4}}$\newauthor
 and
A.~Dunning${^{4}}$
\\
$^{1}$International Centre for Radio Astronomy Research, Curtin University, Bentley, WA 6102, Australia \\
$^{2}$Department of Astronomy and Astrophysics, University of California, Santa Cruz, CA 95064, USA\\
$^{3}$Kavli Institute for the Physics and Mathematics of the Universe, 5-1-5 Kashiwanoha, Kashiwa 277-8583, Japan.\\
$^4$CSIRO Astronomy and Space Science, PO Box 76, Epping, NSW 1710, Australia.
}

\date{Accepted XXX. Received YYY; in original form ZZZ}

\pubyear{2020}

\begin{document}

\label{firstpage}
\pagerange{\pageref{firstpage}--\pageref{lastpage}}
\maketitle

\begin{abstract}
We develop a sophisticated model of FRB observations, accounting for the intrinsic cosmological gas distribution and host galaxy contributions, and give the most detailed account yet of observational biases due to burst width, dispersion measure, and the exact telescope beamshape. Our results offer a significant increase in both accuracy and precision beyond those previously obtained.
Using results from ASKAP and Parkes, we present our best-fit FRB population parameters in a companion paper. Here, we consider in detail the expected and fitted distributions in redshift, dispersion measure, and signal-to-noise. We estimate that the unlocalised ASKAP FRBs arise from $z<0.5$, with between a third and a half within $z<0.1$. Our predicted source-counts (``logN--logS'') distribution confirms previous indications of a steepening index near the Parkes detection threshold of $1$\,Jy\,ms. We find no evidence for a minimum FRB energy, and rule out $E_{\rm min} > 10^{39.0}$\,erg at 90\% C.L.
Importantly, we find that above a certain DM, observational biases cause the Macquart (DM--z) relation to become inverted, implying that the highest-DM events detected in the unlocalised Parkes and ASKAP samples are unlikely to be the most distant. More localized FRBs will be required to quantitatively estimate this effect, though its cause is a well-understood observational bias. Works assuming a 1--1 DM--z relation may therefore derive erroneous results.
Our analysis of errors suggests that limiting factors in our analysis are understanding of FRB spectral behaviour, sensitivity response of search experiments, and the treatment of the repeating population and luminosity function.
\end{abstract}

\begin{keywords}
transients: fast radio bursts -- methods: statistical
\end{keywords}



\input{introduction}

\input{Ingredients1_z_dm}

\input{Ingredients2_frb_pop}

\input{Ingredients3_det_thresh}

\input{Methodology}

\input{Surveys}

\input{Results1_initial}

\input{Results2_pzdm}

\input{Results3_pzgdm}

\input{DiscussionConclusion}

\section*{Acknowledgements}

This research has made use of NASA's Astrophysics Data System Bibliographic Services. This research made use of Python libraries \textsc{Matplotlib} \citep{Matplotlib2007}, \textsc{NumPy} \citep{Numpy2011}, and \textsc{SciPy} \citep{SciPy2019}. This work was performed on the gSTAR national facility at Swinburne University of Technology. gSTAR is funded by Swinburne and the Australian Government’s Education Investment Fund. This work was supported by resources provided by the Pawsey Supercomputing Centre with funding from the Australian Government and the Government of Western Australia. This research was partially supported by the Australian Government through the Australian Research Council's Discovery Projects funding scheme (projects DP180100857 and DP210102103).

\section*{Data Availability}

The data underlying this article will be shared on reasonable request to the corresponding author.


\bibliographystyle{mnras}
\bibliography{bibliography}

\appendix

\input{appendix}

\bsp	
\label{lastpage}
\end{document}

%% file: introduction.tex
\section{Introduction}

Fast radio bursts (FRBs) are radio transients of millisecond duration and extragalactic origin \citep{Lorimer2007,Thornton2013}. Their progenitors are unknown, with very many production mechanisms propoed \citep{Platts2018}.
FRB surveys are providing increasingly large statistics with which to study the FRB population \citep{Bhandarietal2018,Shannonetal2018,CHIME_catalog1_2021}, including a handful of localised FRBs \citep{Tendulkar2017,Bannister2019,Prochaska2019,RaviNature2019,MarcoteRepeaterLocalisation2020}. Furthermore, \citet{Macquart2020} have used localised FRBs as probes of the cosmological distribution of ionised gas, illustrating their utility for cosmological studies \citep{McQuinn2014,MasuiSigurdson2015,Madhavacheril2019,CalebHelium2019}. Of the key questions surrounding FRBs, this work focuses on FRB population statistics.

Studies of the FRB population are important both for understanding the nature of FRBs themselves, and their use as cosmological probes. Typical fitted parameters include the FRB luminosity function, e.g.\ minimum and maximum energies, and its shape; FRB spectral properties; and the source evolution. Studies including repeating FRBs must also fit the distribution of repetition rates and allow for time-dependence between bursts from a single object.

Linking FRB observations to the underlying true FRB population however is difficult. \citet{Connor2019} review previous methods of studying the FRB population, and emphasise that accurate estimates require accounting for the sensitivity effects of telescope beamshape, intrinsic burst width, and the dispersion measure distribution $p(\DM|z)$ for a given redshift. In short, one must integrate over known or hypothesised intrinsic distributions of these variables, model observational biases, and then attempt to match observations. Doing so improperly will produce biased results.

In what is usually seen as a different line of inquiry, cosmological studies using FRBs take advantage of their dispersion measure (DM), which integrates the column density of ionised gas along their line of sight. This encodes information on the diffuse gas in voids and galactic halos which is otherwise difficult to study. \citet{Macquart2020} have recently used the localised FRB population to constrain the total baryon density of the Universe and the degree of feedback. In that work, the authors analyse the probability distribution of observed dispersion measures, DM, given the redshift $z$ of identified FRB host galaxies, $p(z,\DM)$. This controls somewhat for the effects of the population of FRBs, which primarily affects the redshift distribution $p(z)$. The authors do note however the potentially biasing effects of the FRB population as observed by FRB surveys, although a comprehensive treatement of such biasing effects is not performed due to the intrinsic error in using a small sample size (5--7).
Many cosmological studies (such as helium reionisation) require many FRBs to both exist and be detectable at a redshift of $z\sim4$ \citep{CalebHelium2019}, which is well beyond the most distant localized FRB to date at $z\sim 0.6$ \citep{Law2020VLAlocalization}.

Fundamentally, both lines of inquiry aim to study the intrinsic distribution of FRBs in $z$--DM space, $p(\DM,z)$. The only difference is the aspect of interest: population studies try to resolve a redshift distribution $p(z)$ and treat the distribution $p(\DM|z)$ as a nuissance distribution, while cosmological studies aim to resolve $p(\DM|z)$ and attempt to remove the $p(z)$ factor. They are thus fundamentally coupled problems. Unbiased estimates of the cosmological distribution of ionised gas require knowing the FRB population and the consequent biasing effects on measured dispersion measures; and understanding the FRB population requires knowing the dispersion-measure distribution and its biasing effects on the measured luminosity function.

\citet{Calebetal2016} provide the first comprehensive model of observational biases on a simulated burst population, and {\sc FRBPOPPY} \citep{Gardenier2019FRBPOPPY} is being developed to include such effects. To date however, only \citet{Luo2020} have used this approach to fit population parameters. The authors study a sample of FRBs detected in the $\sim$1\,GHz band from Parkes, the Upgraded Molonglo Synthesis Telescope (UTMOST), the Australian Square Kilometre Array Pathfinder (ASKAP), Arecibo, and the Greenbank Telescope (GBT).
The authors evaluate the validity of their model using {\sc MULTINEST} \citep{MultinestFeroz2009}, which applies a Bayesian framework, and find a peak FRB luminosity $L^*$ of $2.9_{-1.7}^{+11.9} 10^{44}$\,erg\,s$^{-1}$, a differential power-law index of $-1.79_{-0.35}^{0.31}$, and a volumetric rate of 
$3.7_{-2.4}^{+5.7}\,10^{4}$\,Gpc$^{-3}$\,yr$^{-1}$ above $10^{42}$\,erg\,s$^{-1}$.

In this work, we significantly improve upon FRB population models in the following manner:
\begin{itemize}
    \item using an unbiased sample of FRBs from ASKAP and Parkes;
    \item using seven localised FRBs detected by ASKAP;
    \item correctly accounting for the full telescope beamshape;
    \item using the measured signal-to-noise ratio in probability estimates;
    \item including the intrinsic spread in the cosmological DM contribution due to large-scale structure and galaxy halos;
    \item and allowing for redshift evolution of the FRB event rate per comoving volume.
\end{itemize}
As with other population models, to make the problem tractable, we assume that the cosmological rate evolution, host galaxy DM contribution, burst width, and burst energy distributions are all independent; and that FRB observations are random and uncorrelated, i.e.\ we do not model rapidly repeating FRBs.

We begin by describing the ingredients to our model: a model of the DM distribution of FRBs as a function of redshift (Section~\ref{sec:ingredients1}), based on the model of \citet{Macquart2020}; the intrinsic properties of FRBs, such as their burst width and the luminosity function, using a standard power-law description (Section~\ref{sec:ingredients2}); and the influence of observational effects such as beamshape and search sensitivity (Section~\ref{sec:ingredients3}). The method to combine these to calculate the expected ``z-DM'' distribution for observed FRB surveys is given in Section~\ref{sec:methodology}. In Section~\ref{sec:survey_results}, we describe the data from ASKAP and Parkes to which we fit our model using maximum-likelihood methods. 
The best-fit FRB population parameters, and their uncertainties, are given in a companion paper \citep{James2021Lett}. In this work, we present detailed comparisons to the observed DM, redshift, and signal-to-noise ratio distributions in Section~\ref{sec:results2}, where we test for goodness-of-fit and search for deviations from expectations. Section~\ref{sec:results3_zdm} shows our estimates for the expected z--DM distribution of FRBs detected by ASKAP and Parkes. We summarize our results in Section~\ref{sec:conclusion}. We attach in appendices a discussion of neglected effects in our modelling, and extra data for alternative source evolution scenarios.

%% file: Ingredients1_z_dm.tex
\section{Dispersion measure distribution}
\label{sec:ingredients1}

The distribution of dispersion measure, DM, of FRBs from a given redshift $z$, $p({\rm DM}|z)$, is of both intrinsic interest, and is a nuissance factor in calculating the properties of the FRB population itself. Here, we use the method and parameters of \citet{Macquart2020}. We model the DM of an FRB as
\begin{eqnarray}
\DM & = & \DM_{\rm ISM}+ \DM_{\rm halo} + \DM_{\rm cosmic} + \DM_{\rm host},
\end{eqnarray}
with respective contributions from the Milky Way's interstellar medium (ISM), it's halo, the cosmological distribution of ionised gas, and the FRB host. In this work, we divide this into an `extragalactic' contribution,
\begin{eqnarray}
\DM_{\rm EG} & \equiv & \DM_{\rm cosmic} + \DM_{\rm host}, \label{eq:DMEG}
\end{eqnarray}
and a `local' contribution,
\begin{eqnarray}
\DM_{\rm local} & \equiv & \DM_{\rm ISM}+ \DM_{\rm halo}. \label{eq:DMlocal}
\end{eqnarray}
The `local' contribution is subtracted from FRB observations, and thus all comparisons between expectations and measurements are made in terns of $\DM_{\rm EG}$. This model slightly differs from that in \citet{Macquart2020}, who model both $\DM_{\rm host}$ and $\DM_{\rm halo}$ using the same nuisance term, $DM_X$. The distinction becomes important at large redshifts.

\subsection{DM$_{\rm ISM}$}
\label{sec:dm_mw}

We use the NE2001 model \citep{CordesLazio01},\footnote{Ben Bar-Or, J.~Prochaska, available at https://readthedocs.org/projects/ne2001/} to estimate the Galactic contribution to dispersion measure. Since DM is an ingredient in the calculation of detection efficiency (see Section~\ref{sec:efficiency}), the full integral for the FRB rate extends over the pointing direction as a function of Galactic coordinates, as discussed in Section~\ref{sec:integral_galactic} and Eq.~\ref{eq:integral_galactic}. Since most FRBs and FRB searches have been at high Galactic latitudes however, we use the mean value $\overline{\DM}_{\rm ISM}$ 
to calculate the sensitivity for each survey, 
while using the individual values $\DMISM$
when calculating FRB likelihoods.

\subsection{DM$_{\rm halo}$}
\label{sec:dm_halo}

The exact contribution of the Milky Way halo to DM is uncertain, with estimates of order 10--80\,pc\,cm$^{-3}$ \citep{ProchaskaZheng2019,KeatingPen2020}. FRBs have been observed down to a DM of little more than 110\,pc\,cm$^{-3}$ \citep{CHIME2019a} and 114\,pc\,cm$^{-3}$ \citep{Shannonetal2018}, favouring the middle of this range and consistent with 
current estimations based on the full set of observed DMs \citep{Platts2020}.
We therefore use a value of DM$_{\rm halo}=50$\,pc\,cm$^{-3}$ in our default model.

Deviations between our assumed values for DM$_{\rm ISM}$ and DM$_{\rm halo}$ will be absorbed into our model for the host galaxy contribution.

\subsection{Cosmological DM}
\label{sec:dm_cosmic}

We caution that symbols $E$, $F$, and $\alpha$ have different definitions in this section than in the remainder of this work. The notation regarding DM$_{\rm cosmic}$ is derived from \citet{Macquart2020}, and we preserve it for ease of reference to that work.

The `cosmological' contribution to DM, DM$_{\rm cosmic}$, can be understood as the DM incurred when an FRB is emitted at $z$ at a random point in the Universe and propagates until the current epoch, $z=0$. A parameterisation based on detailed simulations \citep{McQuinn2014}
is given in \citet{Macquart2020}, as a function of burst redshift $z$, as
\begin{eqnarray}
{\rm DM_{\rm cosmic}} & = & \left< {\rm DM}_{\rm cosmic} \right> \Delta_{\rm DM} \\
p({\rm DM_{\rm cosmic}} | z) & = & \frac{ p(\Delta_{\rm DM}|z)}{\left< {\rm DM}_{\rm cosmic} \right>}. \label{eq:pcosmic}
\end{eqnarray}

The expected value $\left< {\rm DM}_{\rm cosmic} \right>$ is calculated as per \citet{Kunihito2003,Inoue2004}:
\begin{eqnarray}
\left< {\rm DM}_{\rm cosmic} \right> & = & \int_0^z \frac{c \bar{n}_e(z^\prime) dz^\prime}{H_0 (1+z^\prime)^2 E(z)},\\
E(z) & = & \sqrt{\Omega_m(1+z^\prime)^3 + \Omega_\Lambda}, \label{eq:Ez}
\end{eqnarray}
using the mean density of ions, $\bar{n}_e$, and cosmological parameters relevant for the range $0 \le z \le 5$: $H_0=67.4$\,km\,s$^{-1}$\,Mpc$^{-1}$, and matter and dark energy densities $\Omega_m=0.315$ and $\Omega_\Lambda=0.685$ for a critical density Universe. See \citet{Macquart2020} and references contained therein for further details --- cosmological parameters are taken from \citet{PlanckCosmology2018}.

The probability of deviations from the mean, $p(\Delta_{\rm DM})$, is given by
\begin{eqnarray}
p(\Delta_{\rm DM}) & = & A \Delta^{-\beta}_{\rm DM} \exp \left[- \frac{(\Delta^{-\alpha}_{\rm DM} - C_0)^2}{2 \alpha^2 \sigma_{\rm DM}^2} \right],
\end{eqnarray}
with $\alpha=3$, $\beta=3$, and $C_0$ being numerically tuned such that the expectation value of the distribution is unity. The degree of feedback $F$ is reflected in the choice of $\sigma_{\rm DM}=F z^{-0.5}$. 
In this work, we fix $F=0.32$ based on results of \citep{Macquart2020}.

The resulting distribution of DM$_{\rm cosmic}$, $p({\rm DM}_{\rm cosmic}|z)$ is shown in Figure~\ref{fig:dm_z_basic}.

\subsection{DM$_{\rm host}$}
\label{sec:dm_host}

The contribution of the FRB host galaxy (including the local environs of the FRB itself) to DM is highly uncertain. Some FRBs, most notably FRB~121102
and FRB~190608, show a large excess DM beyond what is expected from cosmological and MW contributions, which cannot be explained by passage through an intervening galaxy along the line-of-sight \citep{Spitler2014,Chatterjee2017,Tendulkar2017,Hardy2017,Chittidi2021}. 
Yet as noted in Section~\ref{sec:dm_halo}, many FRBs do not allow for a great deal of excess DM. \citet{Macquart2020} generically model this large spread using a log-normal distribution
\begin{eqnarray}
p({\rm DM}_{\rm host}^\prime) = \frac{1}{{\rm DM_{\rm host}^\prime}} \frac{1}{\sigma_{\rm host} \sqrt{2 \pi}} 
\exp \left[ -\frac{(\log {\rm DM}^\prime_{\rm host}-\mu_{\rm host})^2}{2 \sigma_{\rm host}^2} \right] \label{eq:phost}.
\end{eqnarray}
We also correct the host contribution for redshift via
\begin{eqnarray}
{\rm DM}_{\rm host} & = & \frac{{\rm DM}_{\rm host}^\prime}{1+z}.
\end{eqnarray}

In this work, we use $\mu_{\rm host}$ and $\sigma_{\rm host}$ as free parameters. Thus uncertainties in other DM contributions --- including from our assumed value of feedback $F$ --- will be absorbed into these quantities.

\subsection{The intrinsic z--DM grid}

\begin{figure} 
    \centering
    \includegraphics[width=\linewidth]{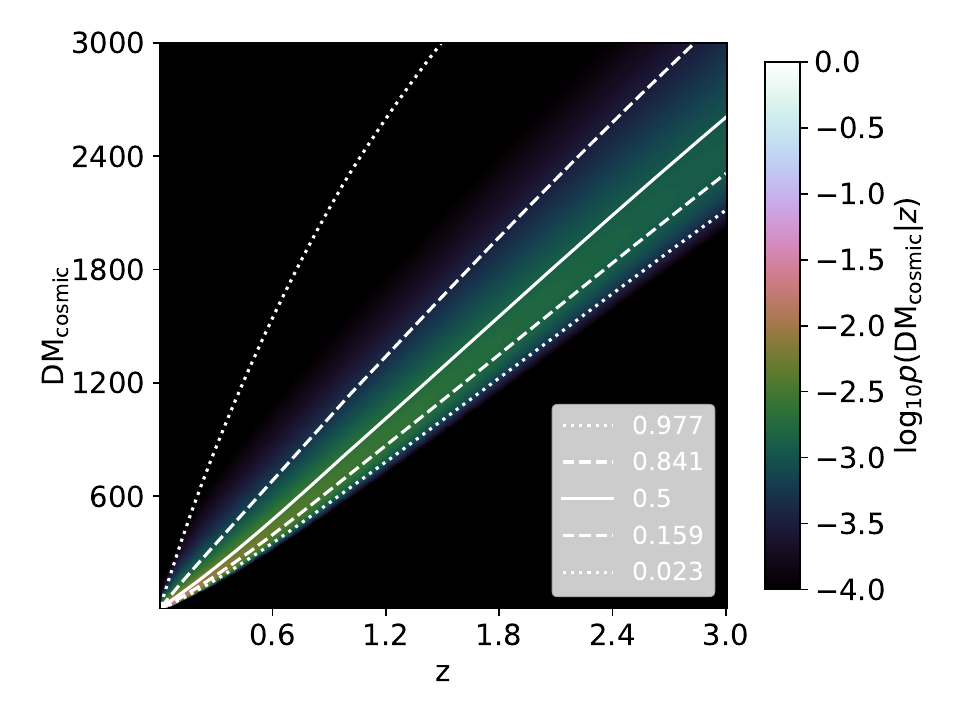}
    \includegraphics[width=\linewidth]{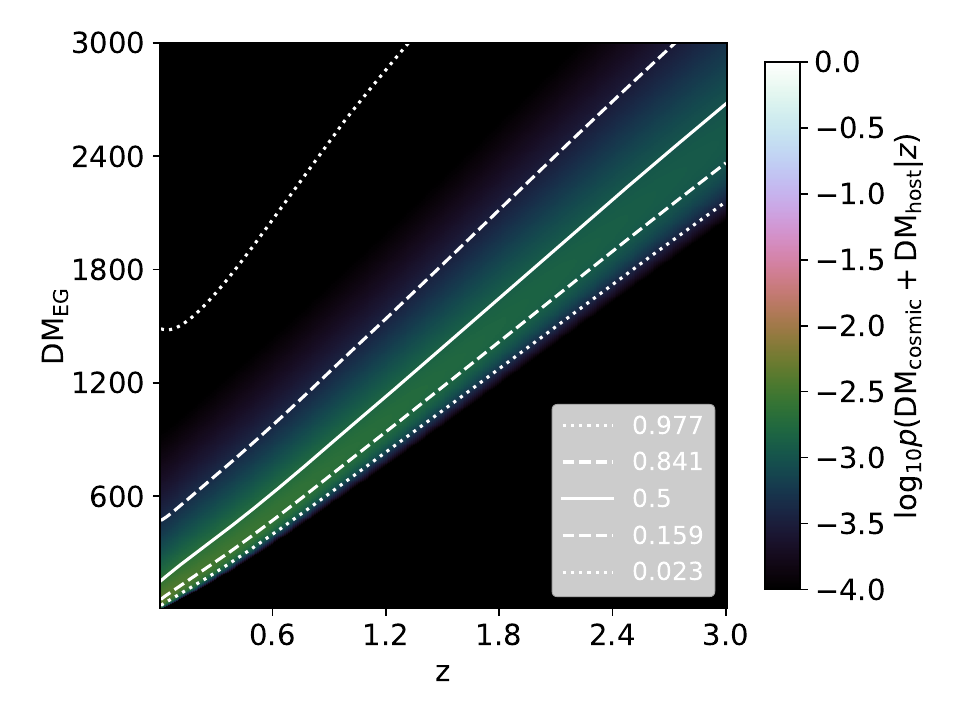}
    \caption{Distribution $p({\rm DM} |z)$ of observation-independent DM parameters, $DM_{\rm cosmic}$ only (top), and also including the best-fit distribution for $DM_{\rm host}$ derived in \citet{James2021Lett} (bottom), as a function of redshift $z$, showing contours.}
    \label{fig:dm_z_basic}
\end{figure}

The probability distribution of observation-independent factors, ${\rm DM}_{\rm cosmic} + {\rm DM}_{\rm host}$, is given in Figure~\ref{fig:dm_z_basic}. In this work, a linear grid in both $z$ and DM space is used, with 1200 DM points spaced from 0--7000 in intervals of 5\,pc\,cm$^{-3}$, and 500 in redshift from 0.01--5. FRBs have their nominal local contributions, ${\rm DM}_{\rm local}$ subtracted from their observed values of DM prior to evaluating their likelihood on this grid. In this model, only a small fraction of FRBs will have a DM very much larger than the mean. 
In particular, for $z<1.5$, the majority of the spread in DM comes from the host galaxy, rather than the cosmological contribution.

%% file: Ingredients2_frb_pop.tex
\section{FRB population}
\label{sec:ingredients2}

\subsection{Energetics}

Our model of the FRB population $\Phi$ is consistent with that used in the literature. We adopt a power-law distribution of burst energies $E$ between $E_{\rm min}$ and $E_{\rm max}$ with integral slope $\gamma$. We use `burst energy' as the isotropic equivalent energy at $1.3$\,GHz, and assume an effective bandwidth of 1\,GHz when converting between `per Hz' and total quantities. If FRBs are beamed into a characteristic solid angle $\Omega_{\rm FRB}$, then all energies in this work should be scaled down by a factor $\Omega_{\rm FRB} (4 \pi)^{-1}$ --- but the true FRB rate will increase by an identical factor.

In this model, the probability of a burst occurring above a threshold $E_{\rm th}$ is a piecewise function,
\begin{eqnarray}
p(E>E_{\rm th}) & = & 1 ~~~(E < E_{\rm min}) \nonumber \\
p(E>E_{\rm th}) & = & 0 ~~~(E > E_{\rm max}) \nonumber \\
p(E>E_{\rm th}) & = & \frac{ \left( \frac{E_{\rm th}}{E_{\rm min}}\right)^\gamma - \left(\frac{E_{\rm max}}{E_{\rm min}}\right)^\gamma  }{1-\left(\frac{E_{\rm max}}{E_{\rm min}}\right)^\gamma} ~{\rm otherwise}. \label{eq:integral_energy_distribution}
\end{eqnarray}
Observations show no evidence for a minimum FRB energy, and indeed the event rate is generally insensitive to $E_{\rm min}$ for $\gamma>-1.5$ \citep{Macquart2018b}. Thus we use a very low value of $E_{\rm min}=10^{30}$\,erg, which is several orders of magnitude below the minimum detected burst energy of all known FRBs --- including SGR 1935+2154 at $10^{34}$--$10^{35}$\,erg \citep{Magnetar_CHIME,Magnetar_STARE2} --- and treat this as a fixed parameter. We re-examine this assumption in Section~\ref{sec:emin}. 

Several other authors have used a Schechter function to model the FRB luminosity function \citep{Lu2019,Luo2020}, which adds an exponential cut-off of the form $\exp(-E/E_{\rm max})$ to the luminosity function of Eq.~\ref{eq:differential_energy_distribution}. This is neither observationally nor theoretically motivated --- the form of the Schechter function is used to model galaxy luminosities --- and adds computational complexity, but avoids the unphysicality of a sharp cut-off. A preliminary investigation showed that using the Schechter function provided effectively identical fits to the data. We therefore advocate for the simple power-law model out of simplicity.

To calculate $E_{\rm th}$, we convert between FRB energy $E$ and observable fluence $F$ using
\begin{eqnarray}
E(F) & = & \frac{4 \pi D_L^2(z)}{(1+z)^{2+\alpha}} \Delta \nu F \label{eq:F_to_E},
\end{eqnarray}
where $\alpha$ is the spectral index ($F \propto \nu^{\alpha}$), and $\Delta \nu$ the bandwidth (here we use 1\,GHz). \citet{Macquart2019a} fit $\alpha$ to 23 FRBs detected by ASKAP in Fly's Eye mode, finding $\alpha=-1.5_{-0.3}^{+0.2}$. Thus we use a default value of $\alpha=-1.5$. We return to the interpretation of $\alpha$ shortly.

\subsection{Population evolution}
\label{sec:population_evolution}

The rate of FRBs per comoving volume will likely be a function redshift. While FRB host galaxies do not appear to be drawn from a population sampled proportionally to their star-forming activity
\citep{Safarzadeh2021}, 
they certainly are not exclusively
associated with very old galaxies in which star-forming activity has ceased \citep{Bhandari2020a,Heintz2020}. 
We therefore adopt the approach of \citet{Macquart2018b} and generically model the population evolution of FRBs as being to some power of the star-formation rate, i.e.\ 
\begin{eqnarray}
\Phi(z) & = & \frac{\Phi_0}{1+z} \left( \frac{{\rm SFR}(z)}{{\rm SFR}(0)} \right)^n, \label{eq:phiz}
\end{eqnarray}
with $\Phi_0$ --- and hence $\Phi(z)$ --- taking the units of bursts per proper time per comoving volume, i.e.\ bursts yr$^{-1}$ Mpc$^{-3}$. The factor of $(1+z)^{-1}$ converts between proper time in the emission and observer frames. We take SFR$(z)$ from \citet{MadauDickinson_SFR},
\begin{eqnarray}
{\rm SFR}(z) & = & 1.0025738 \frac{(1+z)^{2.7}}{1 + \left(\frac{1+z}{2.9}\right)^{5.6}}. \label{eq:sfr_n}
\end{eqnarray}
This model is useful in that $n$ can be scaled as a smooth parameter. However it does not accurately model the source evolution should FRB progenitors originate from e.g.\ binary mergers with long delay times, as investigated by \citet{Cao2018delayedmergers}.

The total FRB rate in a given redshift interval $dz$ and sky area $d\Omega$ will also be proportional to the total comoving volume $dV$,
\begin{eqnarray}
\frac{dV}{d\Omega dz} & = & D_H \frac{(1+z)^2 D_A^2(z)}{E(z)}, \label{eq:comoving_volume}
\end{eqnarray}
for angular diameter distance $D_A$, Hubble distance $D_H=c/H_0$, and scale factor $E(z)$ from Eq.~\ref{eq:Ez}.

\begin{figure}
    \centering
    \includegraphics[width=\linewidth]{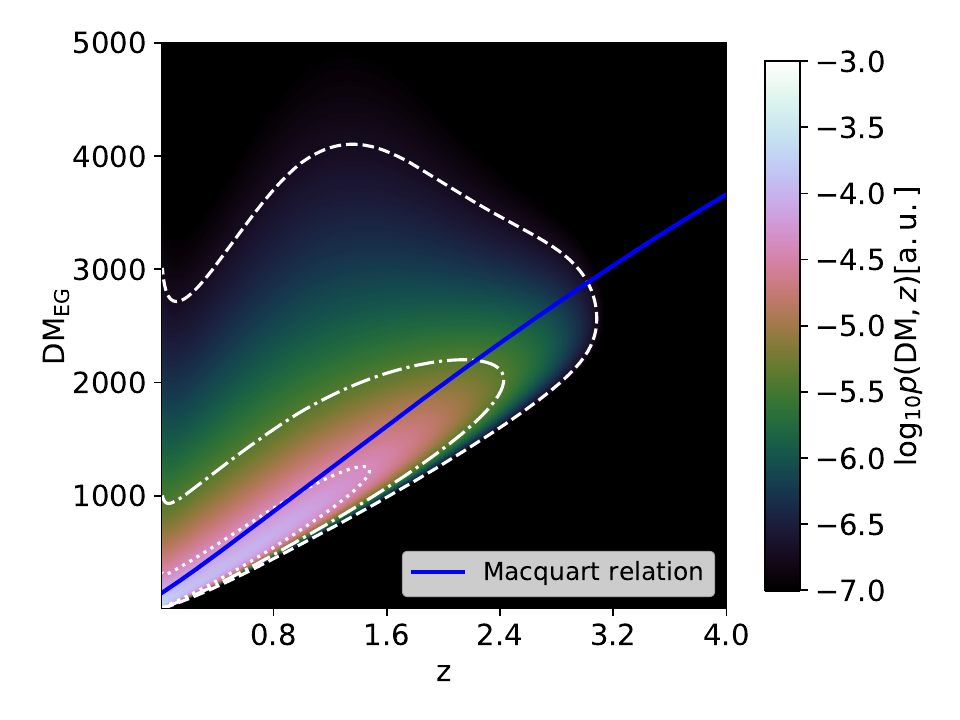}
    \caption{Relative rate of FRB detections in $z$--DM space when ignoring beamshape and burst width effects for the best-fit FRB population parameters presented in \citet{James2021Lett}. All observational biases from Section~\ref{sec:ingredients3} have been ignored, and a constant detection threshold of 1\,Jy\,ms is used. The Macquart relation, approximated by DM=$\left< {\rm DM}_{\rm cosmic} \right>$+$\exp(\mu_{\rm host})$, is also shown. The contours represent 50\% (dotted), 90\% (dash-dot), and 99\% (dashed) of the FRB population.}
    \label{fig:basic_F0}
\end{figure}

Applying this model with the best-fit FRB population parameters derived in our accompanying work to the DM-distribution of Section~\ref{sec:ingredients1} with a nominal threshold of 1\,Jy\,ms produces the distribution of FRBs shown in Figure~\ref{fig:basic_F0}. This ignores the important observational biases to be introduced in Section~\ref{sec:ingredients3}, and hence a quantitative analysis of the implications are left to Section~\ref{sec:results2}.  However, it is clear that while at least 90\% of FRBs will follow a 1--1 DM-z relation \citep[the Macquart relation][]{Macquart2020}, a significant minority will lie well above the DM--z curve. Indeed the highest DM events in a large sample are not likely to be the most distant. Consider DM$\ge3000$\,pc\,cm$^{-3}$ in Figure~\ref{fig:basic_F0}. Such events can be produced by $z\sim$3.2 lying on the Macquart relation. However, they must be the most intrinsically luminous FRBs to be detectable. At $z\sim$1.6, observations probe a factor of $\sim 4$ further down the energy distribution, allowing a greater number of events to be visible, and its high-DM tail
may dominate the DM$\ge3000$\,pc\,cm$^{-3}$ event rate. This effect becomes more important for steeper luminosity distributions (large negative $\gamma$) --- this plot uses $\gamma = -1.2$.

\subsection{Interpretation of $\alpha$}
\label{sec:alpha}

Many FRBs have a limited band occupancy \citep[originally noted for FRB~121102 by ][]{Law2017}, in which case the notion of a spectral index for an individual FRB has little meaning. In this case, the results of \citet{Macquart2019a} can be interpreted as meaning either that there are more low-frequency FRBs, or that low-frequency FRBs are stronger. For an experiment with bandwidth similar to or less than that of the FRB bandwidth (which is the case with the data used here --- see Section~\ref{sec:survey_results}), the latter interpretation behaves identically to that of broadband bursts defined by a spectral index. However, the interpretation of an FRB population with a frequency-dependent rate does not. We denote this interpretation as the `rate interpretation' of $\alpha$, and that of \citet{Macquart2019a} as the `spectral index' interpretation.

Under the spectral index interpretation of $F(\nu) \sim \nu^\alpha$, a negative $\alpha$ increases the detection threshold $E_{\rm th}$ at high $z$ due to the k-correction factor of $(1+z)^{-\alpha}$ through Eq.~\ref{eq:F_to_E}. This in turn decreases the rate by a factor $(1+z)^{\gamma \alpha}$ when $E_{\rm th} << E_{\rm max}$ through Eq.~\ref{eq:integral_energy_distribution}. Under the rate interpretation, the FRB population itself behaves as $\Phi \sim \Phi(z,\nu) = \Phi(z) \nu^\alpha$, and the k-correction therefore directly changes the rate, adding an additional factor of $(1+z)^\alpha$ to Eq.~\ref{eq:phiz}. Therefore, when $\gamma=-1$, and $E_{\rm th} << E_{\rm max}$, the two interpretations are identical. The situation becomes less simple near $E_{\rm max}$, which is frequency dependent under the spectral index interpretation, and constant under the rate interpretation --- and the true behaviour may be more complicated than either result.

Ultimately, we expect further observational data to be required to discriminate between the two scenarios, and consider both equally plausible for the time being. In this work, we present results using the spectral-index interpretation, but give additional data for the rate interpretation when constraining FRB population parameters.

%% file: Ingredients3_det_thresh.tex
\section{Detection threshold --- observational biases}
\label{sec:ingredients3}

FRB surveys usually calculate the fluence threshold above which FRBs would be detected using the radiometer equation, referenced to a 1\,ms duration burst, using the sensitivity of the telescope at beam centre. This readily calculable value represents an unrealistic ideal. Bursts of longer duration will be harder to detect due to increased noise, while those viewed away from beam centre will be seen at less sensitivity. Furthermore, incoherent dedispersion searches will not perfectly match the shape of an FRB to the time--frequency resolution of the search, resulting in a lower detection efficiency.

In this work, we model the effective fluence threshold $F_{\rm th}$ as a function of nominal fluence threshold at 1\,ms $F_1$, beam sensitivity $B$ (normalized to a maximum of 1), 
and an efficiency factor due to burst duration, $\eta$, as
\begin{eqnarray}
F_{\rm th} & = & \frac{F_1}{\eta B}. \label{eq:Fth}
\end{eqnarray}
This results in a theoretical distribution of bursts in z--DM space, $p(z,{\rm DM})$, as
\begin{eqnarray}
p(z,{\rm DM}) = \int dB \int d \eta p(z,{\rm DM} | F_{\rm th}(\eta,B)) \Omega(B) p(\eta), \label{eq:dm_eta_integral}
\end{eqnarray}
where $p(z,{\rm DM}|F_{\rm th})$ is the distribution at a fixed threshold (Figure~\ref{fig:basic_F0}), 
$\Omega(B)$ is the region of sky at which the beam sensitivity is $B$, and $p(\eta)$ is the probability that burst properties lead to a total detection efficiency $\eta$. The effects of these two factors are investigated in the following sections.

\subsection{Beamshape}
\label{sec:beamshape}

A telescope's beamshape is usually represented as the relative sensitivity $B$ as a function of the sky position $\Omega$ relative to boresight, $B(\Omega)$, such that $B(0)=1$. The beamshape is often approximated as a Gaussian or Airy function, although precise measurements of $B$ can become important when attempting to localise FRBs detected in multiple beams, or estimating the relative rate of single- vs multiple-beam detections \citep{Vedantham2016,Macquart2018b}.

For the purpose of estimating the number of FRBs detected however, the `inverse beamshape', $\Omega(B)$, which describes the amount of sky $\Omega$ viewed at any given sensitivity $B$, becomes more relevant \citep{James2019cSensitivity}. Most calculations of FRB rates have characterised a telescope beam as viewing out to the FWHM at full sensitivity, i.e.\ $\Omega(B) = \Omega_{\rm FWHM} \delta(B-1)$ \citep[e.g.][]{Thornton2013,Bhandarietal2018}. Others have used a Gaussian approximation for the beamshape \citep[e.g.][]{Lawrence2017_poisson}, which is equivalent to $\Omega(B) d \log B = {\rm const}$. We here analyse the sufficiency of these approximations, using for the Gaussian approximation $\sigma=({\rm FWHM}/2) (2 \log 2)^{0.5}$, where the full width at half maximum (FWHM) assumes an Airy disk, i.e.\ HPBW=$1.22 \lambda/D$ for wavelength at central frequency $\lambda$ and dish diameter $D$.

ASKAP FRB observations have varied the observation frequency and configuration of beams formed from ASKAP's phased array feeds (PAFs). However, the majority of both fly's eye and incoherent sum observations have used the `closepack36' configuration at a central frequency of 1.296\,GHz. We therefore use the beamshape derived in \cite{James2019cSensitivity}. In the case of the Parkes multibeam, we use a central frequency of 1.382\,GHz, and the simulations of K.~Bannister \citep[published as][]{Vedantham2016} and A.~Dunning \citep[referenced as `private communication' by ][]{Macquart2018a}), which produce equivalent results for $\Omega(B)$. This also allows us to conclude that while the ability to localise FRBs detected in multiple beams may be limited by systematic uncertainties in the beamshape \citep{Macquart2018a}, the inverse beamshape $B(\Omega)$ is robust against such certainties, since it does not care about where on the sky any given patch of sensitivity is located.

\begin{figure}
    \centering
    \includegraphics[width=\columnwidth]{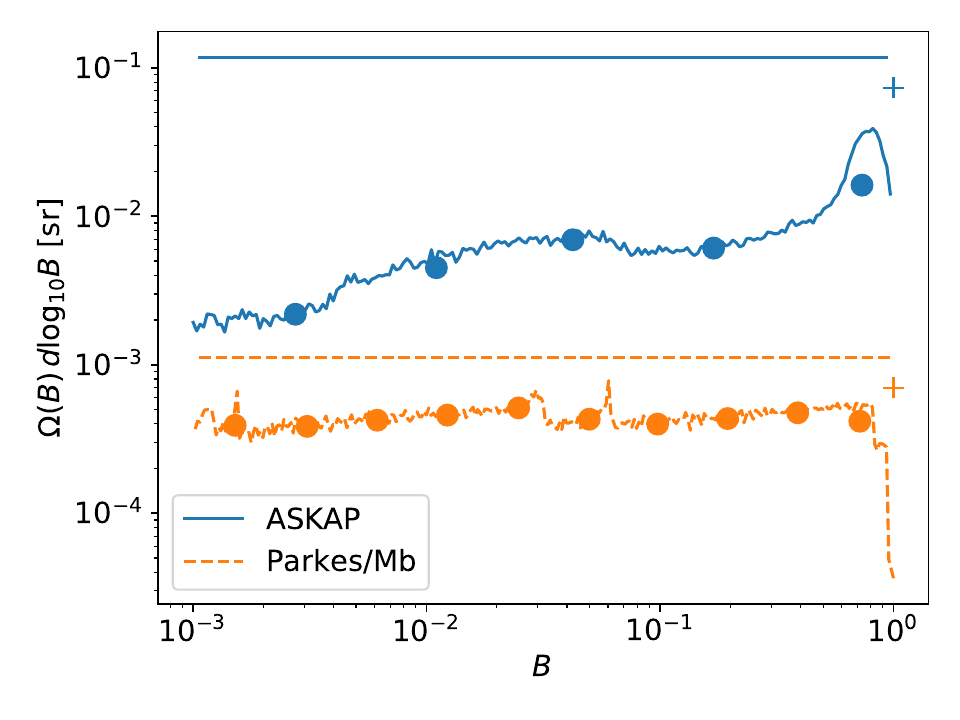}
    \caption{Inverse beamshape $\Omega(B)$ as a function of beam sensitivity $B$ for ASKAP (closepack36 configuration at 1.296\,GHz) and the Parkes Multibeam (1.382\,GHz). Shown are the best measurements (lines), FWHM approximation (crosses), the numerical approximations used in this work (circles), and the Gaussian beamshape (flat horizontal lines). The points for the FWHM and numerical approximations represent $\delta$ functions --- in the FWHM case, the true amplitude is shown, while in the numerical approximations, the amplitudes are renormalised by their spacing in $\log B$ for comparability with the best measurements.}
    \label{fig:omega_B}
\end{figure}

\begin{figure}
    \centering
    \includegraphics[width=\columnwidth]{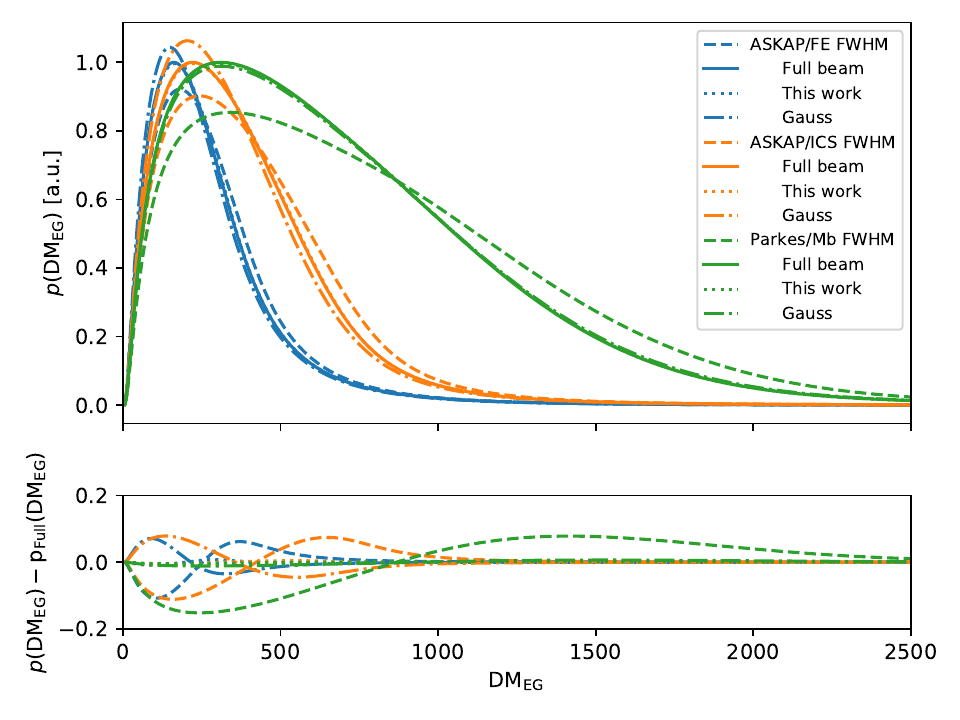}
    \caption{
    Top: dependence of expected FRB DM distributions (pc\,cm$^{-3}$) on beamshape models for three different FRB surveys. The beamshape models considered are shown in Figure~\ref{fig:omega_B}: the`FWHM' approximation, the `Full' beamshape $\Omega(B)$, a numerical approximation used in `this work', and a Gaussian beamshape. All curves for each telescope are normalised such that the distribution for the full beamshapes peaks at unity. Bottom: the difference between the `Full' beamshape and that found when using the `FWHM', `Approx.', and Gaussian approximations.}
    \label{fig:beam_approx}
\end{figure}

Figure \ref{fig:omega_B} shows the resulting `inverse beamshape' function $\Omega(B)$. This is compared to the equivalent $\Omega(B)$ when using the Gaussian and $\Omega_{\rm FWHM}$ approximations. Since implementing the full function $\Omega(B)$ in the calculation of the z--DM distribution is numerically expensive, we investigate the accuracy of reducing $\Omega(B)$ to a small number of values.
A set of such values is also shown in Figure~\ref{fig:omega_B}. The accuracy of all approximations is assessed against the full beamshape, by comparing the total predicted number of events and the mean value of DM to those calculated for the full beamshape, and also assessing the maximum difference between the $p({\rm DM})$ curves, as shown in Figure~\ref{fig:beam_approx}. Table~\ref{tab:beam_approx} lists the resulting errors. This is evaluated for the best-fit set of parameters found in our accompanying paper --- however, a brief investigation has shown that results are not sensitive to the assumed parameters within a reasonable range.

\begin{table} 
    \centering
    \caption{Percentage errors in the total FRB rate and mean DM value,  $\overline {\rm DM}$, and maximum deviation $|\delta {\rm DM}|_{\rm max}$, when using different beam approximations when compared to that found for the full beamshape function $\Omega(B)$, for each of three FRB surveys considered.}
    \label{tab:beam_approx}
    \begin{tabular}{c c |c c c}
    \hline
   Survey &  Approximation & Rate & $\overline {\rm DM}$ & $|\delta {\rm DM}|_{\rm max}$\\
   \hline
\multirow{3}{*}{ASKAP/FE}      & FWHM  & +76 & +6 & 0.1 \\
       & This work & +17 & +0.2 & 0.009 \\ 
    &    Gauss & +513 & -3 & 0.07 \\
    \hline
\multirow{3}{*}{ASKAP/ICS}      & FWHM  & +73 & +8 & 0.11 \\
       & This work & +5.8 &  +0.2 & 0.005 \\ 
    &    Gauss & +440 &  -4 & 0.08\\
        \hline
\multirow{3}{*}{Parkes/Mb}      & FWHM  & +20 & +14 & 0.15  \\
       & This work & +2.8 & +0.2 & 0.005 \\ 
    &    Gauss & +160 & +1 & 0.011 \\
    \hline
    \end{tabular}
\end{table}

We find that using five values of $B$ for ASKAP, and ten for Parkes, achieves an approximate DM distribution with an error in $p({\rm DM})$ of less than 1\%, and an error in the mean value $\overline{\rm DM}$ of only 0.2\%. In contrast, using the FWHM approximation, which is the standard in the current literature, results in $\mathcal{O}\sim$10\% deviations in the DM distribution and its mean value, and pushes the mean towards higher values. The total expected detection rate found when using the FWHM approximation is almost double that when found when using the full ASKAP beam, but there is only a 20\% excess for Parkes. When assuming Gaussian beams, a huge excess in the total rate of ASKAP bursts is predicted, since this does not account for the closely packed, and thus overlapping, beams. We note that uncertainties in the true ASKAP beamshape due to the calibration procedure \citep[see][]{James2019cSensitivity} are less than the errors introduced by our numerical approximation.

In the case of Parkes/Mb, the excess rate when using a Gaussian beam is due to outer beams being less sensitive than the central beam at which the sensitivity is usually calculated. However, the Gaussian beam approximation accurately estimates $\overline{\rm DM}$ and the shape of $p({\rm DM})$ distribution. This suggests that even very complex beamshapes, such as that of the Canadian Hydrogen Intensity Mapping Experiment (CHIME), could be included in our model in a relatively simplified manner.

\subsection{Detection efficiency}
\label{sec:efficiency}

We model the threshold at which an FRB of fixed fluence $F$ can be detected as scaling with the square root of its effective width, $w_{\rm eff}$, relative to the nominal width of 1\,ms, using an efficiency factor $\eta$:
\begin{eqnarray}
\eta & \equiv & \frac{F_1}{F_{\rm th}(w_{\rm eff})} \\
& = & \left( \frac{w_{\rm eff}}{1\,{\rm ms}}\right)^{-0.5}. \label{eq:eta}
\end{eqnarray}
The effective width is modelled as per \citet{Cordes_McLaughlin_2003}, being a function of its intrinsic duration $w_{\rm int}$, scattered width $w_{\rm scat}$, DM smearing within each frequency channel $w_{\rm DM}$, and the time-resolution of the search $w_{\rm samp}$:
\begin{eqnarray}
w_{\rm eff} & = & \sqrt{w_{\rm int}^2 + w_{\rm scat}^2 + w_{\rm DM}^2 + w_{\rm samp}^2}. \label{eq:effective_width}
\end{eqnarray}
Often, the scattered width and intrinsic width are indistinguishable, and their separation only becomes important for telescopes observing at different frequencies. We therefore define the `incident' width, $w_{\rm inc}$, as
\begin{eqnarray}
w_{\rm inc}^2 & = & w_{\rm int}^2 + w_{\rm scat}^2. \label{eq:incident_width}
\end{eqnarray}
An alternative model is presented by \citet{Arcus2020}, which is based on fits to simulated ASKAP and Parkes FRBs. Since it is not clear how the fit parameters translate to the general case, and 
because we wish to present a broadly scalable model, we do not use their formulation. We remark rather that the widely used model of Eq.~\ref{eq:effective_width} should be investigated further.

In order to model the distribution of $w_{\rm inc}$, $p(w_{\rm inc})$, we use a log-normal distribution,
\begin{eqnarray}
p(w_{\rm inc}) d w_{\rm inc} = \frac{w_{\rm inc}}{\log \sigma_w (2 \pi)^{0.5}} e^{- \frac{1}{2} \left( \frac{\log w_{\rm inc} - \log \mu_{w}}{\log \sigma_{w}} \right)^2}. \label{eq:widthlognormal}
\end{eqnarray}
We do not include any DM or $z$ dependence in the width distribution --- see Appendix~\ref{sec:scattering_appendix} for further discussion on this topic.

Unlike \citet{Luo2020}, we do not include fits of the model parameters $\mu_w$ and $\sigma_w$ as part of our general fitting process. Rather, we use the low correlation between $\mu_w$, $\sigma_w$ and other parameters to first fit for these values using a preliminary parameter set, and then check that the fit is still valid for the final parameter set presented in our companion paper \citep{James2021Lett}.

\begin{figure} 
    \centering
    \includegraphics[width=\columnwidth]{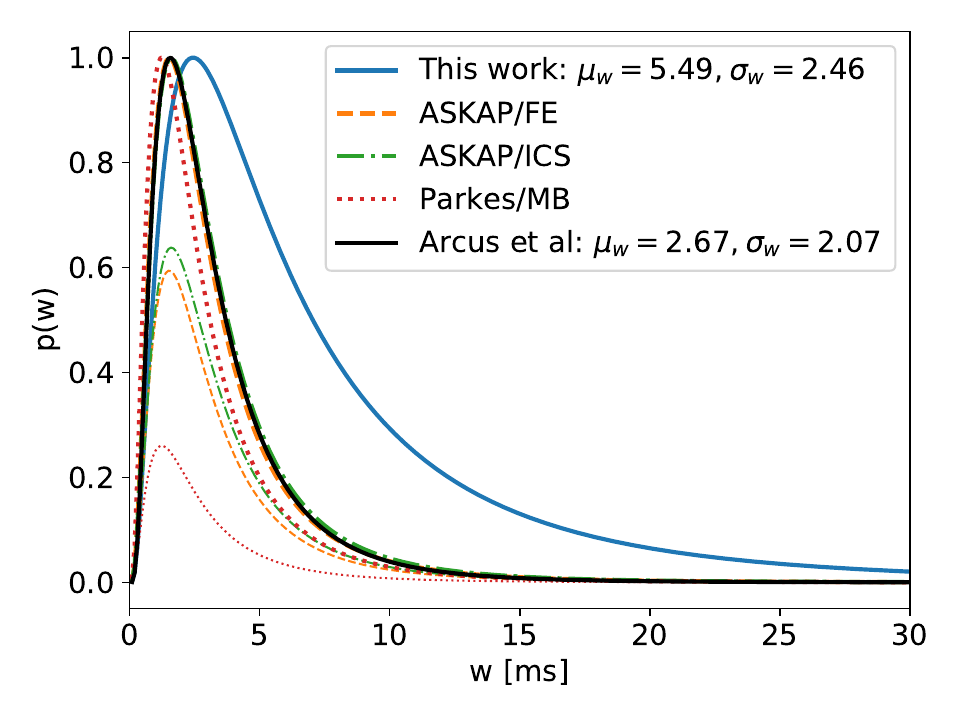}
    \caption{The effect of intrinsic burst width on survey sensitivity. Blue: intrinsic distribution of burst widths $w_{\rm int}$ (peak set to unity). Orange/green/red: the simulated width distributions expected from the three FRB surveys after accounting for observational bias.  Two normalisations are shown: the upper curves have been normalised to a peak probability of unity, for comparison with the fit to observed values by \citet{Arcus2020} (shown in black); while the lower curves are normalised relative to the rate for $w=0$, and thus illustrate the relative reduction in event rate as a function of intrinsic width. The upper curves for ASKAP/FE and ASKAP/ICS coincide almost identically with the results of \citet{Arcus2020}.}
    \label{fig:width_effect}
\end{figure}

\begin{table} 
    \centering
        \caption{The effects of different assumptions on the intrinsic burst width distribution at $1.3$\,GHz for three FRB surveys. Given are the total detection rate (normalised to a rate of unity at $w=0$), and mean DM $\overline{\rm DM}$. The width distributions are parameterised via Eq.~\ref{eq:widthlognormal}, with given values of $\mu_w$ and $\sigma_w$ corresponding to no intrinsic width, the observed width distribution of \citet{Arcus2020}, when accounting for observational bias to find the intrinsic distribution, and when numerically approximating ($\sim$) the intrinsic distribution with a few characteristic values.}
    \label{tab:widths}
    \begin{tabular}{r| c c | c c c }
    \hline
    Parameter       & $\mu_w$ & $\sigma_w$ & ASKAP/FE & ASKAP/ICS & Parkes \\
    \hline
\multirow{4}{*}{Rate}   & 0     & 0      &   1     & 1     & 1 \\
                    & 2.67  & 2.07 & 0.46  & 0.51  & 0.20 \\
                    & 5.49  & 2.46  & 0.27   & 0.30  & 0.11 \\
                    & $\sim$5.49 & $\sim$2.46  & 0.27   & 0.30  & 0.11 \\
            \hline
\multirow{4}{*}{$\overline{\rm DM}$}    & 0     & 0      &   263     & 371 & 488 \\
                    & 2.67  & 2.07  & 286  & 397 & 724 \\
                    & 5.49  & 2.46  & 293  & 401 & 726 \\
                    & $\sim$5.49  & $\sim$2.46  & 292 & 400 & 724 \\
                    \hline
    \end{tabular}
\end{table}

\begin{figure} 
    \centering
    \includegraphics[width=0.49\textwidth]{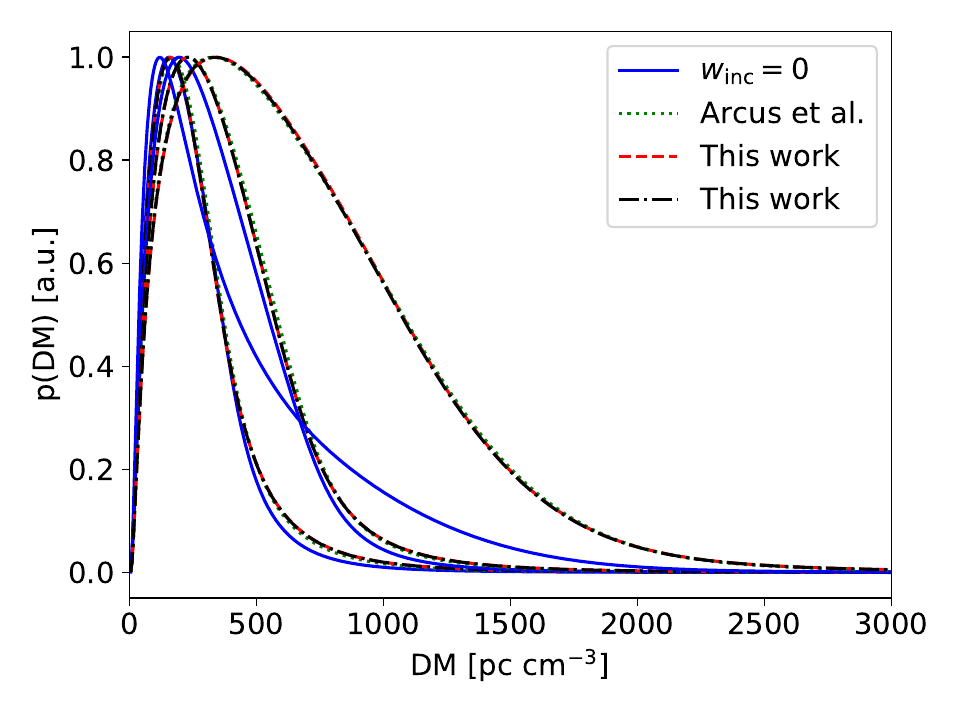}
    \caption{Effect of using different distributions of the intrinsic width $w_{\rm int}$ on the expected DM distribution of FRBs, $p({\rm DM})$, from the FRB surveys considered here: ASKAP/FE (left-most), ASKAP/ICS (centre), and Parkes/Mb (right-most). Assuming no intrinsic width (blue solid), using a log-normal distribution of widths with parameters from \citet{Arcus2020} (green dotted), using a log-normal distribution with parameters derived in this work (red dashed), and numerically approximating the latter (black dot-dashed). The distributions are normalised to a peak of unity for illustrative purposes.}
    \label{fig:dm_width}
\end{figure}

\citet{Arcus2020} use Eq.~\ref{eq:widthlognormal} to model the observed width distribution $p(w_{\rm inc})$ of ASKAP and Parkes FRBs, finding $\mu_w = 2.67$\,ms and $\sigma_w = 2.07$. We instead use Eq.~\ref{eq:widthlognormal} to model the intrinsic width distribution, and vary $\mu_w,\sigma_w$ until the simulated width distribution of the ASKAP/FE survey matches the parameterisation of observed widths by \citet{Arcus2020}. We obtain $\mu_w = 5.49$ and $\sigma_w = 2.46$, and then proceed to use these values in further calculations to optimise FRB population parameters. 
This is shown in Figure~\ref{fig:width_effect}, with total rates, and mean estimated DM values, given in Table~\ref{tab:widths}.
Finally, we re-evaluate the fits using the final optimal set of parameters we present in our companion paper \citep{James2021Lett}. 

Comparing the intrinsic (blue) and observed (black, coloured) distributions in Figure~\ref{fig:width_effect}, modelling the intrinsic FRB rate, and accounting for observational bias, correctly reproduces the observed FRB width distribution as estimated by \citet{Arcus2020}. The effect of observational bias is clearly reflected in the total expected FRB rate, with high-width bursts much less likely to be detected. The magnitude of this effect, shown in Table~\ref{tab:widths}, depends on the significance of the other terms in Eq.~\ref{eq:effective_width}. When the time resolution is poor ($w_{\rm samp}$ large), the effect of a large intrinsic width is less --- thus the reduction in rate for ASKAP is less than that for Parkes surveys. A second-order effect is that more-sensitive surveys, which probe further into the Universe and see FRBs with on-average higher DMs, are also less sensitive to $w_{\rm int}$, since $w_{\rm DM}$ is larger. Hence the reduction in rate for ASKAP/ICS is slightly lower than for ASKAP/FE, while the greatest effect is for the Parkes/Mb survey, where the intrinsic FRB width reduces the number of detected FRBs by a factor of 10.

A consequence of this is that the true number of high-width FRBs will be very difficult to estimate, so that our log-normal model is effectively untested beyond $w=10$\,ms. Thus while we estimate that ASKAP/FE and ASKAP/ICS miss $\sim70$\% of FRBs due to their intrinsic width, and Parkes/Mb 90\%, we do not consider this quantitatively reliable --- there may be virtually any number of high-width events remaining to be detected. We note that \citet{CHIME_catalog1_2021} find evidence for an intrinsically large population of FRBs with scattering width at 600\,MHz between 10\,ms and 100\,ms (approximately between 4\,ms and 40\,ms after scaling to ASKAP/Parkes frequencies using $\tau_{\rm scat} \sim \nu^{-4}$) which is also poorly constrained due to selection effects.

Whatever the lost rate, losses will preferentially arise from the nearby Universe where $w_{\rm DM}$ is low. Including the width distribution therefore increases the mean expected DM, $\overline{\rm DM}$. This effect is small ($\sim 10$\%) for ASKAP observations, but more significant for Parkes ($\sim 30$\%).

Finally, we note that while including the width distribution is clearly very important, the details matter less. Using the parameters of \citet{Arcus2020}, or approximating the true distribution with a small number of points for computational efficiency, produces an almost identical value of $\overline{\rm DM}$. This also means that the loss of efficiency to high-width bursts for the {\sc HEIMDALL}\footnote{ https://sourceforge.net/p/heimdall-astro/wiki/Home/} search pipeline found by \citet{Gupta2021} --- which is commonly used in Parkes FRB searches --- is insignificant to our modelling.

\subsection{Numerical implementation}

The integrals over $B$ and $\eta$ in Eq.~\ref{eq:dm_eta_integral} are numerically expensive. Furthermore, we have shown above that approximating the beamshape with 5--10 values, and the width distribution with five, allows for a very good approximation to the continuous distributions. Therefore, we approximate these continuous distributions with these discrete distributions, i.e.\
\begin{eqnarray}
\Omega(B) & \sim & \sum_i^{N_B} \Omega_i \delta(B-B_i)  \label{eq:discrete_B} \\
p(\eta(DM,w)) & \sim & \sum_i^{N_w} p_i \delta(\eta(DM,w_w)-\eta(DM,w_i)), \label{eq:discrete_eta}
\end{eqnarray}
recalling that $\eta$ is a function of both DM and $w$ through Eq.~\ref{eq:eta} and \ref{eq:effective_width}. Therefore, the z--DM distribution of Eq.~\ref{eq:dm_eta_integral} becomes a weighted sum over individual distributions at fixed observational thresholds,
\begin{equation}
p(z,{\rm DM}) = \sum_{i=1}^{N_B} \sum_{j=1}^{N_w} p\left(z,{\rm DM},F_{\rm th}(B_i, \eta({\rm DM},w_j)) \right) \Omega_i p_j. \label{eq:dm_eta_sum}
\end{equation}
This implementation is available at GitHub \citep{frb,zdm}.

%% file: Methodology.tex
\section{Methodology}
\label{sec:methodology}

The ingredients described above are implemented in {\sc Python}. Here, we describe its effective implementation, which is identical in approach to the method proposed by \citet{Connor2019} and implemented by \citet{Luo2020}, even if the exact details differ.

We consider FRB data from some number of independent FRB surveys. The total likelihood ${\mathcal L}$ of the outcome of multiple independent FRB surveys is simply a product of their individual likelihoods, ${\mathcal L}_i$,
\begin{eqnarray}
{\mathcal L} = \prod_i^{N_{\rm surveys}} {\mathcal L}_i, \label{eq:sum_over_surveys}
\end{eqnarray}
for surveys $i=1\ldots N_{\rm surveys}$. We describe ${\mathcal L}_i$ as the product of three independent terms:
\begin{eqnarray}
{\mathcal L}_i = p_n(N_i) \prod_{j=1}^{N_i} p_{\rm dmz}({\rm DM}_{j},z_{j}) p_s(s_{j}|{\rm DM}_{j}, z_{j}). \label{eq:sum_over_bursts}
\end{eqnarray}
Here, $p_n$ is the probability of detecting $N_i$ FRBs in survey $i$, while $p_{\rm dmz}$ is the likelihood of the $j^{\rm th}$ FRB from survey $i$ being detected with dispersion measure DM$_{j}$ and, when applicable, at redshift $z_{j}$. We also include the probability $p_s$ of an FRB being detected with fluence $F$ a factor $s$ above the fluence threshold $F_{\rm th}$ given it was observed with properties ${\rm DM}_{j},z_{j}$. Each of these terms is described independently below.

It is also possible to separate further terms in Eq.~\ref{eq:sum_over_bursts}. As noted by \citet{Vedantham2016}, for a multibeam instrument, the relative likelihood of a single- vs multi-beam detection, and the relative likelihood of detection in different beams of varying sensitivity, are functions of the FRB fluence distribution. Such measures are only relevant when the true FRB fluence $F$ cannot be resolved, as is common with FRB detections by the Parkes multibeam. When $F$ can be reconstructed, as is the case for CRAFT FRB detections with ASKAP \citep{Shannonetal2018}, then the survey acceptance to that particular FRB can be calculated exactly, and $p(s)$ in Eq.~\ref{eq:sum_over_bursts} can be written in terms of $p(F)$. Similarly, we will not include the observed value of an FRB's width when evaluating $p_{\rm dmz}$ and $p_s$, with only the overall width distribution being accounted for. We consider that adding such terms will yield only a small increase in analytic power for resolving the FRB population, at the cost of a large increase in complexity. Thus they are ignored --- however we do acknowledge that we are discarding a small amount of information by doing so.

\subsection{Probability of $N$ detections, $p_n(N)$}

Ignoring the correlations caused by repeating FRBs (see the discussion in Section~\ref{sec:repeater_discussion}), the total number of observed FRBs in survey $i$, $N_i$ comes from a Poisson distribution,
\begin{eqnarray}
p_n(N_i) & = & \frac{\left<N_i\right>^{N_i} \exp(-\left<N_i\right>)}{N_i!}, \label{eq:ni_given_n}
\end{eqnarray}
where $\left<N_i\right>$ is the expectation value of $N_i$. The calculation of $\left<N_i\right>$ is the heart of the problem that we address in this work, since it must necessarily incorporate all relevant properties that affect the detection rate.

Combining the dependencies in Sections \ref{sec:ingredients1}--\ref{sec:ingredients3}, $\left<N_i\right>$ is calculated as
\begin{eqnarray}
\left<N_i\right> & = & T_i R_i, \\
R_i & = & \int dz \, \Phi(z) \frac{d V(z)}{d \Omega dz} \int d\DM p(\DM |z) \label{eq:Ni}  \\
&& \int dB \, \Omega(B) \int dw p(w)  p(E>E_{\rm th}(B,w,z,\DM)). \nonumber
\end{eqnarray}
Here, $T_i$ is the survey duration, which multiplies a rate $R_i$ to produce the total expected number of bursts. $\Phi(z)$ is the FRB source evolution function (Eq.~\ref{eq:phiz}), $d V(z)/d\Omega/dz$ is the comoving volume per steradian per redshift interval from Eq.~\ref{eq:comoving_volume}, $p(\DM|z)$ is the extragalactic DM distribution found by convolving Eq.~\ref{eq:pcosmic} and \ref{eq:phost} (shown in Figure~\ref{fig:dm_z_basic}); $\Omega(B)$ is the beamshape discussed in Section~\ref{sec:beamshape} and approximated as per Eq.~\ref{eq:discrete_B}; $p(w)$ is the width distribution of Eq.~\ref{eq:widthlognormal}, discretised as per Eq.~\ref{eq:discrete_eta}; and $p(E>E_{\rm th})$ is the cumulative energy function of Eq.~\ref{eq:integral_energy_distribution}. The dependency of this threshold $E_{\rm th}$ on the parameters $(B,w,z,\DM)$ is encapsulated in Eq.~\ref{eq:F_to_E}, Eq.~\ref{eq:Fth}, and Eq.~\ref{eq:eta}.

For some surveys, no controlled survey time $T_i$ is available, and this term is simply set to unity in Eq.~\ref{eq:sum_over_bursts}. However, $R_i$ can be calculated regardless of knowledge of $T_i$. For those surveys with known $T_i$, the most likely value of the lead constant in the population function of Eq.~\ref{eq:sfr_n}, $\Phi_0$, can be estimated without recalculating the integral of Eq.~\ref{eq:Ni}.

\subsection{Calculating $p_{\rm dmz}$}

The probability of an FRB being observed with a given dispersion measure DM and redshift $z$ is given by the appropriate integrand of Eq.~\ref{eq:Ni},
\begin{eqnarray}
p_{\rm dmz} & = &  R_i^{-1} \Phi(z) \frac{d V(z)}{d \Omega dz} p(\DM |z) \int dB \Omega(B) \\
&& \int dw p(w)  p(E>E_{\rm th}(B,w,z,\DM)) \label{eq:pzdm}.
\end{eqnarray}
For FRBs with no measured host redshift, the relevant quantity is
\begin{eqnarray}
p_{\rm dm} & = & \int p_{\rm dmz} dz,
\end{eqnarray}
and $p_{\rm dm}$ replaces $p_{\rm dmz}$ in Eq.~\ref{eq:sum_over_bursts}. The rate $R_i$ is used as a normalising factor in Eq.~\ref{eq:pzdm}, so that
\begin{eqnarray}
\int \int p_{\rm dmz} d{\rm DM} dz = \int p_{\rm dm} d{\rm DM} =  1.
\end{eqnarray}
The shape of $p_{\rm dmz}$ in $z$--DM space is a primary quantity of interest in this work.

\subsection{Calculating $p_s({\rm s}$)}

The measured fluence $F$ of an FRB also holds information on the FRB population. However, in many telescope systems --- and notably for Parkes \citep{Macquart2018a} --- $F$ is not directly measured, since the location of the FRB in the beam is not known. Furthermore, we are interested in $p(F|F_{\rm th})$, i.e.\ the probability of measuring $F$ given an FRB has been detected at threshold $F_{\rm th}$, which itself has complex dependency through Eq.~\ref{eq:Fth}.

This difficulty can be overcome by noting that the signal-to-noise ratio, SNR, is a readily observable parameter for an FRB, and most FRB-hunting systems use a well-defined threshold SNR, SNR$_{\rm th}$, to distinguish FRBs from noise. As per \citet{Jamesetal2019a}, who base their work on \citet{Crawford1970}, we define
\begin{eqnarray}
s & \equiv & \frac{\rm SNR}{\rm SNR_{\rm th}} \label{eq:SNRratio} \\
& = & \frac{F}{F_{\rm th}} \label{eq:Fratio}
\end{eqnarray}
where $F_{\rm th}$ is the fluence threshold to that FRB. As detailed in Section~\ref{sec:ingredients3}, $F_{\rm th}$ is a function of the burst DM, width, and the location in which it is observed by the telescope's beam, so that neither term in Eq.~\ref{eq:Fratio} is known. However, the ratio is preserved in the measurable quantities of Eq.~\ref{eq:SNRratio}, making $s$ a very useful observable.

The probability $p(s)$ of observing $s$ in the range $s$ to $s+ds$ given that an FRB has already been observed is
\begin{eqnarray}
p_s(s|{\rm DM},z) & = & \frac{R_i^{-1}}{p_{\rm dmz}}  \Phi(z) \frac{d V(z)}{d \Omega dz} p(\DM |z)  \nonumber \\
&& \int dB \Omega(B) \int dw p(w) \frac{dp(s F_{\rm th})}{ds}, \nonumber \\
p_s(s|{\rm DM}) & = &  \frac{R_i^{-1}}{p_{\rm dm}} \int \Phi(z) \frac{d V(z)}{d \Omega dz} p(\DM |z) dz    \nonumber \\
&&  \int dB \Omega(B) \int dw p(w) \frac{dp(s F_{\rm th})}{ds}  \label{eq:psu}
\end{eqnarray}
for localised and unlocalised FRBs respectively. In the integrands, $p(s F_{\rm th})$ is the probability of detecting a fluence $F=s F_{\rm th}$ in an interval between $s$ and $s+ds$, where $F_{\rm th}$ depends on $B$, $w$, and ${\rm DM}$ as per Eq.~\ref{eq:Fth}.

The probability $p(s F_{\rm th})$ can be found from Eq.~\ref{eq:integral_energy_distribution}. Given that such an FRB has been observed at all, the integral distribution of $E$ given that an FRB has been detected can be found by replacing $E_{\rm th}$ with $E$ and $E_{\rm min}$ with $E_{\th}$. Differentiating by $E$ produces the probability amplitude
\begin{eqnarray}
\frac{dp(E_{\rm obs}|E_{\rm th})}{dE} & = & \frac{\gamma}{E_{\rm th}} \frac{ \left( \frac{E}{E_{\rm th}}\right)^{\gamma-1} }{1-\left(\frac{E_{\rm max}}{E_{\rm th}}\right)^\gamma} \nonumber\\
& = & \frac{\gamma}{E_{\rm th}} \frac{ s^{\gamma-1}}{1-\left(\frac{E_{\rm max}}{E_{\rm th}}\right)^\gamma}. \label{eq:differential_energy_distribution} 
\end{eqnarray}
The value of $E_{\rm th}$ can be found as a function of $z$ by inserting $F_{\rm th}$ into Eq.~\ref{eq:F_to_E}; thus $p(s F_{\rm th})$ is equivalent to $p(s E_{\rm th}(z,F_{\rm th}))$.
Relating $dp/dE$ from Eq.~\ref{eq:differential_energy_distribution} to the required $dp/ds$ of Eq.~\ref{eq:psu} produces
\begin{eqnarray}
\frac{dp(E_{\rm obs}|E_{\rm th})}{ds} & = & \frac{dp(E_{\rm obs}|E_{\rm th})}{dE} \frac{dE}{dF} \frac{dF}{ds} \\
& = & \frac{dp(E_{\rm obs}|E_{\rm th})}{dE} \frac{E_{\rm th}}{F_{\rm th}} F_{\rm th} \\
& = & \gamma \frac{ s^{\gamma-1} }{1-\left(\frac{E_{\rm max}}{E_{\rm th}}\right)^\gamma}. \label{eq:pEobsds}
\end{eqnarray}

Inserting Eq.~\ref{eq:pEobsds} into Eq.~\ref{eq:psu}, and integrating over the appropriate dimensions, produces the desired $p_s$.

\subsection{What about repeating FRBs?}
\label{sec:repeater_discussion}

By writing the individual burst probabilities as being independent in Eq.~\ref{eq:sum_over_bursts}, and assuming that the number of detected bursts follows a Poisson distribution in Eq.~\ref{eq:ni_given_n}, we ignore the potential of FRBs to repeat. While the fraction of the FRB population which is observed to repeat is a current topic of debate, it is certain that many do. Formally, the FRB population described in Section~\ref{sec:ingredients2} represents all \emph{bursts}, rather than all FRB emitters, and the summation of Eq.~\ref{eq:sum_over_bursts} runs over all detected bursts.
The distinction becomes irrelevant for distant, rarely repeating sources for which only ever zero or one bursts will be observed. For bright, nearby repeaters, the probability of having such an object in a survey's field of view is rare, especially when burstiness and/or periodicity is included \citep{Oppermannetal2018,CHIME2020repetition,2020_121102_periodicity1,2020_121102_periodicity2}, and an observation of zero bursts will be more likely than that estimated by Eq.~\ref{eq:ni_given_n}. Conversely, the probability of observing many bursts will also be high, with observations of single bursts being much rarer than otherwise expected. 

We note that the only FRB survey to ever observe an FRB repeat in an unbiased way are the observations by CHIME \citep{CHIME2019b,CHIME2019c,CHIME2020a} --- all other repeating FRBs have been discovered in targeted follow-up observations. This suggests that the majority of FRB observations can safely be classified as being in a `one burst per progenitor' regime, regardless of the true fraction which are actually repeating objects. For these, our approach should be valid. We revisit this assumption in Section~\ref{sec:explanation_repetiton}.

%% file: Surveys.tex
\section{Surveys}
\label{sec:survey_results}

Estimates of the FRB population have been made using data from many telescopes, which are often drawn from {\sc FRBCAT} \citep{frbcat}. Due to the large number of FRBs they have detected and published, results from Parkes and ASKAP remain the most important, and we focus on these instruments here. Other important instruments we wish to examine in future works include the Upgraded Molonglo Observatory Synthesis Telescope (UTMOST), and the  Canadian Hydrogen Intensity Mapping Experiment (CHIME).

The sensitivity of an FRB survey --- and hence the functions $p_N$, $p_{\rm zdm}$, and $p_s$ from Section~\ref{sec:methodology} --- depends on the local contribution to DM, and hence varies with the value of $\DMISM$. Since this fluctuates pointing-by-pointing, in theory these functions must be recalculated for every single pointing direction, which becomes computationally prohibitive. Evaluating $p_{\rm zdm}$ and $p_s$ however for the measured DM$_{\rm EG}$ and $s$ of an FRB is much quicker. This motivates grouping FRB observations not just by telescope, but also by other observational properties, such as Galactic latitude. Here, we use five groups, as described below.

\subsection{Parkes}

All Parkes FRBs published so-far have used the multibeam (`Mb') receiver \citep{staveley1996parkesMultibeam}. However, a new ultra-wideband receiver is now in place \citep{ParkesUWB}, which is being used for FRB searches and follow-up observations. We therefore refer to results from Parkes as ``Parkes/Mb'' to distinguish this from future works.

Of the many Parkes FRB discoveries, we consider only those by the High Time Resolution Universe \citep[HTRU;][]{Keith2010_HTRU,Thornton2013,Champion2016_HTRU,Petroff2014a} and Survey for Pulsars and Extragalactic Radio Bursts \citep[SUPERB;][]{SUPERB1,Bhandarietal2018} collaborations to have an unbiased estimate of observation time, $T_i$. This is because their observation time and pointing directions were pre-determined, and the results published regardless of outcome. Other results suffer from publication bias whereby non-detections are less likely to be published. Thus, while their discovery can contribute to individual FRB likelihoods via $p_{\rm dmz}$ and $p_s$, no well-defined observation time $T_i$ exists for use in Eq.~\ref{eq:Ni}, and the term $p_i(N_i)$ must be neglected in Eq.~\ref{eq:sum_over_bursts}. Thus they cannot contribute to estimates of the total FRB rate.

Since the distribution of DM$_{\rm ISM}$ affects telescope sensitivity, surveys at low Galactic latitudes have significantly reduced sensitivity compared to those at high latitudes, and the full distribution of time spent at each $\DMISM$ must be accounted for. In particular, $\DMISM$ will vary significantly for each pointing at low latitudes, making estimates numerically taxing. We therefore include only FRBs detected at mid ($19.5^{\circ} < |b| < 42^{\circ}$) and high ($42^{\circ} le |b| \le 90^{\circ}$) Galactic latitudes. This criteria leaves 12 FRBs detected in a total of 164.4\,days by HTRU and SUPERB \citep{Bhandarietal2018}, and another 8 FRBs by other groups with no reliable observation time. A full list is given in Table~\ref{tab:parkes_frbs}.

Early searches for FRBs with Parkes used a sparse grid of DMs and arrival times, resulting in sensitivities that would fluctuate by $\pm15$\% \citep{keane2015sensitivities}. This was corrected with the use of {\sc Heimdall}\footnote{http://sourceforge.net/projects/heimdall-astro/}, which has been used to (re)process the data from HTRU and SUPERB. While early HTRU searches extended only to DM=2000\,pc\,cm$^{-3}$ \citep{Thornton2013}, latter searches extended this to 5000\,pc\,cm$^{-3}$; and while the SUPERB `F' pipeline looks for FRBs with DM$\le2000$\,pc\,cm$^{-3}$, the SUPERB `T' pipeline extends the search to 10,000\,pc\,cm$^{-3}$. Thus we treat all Parkes FRB searches as fully covering DM space.

For the Parkes multibeam, nominal sensitivity to a burst at beam centre is $F_1=0.5$\,Jy\,ms to a 1\,ms duration burst \citep{SUPERB1} --- this is an approximation, since different FRB searches used slightly different values of SNR$_{\rm th}$. We neglect the effect of 1-bit sampling with early searches for FRBs with Parkes, which would have slightly degraded the sensitivity of these observations. These and other properties are summarised in Table~\ref{tab:parkes_frbs}.

Two FRBs deserve special mention. Both 010724 and 180309 saturated the primary beam, with their detection SNR being a lower limit on their true SNR in an idealised linear system. In the case of 010724, \citet{Lorimer2007} use the effect of this burst on the absolute normalisation of the data in the analog stage to estimate a flux of 40\,Jy over the time resolution of 5\,ms, i.e.\ a total fluence of 200\,Jy\,ms; we find their other estimate, of 20\,Jy using the multibeam beamshape, to be less reliable for reasons discussed in \citet{Macquart2018a}. Assuming a $1$\,Jy\,ms threshold, this yields $s=200$ for a 5\,ms burst. For 180309, \citet{Oslowski2019} find a SNR of 2416 using data from a simultaneous pulsar folding system, yielding $s=241.6$. Both these numbers have sizeable errors on them, however to leave out the most powerful FRBs would skew the data. An alternative would be to allow lower limits on $s$ to be included in the likelihood analysis, however this has not yet been implemented.

\begin{table*}
    \centering
     \caption{Observational properties of follow-up observations, for ASKAP Fly's Eye (FE) and incoherent sum (ICS) modes \citep{Shannonetal2018,Bannister2019}; the Parkes multibeam (MB) receiver \citep{SUPERB1}; and the Greenbank Telescope's (GBT's) 820\,MHz primary focus and L-band receivers \citep{Kumar2019}. From left to right: the telescope and receiver names, the central frequency $\bar{\nu}$ and total bandwidth $\Delta \nu$, time- and frequency- resolutions $\delta t$ and $\delta \nu$, and nominal sensitivity to a 1\,ms duration burst.}
    \begin{tabular}{c c c c c c c c}
        \hline
    Telescope    & $N_{\rm FRB}$ & $T_{\rm obs}$ [hr] & $\bar{\nu}$   & $\Delta \nu$  & $\delta t$    & $\delta \nu$  & $F_1$ \\
            &    & Mode   & [MHz]         & [MHz]         &   [ms]          & [MHz]           & [Jy\,ms] \\
       \hline
       ASKAP/FE   &  20 & 26,616  & 1296       & 336         & 1.2565        & 1             & 21.9 \\
       ASKAP/ICS   & 7  & N/A  & 1315   & 336       & 1.2565           & 1             & 4.4 \\
       Parkes/Mb   & 13  & 3,946  & 1382         & 337.1            & 0.064         & 0.39          & 0.5 \\
       \hline
    \end{tabular}
    \label{tab:survey_properties}
\end{table*}

\subsection{ASKAP}

The Commensal Real-time ASKAP Fast Transients (CRAFT) group have performed several FRB surveys with ASKAP. The majority of ASKAP FRBs have been observed in single-antenna (``Flye's Eye'', or ``FE'') mode during the `lat50' survey, i.e.\ while observing Galactic latitudes of $|b|=50^{\circ} \pm5^{\circ}$  \citep{Bannisteretal2017,Shannonetal2018}. Twenty FRBs have been initially reported \citep{Macquart2019a,Bhandarietal2019}, with a total recorded data time of $1274.6$ antenna-days duration \citep{James2019cSensitivity}. A further six FRBs have been detected in a variety of surveys \citep{Macquart2019a,Bhandarietal2019,Qiu2019}, of which four satisfy the Galactic latitude requirement. These are listed in Table~\ref{tab:ASKAP_FE_FRBs}. We err on the side of caution and do not assume a known observation time for this last category, since several other FRB searches outside the lat50 survey have been performed and were not reported. As with Parkes, $p_i(N_i)$ in Eq.~\ref{eq:sum_over_bursts} is only evaluated for the former category.

All ASKAP/FE searches have used the same time/spectral resolutions and almost identical frequency ranges, as given in Table~\ref{tab:survey_properties}. The beamshape and threshold for this survey are given in \citet{James2019cSensitivity}.

ASKAP has recently been observing in incoherent sum mode (ICS), with voltage buffers used in offline analysis to localise FRBs to sub-arcsecond precision \citep{Bannister2019}. Follow-up observations with radio and optical instruments have determined the redshifts of the host galaxies of each FRB, allowing the DM--$z$ grid to be directly probed for the first time. This mode has undergone an extended period of commissioning, with the number of telescopes, observation frequency, and time resolution of the search all varying. The total observation time is difficult to estimate, again precluding the use of observation time in this survey's likelihood calculation. A comprehensive analysis would involve recalculating the $z$--DM grid for each and every observed burst. For reasons of computational efficiency, we instead use mean observation parameters to evaluate the likelihood on this grid. This precludes the use of FRB~191001, which was detected at a lower frequency during commensal observations \citep{Bhandari2020b}. The remaining seven FRBs used are given in Table~\ref{tab:ASKAP_ICS_FRBs}.

\begin{table}
\centering
\caption{Properties of fast radio bursts detected by the Parkes multibeam system, and used in this work. Given is the original FRB designation; measured total DM and DM$_{\rm ISM}$ estimated by the NE2001 \citep{CordesLazio01} in pc\,cm$^{-3}$, and ratio of measured to threshold SNR. Entries with a `$^*$' indicates that this value is approximate; `$^\dag$' are discussed in text.} \label{tab:parkes_frbs}
\begin{tabular}{c c c c c}
\hline
\multicolumn{5}{c}{Parkes: total $T_{\rm obs}=164.4$\,days} \\
\hline
FRB & DM & DM$_{\rm ISM}$ & s & Ref.\ \\
\hline
110214 & 168.9 & 32  & 1.44 & \citet{Petroff_110214_2019} \\
\hline
110220 & 944.4 & 36 & 5.44 & \multirow{4}{*}{\citet{Thornton2013}} \\
110627 & 723 & 48 & 1.22 & \\
110703 & 1103.6 & 33 & 1.78 & \\
120127 & 553.3 & 33 & 1.22 & \\
\hline
090625 & 899.55 & 32 & 2.8 & \multirow{5}{*}{\citet{Champion2016_HTRU}} \\
121002 & 1629.18 & 72 & 1.6 & \\
130626 & 952.4 & 65 & 2 & \\
130628 & 469.88 & 52 & 2.9 & \\
130729 & 861 & 32 & 1.4 & \\
\hline
151230 & 960.4 & 48 & 1.7 & \multirow{2}{*}{\citet{Bhandarietal2018}} \\
160102 & 2596.1 & 36 & 1.6 & \\
\hline
\multicolumn{5}{c}{Parkes: unnormalised observation time} \\
\hline
FRB & DM & DM$_{\rm ISM}$ & s & Ref.\ \\
\hline
010305 & 350 & 44 & 1.02$^*$ & \citet{Zhang2020_010305} \\
010312 & 1187 & 51 & 1.1$^*$ & \citet{Zhang2019_010312} \\
010724 & 375 & 45 & 200$^{\dag}$ & \citet{Lorimer2007} \\
131104 & 779 & 71 & 3.06 & \citet{Ravi2015} \\
140514 & 562.7 & 36 & 1.6 & \citet{Petroff2015} \\
150807 & 266.5 & 38 & 5$^*$ & \citet{RaviScience2016} \\
\hline
180309 & 263.52 & 46 & 241.6$^{\dag}$ & \multirow{2}{*}{\citet{Oslowski2019}} \\
180311 & 1570.9 & 46 & 1.15 &  \\
\hline
\end{tabular}
\end{table}

\begin{table}
\centering
\caption{As per Table~\ref{tab:parkes_frbs}, for ASKAP/FE observations.} \label{tab:ASKAP_FE_FRBs}
\begin{tabular}{c c c c c}
\hline
\multicolumn{5}{c}{ASKAP/FE: $T_{\rm obs}=1274.6$\,days} \\
\hline
FRB & DM & DM$_{\rm ISM}$ & s & Ref.\ \\
\hline
170107 & 609.5 & 37 & 1.68 & \citet{Bannisteretal2017} \\
\hline
170416 & 523.2 & 40 & 1.38 & \multirow{19}{*}{\citet{Shannonetal2018}}\\
170428 & 991.7 & 40 & 1.11 & \\
170707 & 235.2 & 38.5 & 1.00 & \\
170712 & 312.8 & 35.9 & 1.34&  \\
170906 & 390.3 & 38.9 &1.79 & \\
171003 & 463.2 & 40.5 &1.45 & \\
171004 & 304.0 & 38.5 &1.15 & \\
171019 & 460.8 & 37 & 2.46 &  \\
171020 & 114.1 & 38.4 &2.05 & \\
171116 & 618.5 & 35.8 &1.24 & \\
171213 & 158.6 & 36.8 &2.64 & \\
171216 & 203.1 & 37.2 &1$^*$ & \\
180110 & 715.7 & 38.8 &3.75 & \\
180119 & 402.7 & 35.6 &1.67 & \\
180128.0 & 441.4 & 32.3 & 1.31 & \\
180128.2 & 495.9 & 40.5 & 1.01&  \\
180130 & 343.5 & 38.7 & 1.08 & \\
180131 & 657.7 & 39.5 & 1.45 & \\
180212 & 167.5 & 30.5 & 1.93 & \\
\hline
\multicolumn{5}{c}{ASKAP/FE: unnormalised time} \\
\hline
FRB & DM & DM$_{\rm ISM}$ & s & Ref.\ \\
\hline
180417 & 474.8 & 26.1 & 1.84 & \citet{Agarwal2019_Virgo} \\
180515 & 355.2 & 32.6 & 1.27 & \citet{Bhandarietal2019} \\
\hline
180324 & 431 & 64 & 1.03 &  \multirow{2}{*}{\citet{Macquart2019a}} \\
180525 & 388.1 & 30.8 & 2.88 & \\
\hline
\end{tabular}
\end{table}

\begin{table}
\centering
\caption{As per Table~\ref{tab:parkes_frbs}, for ASKAP/ICS observations, with redshift $z$ included.} \label{tab:ASKAP_ICS_FRBs}
\begin{tabular}{c c c c c c }
\hline
\multicolumn{5}{c}{ASKAP/ICS: unnormalised time} \\
\hline
FRB & DM & DM$_{\rm ISM}$ & s & z & Ref.\ \\
\hline
180924 & 362.4 & 40.5 & 2.34 & 0.3214 & \citet{Bannister2019} \\
181112 & 589.0 & 40.2 & 2.14 & 0.4755 &  \citet{Prochaska2019} \\
\hline
190102 & 364.5 & 57.3 & 1.38 & 0.291 & \multirow{4}{*}{\citet{Macquart2020}}\\
190608 & 339.5 & 37.2 & 1.79 & 0.1178 & \\
190611.2 & 322.2 & 57.6 & 1.03 & 0.378 &  \\
190711 & 594.6 & 56.6 & 2.64 & 0.522 & \\
\hline
190714 & 504.7 & 38.5 & 1.19 & 0.209 & \citet{Heintz2020} \\
\hline
\end{tabular}
\end{table}

%% file: Results1_initial.tex
\section{Initial results}
\label{sec:results1}

\subsection{Calculations}

We use a brute-force approach to find the best-fit values of  $E_{\rm max}, \alpha, \gamma, n, \mu_{\rm host},\sigma_{\rm host}$, and evaluate Eq.~\ref{eq:sum_over_surveys} over a multi-demensional cube of parameter values. The resulting likelihood dependence on each is calculated for single (pairs of) parameters by marginalising over the remaining five (four) parameters. That is, the value taken is the maximum likelihood found when the remaining five (four) values are varied over their full range.

In this work, we use a frequentist approach to setting confidence limits. Confidence intervals are determined using Wilks' theorem,
\begin{eqnarray}
2 \left( \log L_{\rm max} - \log L \right) & \sim & \chi^2_{\rm ndf} \label{eq:Wilks}
\end{eqnarray}
where $\chi^2_{\rm ndf}$ is a chi-square distribution with ndf degrees of freedom, here equal to the number of parameters which have not been marginalised over \citep[either one or two throughout this work;][]{Wilks62,Frederick2006}. The 51 FRBs used in this work should satisfy the large-N limit required for Eq.~\ref{eq:Wilks} to be valid. We do however investigate the validity of this test in an extreme case in Appendix \ref{sec:mc}, and generate Bayesian confidence intervals in Appendix \ref{sec:bayesian}, which turn out to be narrower than those calculated using a frequentist approach.

\subsection{Parameter degeneracy}
\label{sec:degeneracy}

\begin{figure}
    \centering
    \includegraphics[width=0.49\textwidth]{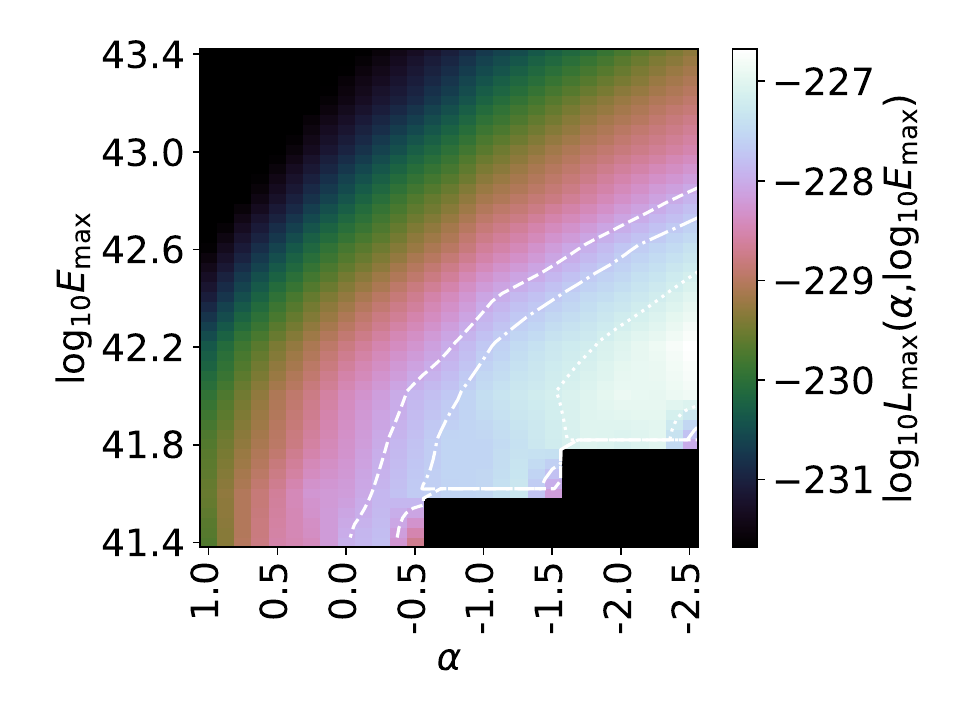} \\
    \includegraphics[width=0.49\textwidth]{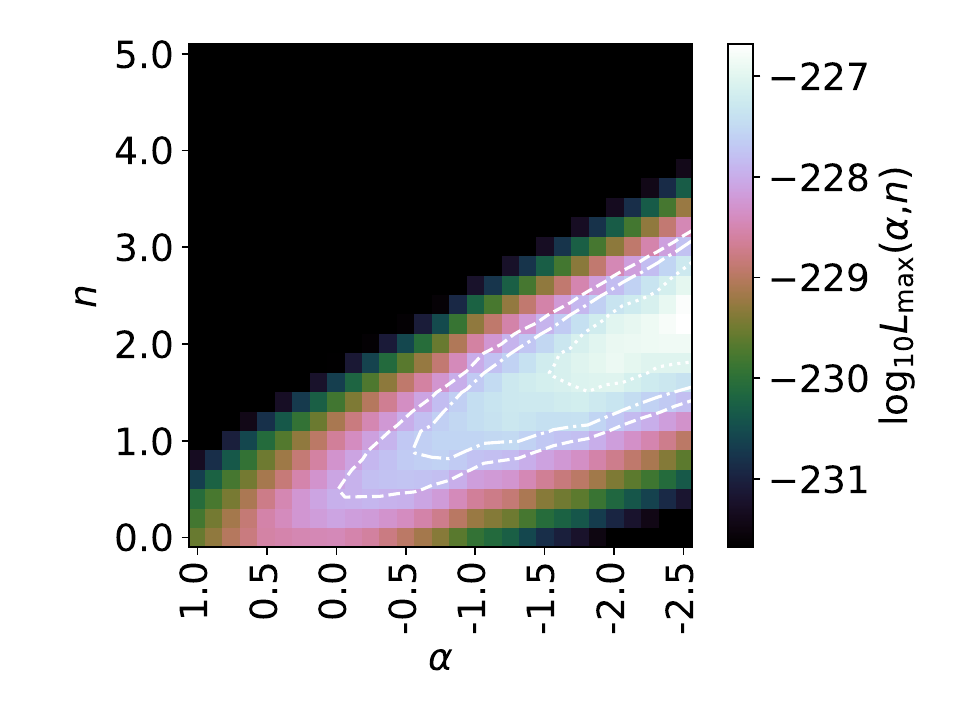}
    \caption{Degeneracy of the marginalised likelihood with $\alpha$. Top: dependence of $E_{\rm max}$, bottom: dependence on $n$. The contours give the 68\%, 90\%, and 95\% confidence limits. The step-like behaviour in $E_{\rm max}$ is due to coarse gridding.}
    \label{fig:degeneracy}
\end{figure}

Fits to data taken at a single frequency suffer from a degeneracy in the fitting parameters between $E_{\rm max}$, $\alpha$, and $n$. This is shown in Figure \ref{fig:degeneracy}, which plots the  variation of the marginalised likelihood over these parameters. The data shows a small preference for large, negative values of $\alpha$ ---
preliminary investigations covered $-8 \le \alpha \le 0$ --- with no distinct maximum being found. For these final computations, we restrict $\alpha$ to the range $-2.5 \le  \alpha \le 1$ for reasons which will be discussed below.

The reason for this degeneracy is that these three parameters are strongly related to the high-DM, high-z cut-off in the observed FRB distribution. This is restricted by the lack of ASKAP/ICS-localised FRBs at high redshifts, and by the number of high-DM FRBs detected by ASKAP/FE and Parkes multibeam observations. Having a steep spectral index (very negative value of $\alpha$) provides a mechanism to reduce the expected number of FRBs via the k-correction, and is both consistent with, and requires, a higher value of $E_{\rm max}$. Similarly, it also allows for a rapidly evolving population with redshift (high $n$), due to suppression by the k-correction. A very large number of high-$z$ events, or large value of $E_{\rm max}$, is excluded when $\alpha$ is near zero.
This degeneracy was also noted by \citet{Lu2019}, who by default use $\alpha=-1.5$.

Without further data at different frequencies, or a prior on any of these three parameters, this degeneracy cannot be broken. All results from the literature on $E_{\rm max}$ and $n$ derive from similar analyses as in this work, albeit with simpler methods, and are therefore not independent. We therefore investigate other evidence for plausible values of $\alpha$, in order to break this near-degeneracy.

There are two broad methods for measuring $\alpha$: searching for spectral dependence in the FRBs detected by a single experiment, and comparing FRB rates between experiments with different frequencies. We discuss results from both below.

\subsection{Spectral dependence from ASKAP FRBs}

In an analysis of 23 ASKAP FRBs detected in the 1147--1483 MHz band, \citet{Macquart2019a} found the summed power was consistent with $\alpha=-1.5^{+0.2}_{-0.3}$.\footnote{The error is in fact almost symmetric, with single-digit rounding resulting in the apparently large asymmetry.} This measurement was possible because of the closely packed ASKAP beams, which allowed the FRB location within the beam to be determined, and hence spectral corrections to be applied due to the frequency-dependent beamshape.

It has been pointed out by the referee of this work that while \citet{Macquart2019a} applied these spectral corrections to observed FRBs, they did not account for such effects to model the FRBs they did not observe. In the spectral index interpretation of $\alpha$, there is no such effect --- all FRBs are broadband, and every measured FRB tends to have the same spectral index. Thus it is impossible to detect a biased fraction, and we use a Gaussian prior of $\alpha=-1.5 \pm 0.3$ for the spectral interpretation of $\alpha$.

In the rate interpretation however, where every FRB is narrow-band, each burst is observed according to the sensitivity at its particular frequency. For experiments with widely spaced beams, the observed sky area usually scales as $\nu^{-2}$, so that far-more low-frequency bursts would be observed even for a frequency-independent rate. A similar effect, but less extreme, would be observed for an intermediate case where FRBs have a distribution of spectral indices that average to $\alpha$, and redder FRBs would preferentially be detected.

The measurements of \citet{Macquart2019a} were made with ASKAP, with closely spaced beams overlapping at the half-power points. To estimate the maximum possible bias from this effect, we perform the following toy calculation. We use a Euclidean source-counts such that rate is proportional to beam power $B^{1.5}$, and assume that ASKAP beams are Gaussian and overlap at half-max near band centre, with width scaling inversely with the frequency range of Table~\ref{tab:survey_properties}. We calculate the frequency-dependent observed rate assuming all FRBs have a very narrow bandwidth (i.e.\ under the rate interpretation), with a true $\alpha=0$. We find this effect would result in 25\% more FRBs at the low-frequency end than the high-frequency end of the ASKAP band, so that one would measure $\alpha=-0.85$, despite the true value being $\alpha=0$ --- i.e.\ the measured value needs to have $0.85$ added to it to to obtain the true value. 
Applying this correction to the results of \citet{Macquart2019a} in the rate interpretation, we use a Gaussian prior of $\alpha=-0.65 \pm 0.3$.

\subsection{Spectral dependence from cross-experimental rates}
\label{sec:boxcar_prior}

FRB spectral dependence can also be derived by comparing detections and non-detections between different instruments, with several results published in the literature.

\citet{Sokolowskietal2018} performed shadowing observations of ASKAP with the  Murchison Widefield Array (MWA) at 170--200\,MHz, with their non-detections being inconsistent with the high fluences predicted by $\alpha=-1.5^{+0.2}_{-0.3}$. Those authors suggest possible explanations as a higher (less negative) value of $\alpha$; missing the emission due to the small 30\,MHz bandwidth and low FRB band occupancy; synchrotron self-absorption in the emission region; and scatter broadening at low frequencies.

Using the then-non-detection of FRBs at 350\,MHz by the Robert C. Byrd Green Bank Telescope, \citet{Chawla_Greenbank_2017} estimate $\alpha > -0.3$. However, the Greenbank result is somewhat dependent on the assumed Parkes rate and FRB scattering distribution, with a constraint of $\alpha > -0.9$ reached under different assumptions --- compatible with the adjusted \citet{Macquart2019a} value in the rate interpretation. \citet{Farah2019} find a relatively low FRB rate using the Upgraded Molonglo Observatory Synthesis Telescope (UTMOST) at 843\,MHz. A value of $\alpha > 0$ would be require for compatibility with the Parkes and ASKAP rates. \citet{Gupta2021} have since shown however that the UTMOST pipeline would tend to under-estimate the SNR of FRBs, which is a likely explanation for the low rate. \citet{CHIME_catalog1_2021} present and analyse the spectral shapes of more than 500 bursts detected by the Canadian Hydrogen Intensity Mapping Experiment (CHIME) FRB system. Their implied all-sky rate is compatible with that of Parkes and ASKAP when assuming $\alpha=0$, although the range of compatible values is not greatly studied, and effects due to different detection efficiencies are not considered.

These results all point to a less negative value of $\alpha$ than found by \citet{Macquart2019a}. We therefore also consider a broad uniform (`boxcar') prior of $-2.5 < \alpha < 1.0$. However, as will be discussed below, we do not consider this prior as equally valid as the Gaussian prior, and thus by default quote the former.

\subsection{Validity of the priors}

We caution against taking the uniform prior as equally valid as the Gaussian prior from the measurements of \citet{Macquart2019a}.  Firstly, of the Parkes, ASKAP, UTMOST, Greenbank, and CHIME rates used by the studies in Section~\ref{sec:boxcar_prior}, only those by CHIME have accounted for telescope beamshape, although beamshape effects are relatively small in ASKAP due to the closely packed beams. Secondly, different telescopes have different response functions to FRB width and DM, and scattering is greater at lower frequencies than at high frequencies. This effect will, in general, reduce the measured rate for low-frequency systems; for a given measured rate, the true rate will be higher, and hence the true value of $\alpha$ will be lower (or more negative) than implied by count rates alone. Thirdly, we emphasise that ASKAP FRBs are absolutely not in the narrow-bandwidth limit used to estimate the bias in the result of \citet{Macquart2019a}. \citet{CHIME_morphology_2021} find 30\% of FRBs to have broadband structure, 60\% to be band-limited, and 10\% to have complicated morphology, indicating that the truth lies in-between the two extreme interpretations of $\alpha$ --- and while the bursts detected by ASKAP do show narrow-band features, these tend to be spread across the spectrum \citep{Shannonetal2018}. Finally, it is also possible that the FRB spectrum is flat below $1$\,GHz, and steep above it --- and all the results discussed in Section~\ref{sec:boxcar_prior} are obtained below $1$\,GHz, and may therefore be considered irrelevant.

Another interesting effect is that all experiments modelled in this work have broader beams at low frequency than at high. Thus, and remaining bias in $\alpha$ due to low-frequency FRBs being more likely to be detected would, to first order, be cancelled when incorporating that same increased chance of detection into FRB beamshape modelling. Indeed, the effect would go in the opposite direction: since Parkes has a wider beam spacing, the positive detection bias to low-frequency FRBs would be even greater than that in the ASKAP measurement of $\alpha$, so that while the true value of $\alpha$ might be less negative than the one found by \citet{Macquart2019a}, the applicable value of $\alpha$ to Parkes measurements may be more negative. Such a difference is clearly a second-order effect however.


\subsection{Comparison of results}

The limits on single FRB population parameters presented in \citet{James2021Lett} are obtained with this approach. We observe that while our prior on $\alpha$ shifts the preferred values of the other parameters by small amounts, it is not a large influence, and primarily acts to limit very strong source evolution models with $n>3$, and no to negative evolution with $n \le 0$.
Table~\ref{tab:other_results} compares these results with a prior on $\alpha$ to those of other authors, as well as a brief summary of which effects are included.

\begin{table*} 
\renewcommand{\arraystretch}{1.3}
    \centering
    \begin{tabular}{c|c c c c c c | c c c c }
        Author & $E_{\rm max}$ & $\alpha$ & $\gamma$ & $n$ & $\mu_{\rm host}$ & $\sigma_{\rm host}$ & Beam & DM|z & $\eta({\rm DM})$ & $f(w)$ \\
        \hline
       \citet{Lu2019}  & $43.1^{+1.1}_{-0.7}{}^{b}$       & $-1.5^a$    & $-0.6 \pm 0.3$         & $0.3^{+1.0}_{-1.1}{}^d$ & N & N & N &  N & N & N \\
       \citet{Luo2020} & $41.16^{+0.51}_{-0.19}{}^{b}$    & $0^a$    & $-0.79_{-0.18}^{+0.16}$ & $0^a$  & $30^{a,c,e}$ & $0.17^{a,c,e}$ & Y & Y & Y & Y\\
       \citet{Arcus2020} & $45.1^a$ & $-2.1^{1.1}_{-1.4}$ & $-0.7_{-0.2}^{+0.2}$ & $1 \pm 1$ & N & N & N & N & Y & Y\\
\citet{Calebetal2016} & 41.2$^g$ & 0$^a$ & N/A & 0,1$^a$ & 100$^a$ & 0$^a$ & Y & Y & Y & Y\\
       \citet{Macquart2020} & N/A & N/A & N/A & N/A & $65^{+95}_{-25}{}^e$ & $0.9_{-0.6}^{+0.9}{}^e$ & N & Y & N & N \\
       This work &   $41.70_{-0.06}^{+0.53}$ & $-1.55^{+0.18}_{-0.18}{}^f$  & $-1.09_{-0.10}^{+0.14}$ &  $1.67_{-0.40}^{+0.25}$  & $130_{-48}^{+66}$   &  $0.53_{-0.11}^{+0.15}$ & Y & Y & Y & Y
    \end{tabular}
    \caption{Summary of results and methods used by other authors in estimating FRB population parameters, compared to those of this work using the spectral interpretation of $\alpha$ with a Gaussian prior. Shown in columns 2--7 are the best-fit parameter values and corresponding uncertainties, converted to the units and notation used in this work. For the sake of comparability, all limits have been converted to 1$\sigma$ standard deviations assuming that uncertainties are Gaussian distributed. Columns 8--11 summarise the different effects that were (Y) or were not (N) included in the modelling: accounting for instrumental beamshape, a distribution in dispersion measure for a given redshift, a DM-dependent telescope threshold, and a distribution of burst widths $f(w)$ and its corresponding effects on sensitivity. Note that a `Y' does not necessarily mean a comparable treatment to this work. ${}^a$ Parameter held fixed by the authors. ${}^{b}$ indicates the use of a Schechter function, with exponential decay after $E_{\rm max}$. ${}^c$ The host galaxy contribution is modelled according to the ``ALGs(YMW16)'' model from \citet{Luo2018}, with approximate mean $\sim0.8$ and standard deviation $0.2$ in $\log_{10}$\,DM, to which is added a `local' contribution with a uniform distribution from 0--50, for a total mean of $30$ and quadrature-added deviation $15$, i.e.\ $0.17$ in $\log_{10}$. ${}^d$ The $(1+z)^\beta$ scaling used in this work is converted to $n$ via $n=2.7\beta$ from Eq.~\ref{eq:sfr_n}. $^e$ These values are not explicitly quoted, and were approximately derived from plots shown in the text. $^f$ This constraint primarily arises from the prior of \citet{Macquart2019a}. $^g$ A log-normal luminosity function was used; here we quote the mean of $10^{38.2}$\,erg plus three standard deviations of $1$ in $\log_{10}$.
    }
    \label{tab:other_results}
\end{table*}

%% file: Results2_pzdm.tex
\section{Results 2: redshift and dispersion measure}
\label{sec:results2}

The best fitting parameter set \citep{James2021Lett} allows a comparison between the expected and observed distributions of FRBs in DM, z, and SNR space. However, each combination of allowed parameters
produces a unique map of the z--DM distribution of FRBs for each telescope. Rather than present a very large number of plots, we use the following approach to identify a finite number of reasonable possibilities.

\begin{table*} 
    \centering
       \caption{Parameter sets used in this work: the best-fit values found in our companion paper \citep{James2021Lett}; sets allowing the parameter in the left-most column to take its minimum and maximum value within 90\% C.L; the best-fitting parameter set when forcing $n=0$, i.e.\ no source evolution; and a set when setting $E_{\rm min}=10^{38.5}$\,erg, but otherwise identical to the best fit.}
    \begin{tabular}{r|c c c c c c c}
    \hline
 Set & $\log_{10} E_{\rm min}$ & $\log_{10} E_{\rm max}$ &$\alpha$ & $\gamma$ &$n$ &$\mu_x$ & $\sigma_x$ \\
 \hline
 Best fit & 30 & 41.7 & -1.55 & -1.09 & 1.67 & 2.11 & 0.53 \\
 \hline
$\log_{10} E_{\rm max}$ & 30.0 & 41.6 & -1.5 & -1.2 & 2.6 & 2.0 & 0.5 \\
      & 30.0 & 42.51 & -1.5 & -1.2 & 1.69 & 2.0 & 0.5 \\
$\alpha$  & 30.0 & 42.0 & -1.88 & -1.05 & 1.8 & 2.0 & 0.5 \\
& 30.0 & 42.0 & -1.2 & -1.1 & 1.48 & 2.0 & 0.5 \\
$\gamma$ & 30.0 & 42.08 & -1.5 & -1.34 & 2.02 & 2.25 & 0.5 \\
      & 30.0 & 41.92 & -1.5 & -0.96 & 1.4 & 2.0 & 0.54 \\
$n$ & 30.0 & 41.8 & -1.5 & -1.0 & 1.11 & 2.25 & 0.5 \\
      & 30.0 & 42.0 & -1.75 & -1.14 & 2.28 & 2.0 & 0.5 \\
$\mu_{\rm host}$ & 30.0 & 42.18 & -1.5 & -1.1 & 1.78 & 1.77 & 0.59 \\
      & 30.0 & 41.8 & -1.5 & -1.1 & 1.47 & 2.41 & 0.63 \\
$\sigma_{\rm host}$ & 30.0 & 42.08 & -1.5 & -1.1 & 1.6 & 2.0 & 0.36 \\
      & 30.0 & 42.0 & -1.5 & -1.1 & 1.6 & 2.0 & 0.81 \\
    \hline
$n=0$ & 30.0 &  41.6 &  -1.25 & -0.9 &  0.  &  2.25 &  0.5 \\
\hline
$E_{\rm min}$ & 39 & 41.7 & -1.55 & -1.09 & 1.67 & 2.11 & 0.53  \\
\hline
    \end{tabular}
    \label{tab:systematic_sets}
\end{table*}

For each parameter, we take its value at both the upper and lower limits of its one-dimensional 90\% C.L., and choose the corresponding values of the other parameters. This results in 12 further parameter sets, which are listed in Table~\ref{tab:systematic_sets}. For comparison, we also include the best-fit parameter set assuming no source evolution, and the upper limit on $E_{\rm min}$ as described in Section~\ref{sec:emin}.

For each of these parameter sets --- including the best-fit set --- we generate the expected FRB distribution in DM--z space, and compare this to the observed values. All the results shown in this section are calculated using the spectral index interpretation of $\alpha$ with the Gaussian prior.

In interpreting the results in this section, we caution that the data to which we compare expectations have been used to determine the FRB population parameters, so the two are not independent. However, given that only eight free parameters (the standard six, plus two for the FRB width distribution) are used to fit multiple observables from three different FRB surveys, good agreement should not be the result of over-fitting, but rather indicate a genuine correspondence between the models used in this work and reality.

\subsection{Observed and predicted distributions}
\label{sec:observed_predicted_z_dm}

\begin{figure*}
    \centering
\begin{overpic}[width=0.45\textwidth]{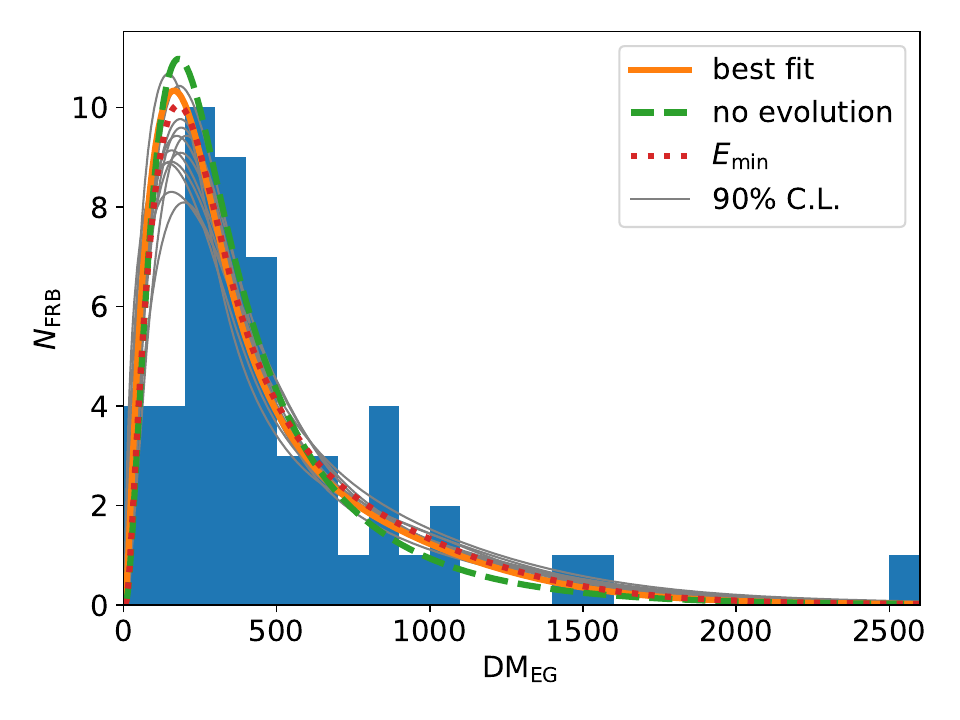}
 \put (30,65) {\Large All}
\end{overpic}
\begin{overpic}[width=0.45\textwidth]{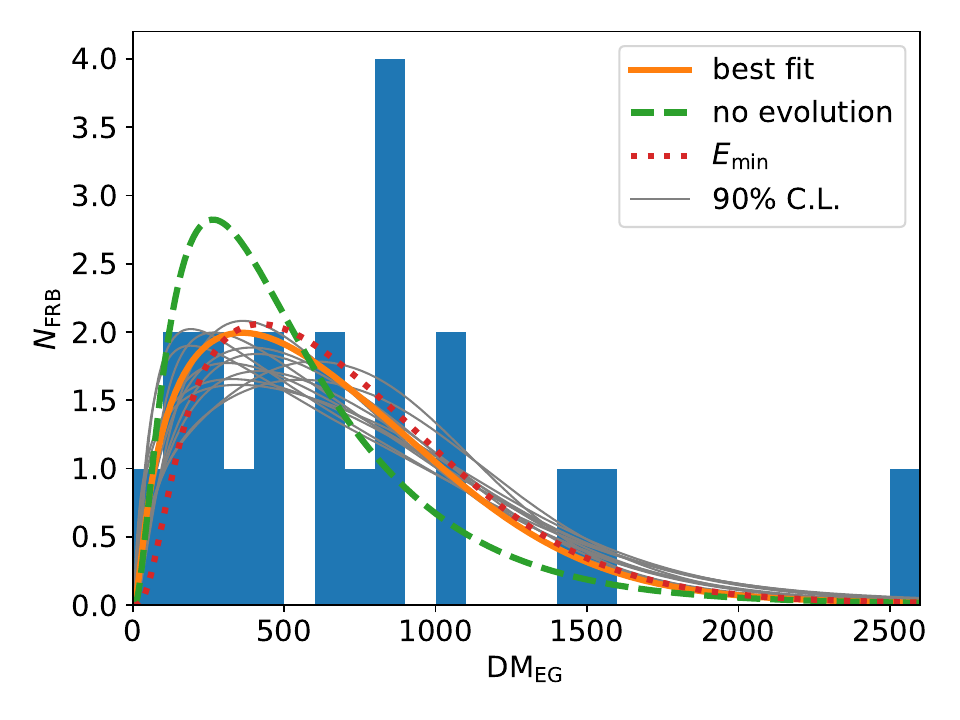}
 \put (15,65) {\Large Parkes/Mb}
\end{overpic}
\begin{overpic}[width=0.45\textwidth]{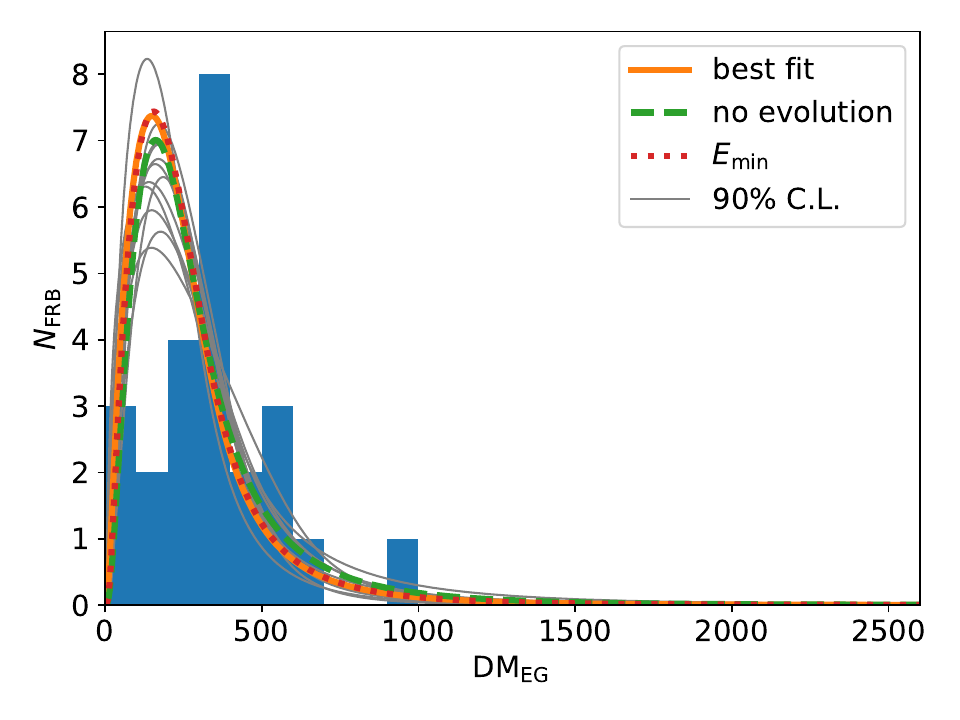}
 \put (30,65) {\Large ASKAP/FE}
\end{overpic} \begin{overpic}[width=0.45\textwidth]{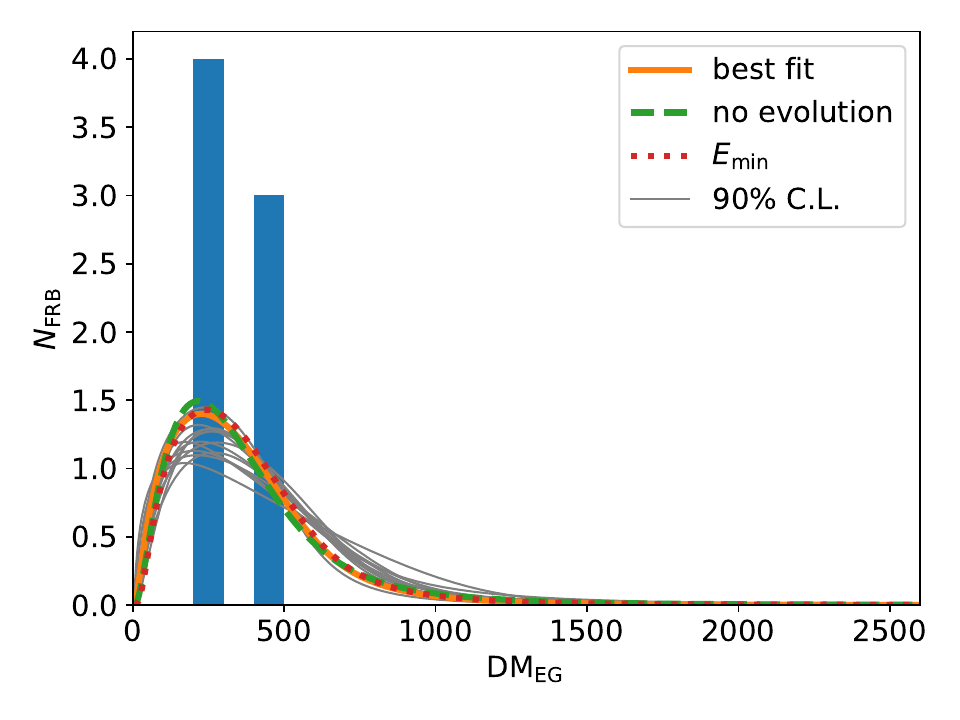}
 \put (30,65) {\Large ASKAP/ICS}
\end{overpic}
\caption{Observed (histogram) and predicted (lines) distributions of DM for all surveys considered (upper left), and the individual Parkes multibeam (upper right), ASKAP Fly's Eye (lower left), and ASKAP ICS (lower right) surveys. Predictions show the best-fit over the entire parameter space, when constrained to no source evolution, when allowing $E_{\rm min}$ to vary, and when varying parameters within their 90\% C.L.}
    \label{fig:all_dm_1D}
\end{figure*}

\begin{figure}
    \centering
\begin{overpic}[width=0.45\textwidth]{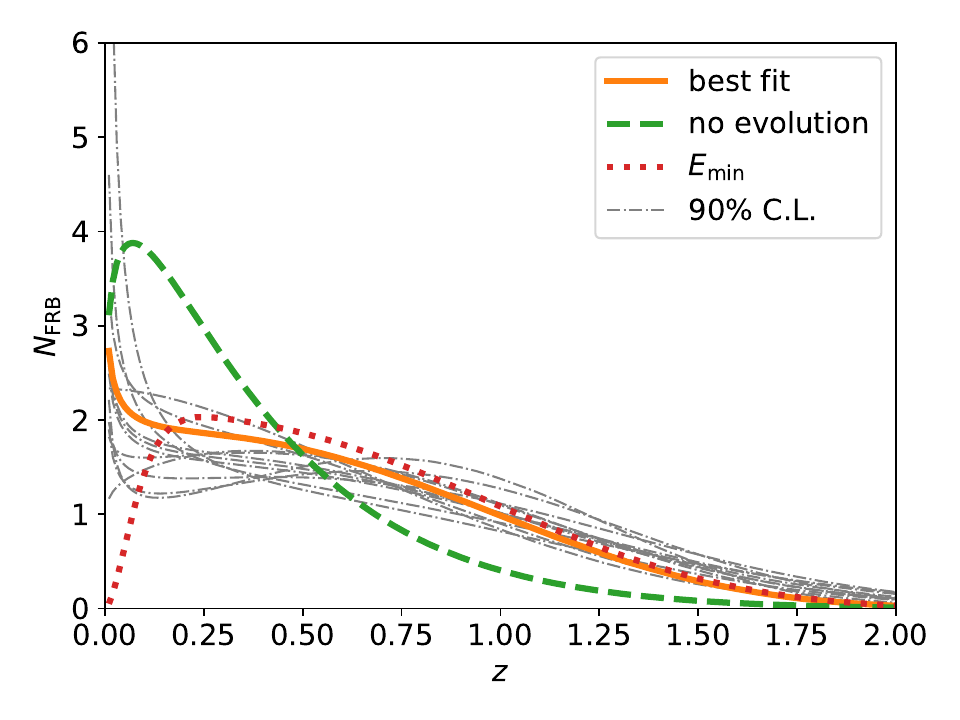}
 \put (20,65) {\Large Parkes/Mb}
\end{overpic}
\begin{overpic}[width=0.45\textwidth]{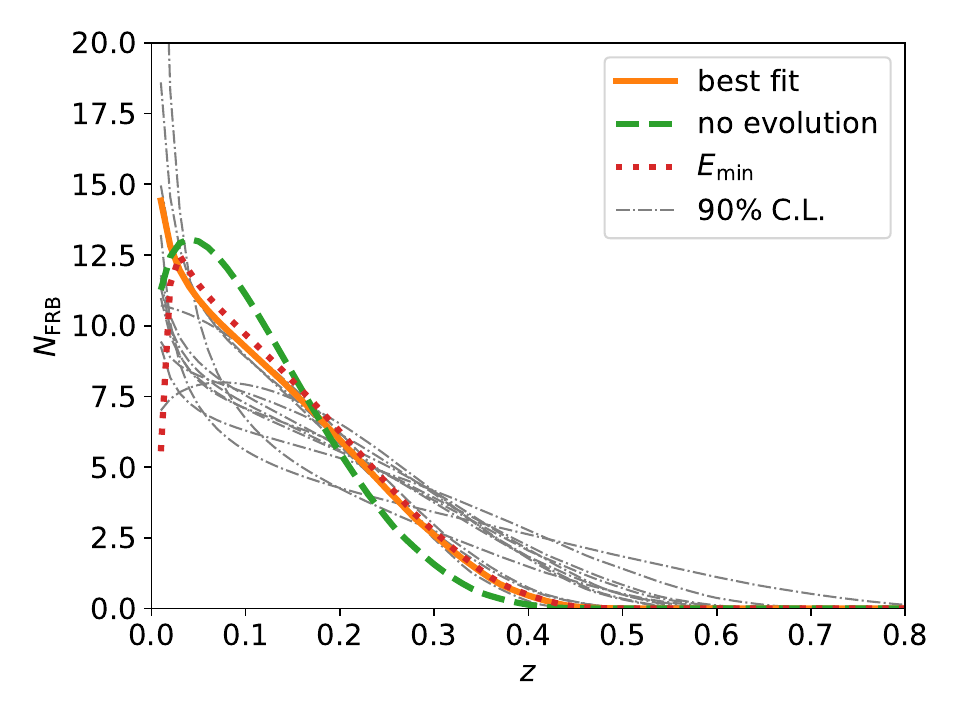}
 \put (20,65) {\Large ASKAP/FE}
\end{overpic} \begin{overpic}[width=0.45\textwidth]{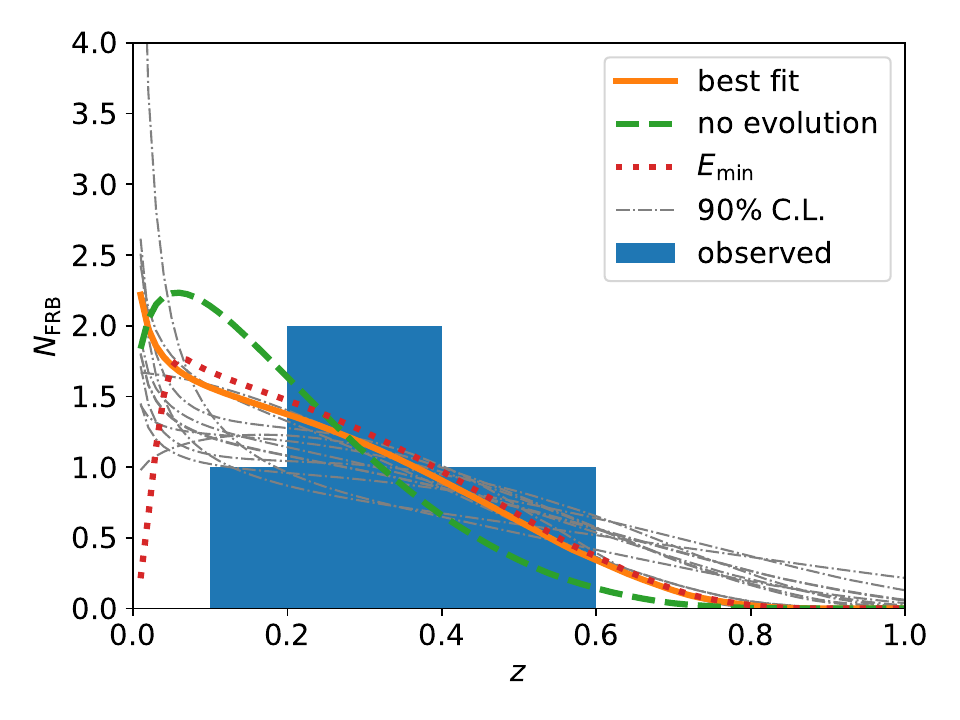}
 \put (20,65) {\Large ASKAP/ICS}
\end{overpic}
\caption{Predicted distributions (lines) of $z$ for Parkes multibeam (top) and ASKAP Fly's Eye (middle) surveys, with the ASKAP ICS results (bottom) also showing the observed values (histogram). Predictions show the best-fit over the entire parameter space, when constrained to no source evolution, when setting $E_{\rm min}=10^{39}$\,erg, and when varying parameters within their 90\% C.L.}
    \label{fig:all_z_1D}
\end{figure}

The predicted DM and $z$ distributions from all fifteen cases --- best-fit, no source evolution, $E_{\rm min}$, and twelve sets reflecting parameter uncertainty --- are shown in Figures~\ref{fig:all_dm_1D} and \ref{fig:all_z_1D} respectively. In the case of the DM distribution, data and predictions are shown individually from each survey described in Section~\ref{sec:survey_results}, and stacked together to allow a better comparison. For the $z$ distribution, the only available data comes from ASKAP/ICS observations --- however, predictions from each individual survey are also shown.

The first question to ask is --- are the best fits indeed good fits? Our best-fitting parameter estimates \citep{James2021Lett} do not necessarily indicate that the modelled DM, z, and SNR distributions are good fits to the data --- merely that they are the best fits given the form of the model used. A by-eye analysis shows that the predictions from the best-fit parameter set are indeed a good match to the data. However, there appears to be an over-prediction of FRBs at low DM and particularly at low redshift. This is a common feature over all the parameter sets within the 90\% error margins, although the degree of peakedness near $z=0$ varies greatly.

We consider four possible explanations for this below.

\subsection{Random fluctuations}
\label{sec:explanation_random}

\begin{figure}
    \centering
    \includegraphics[width=0.95 \columnwidth]{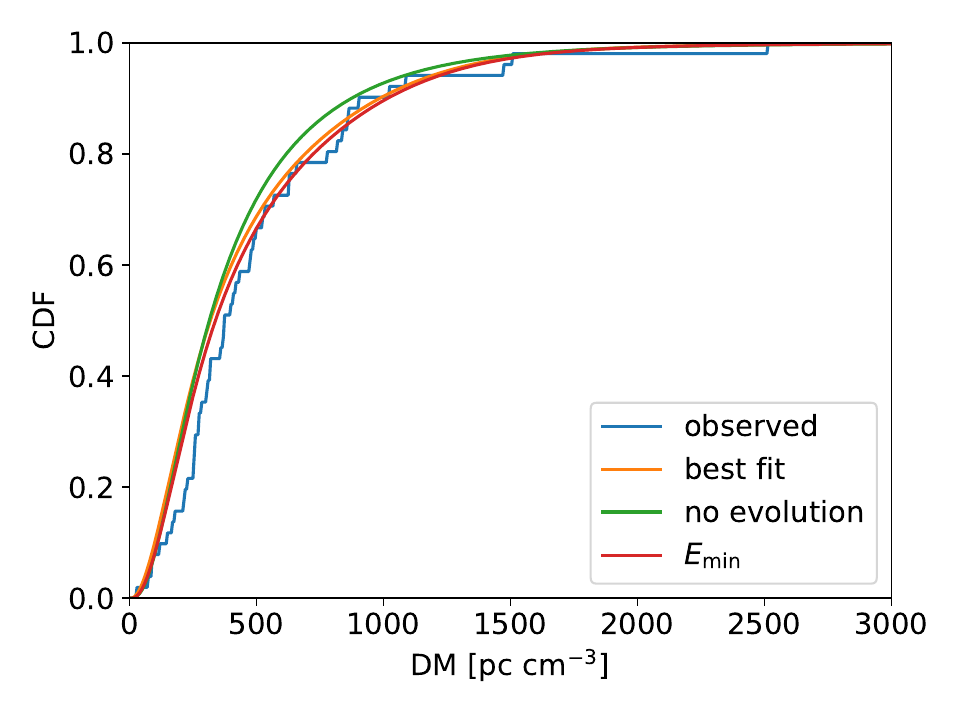}
    \includegraphics[width=0.95 \columnwidth]{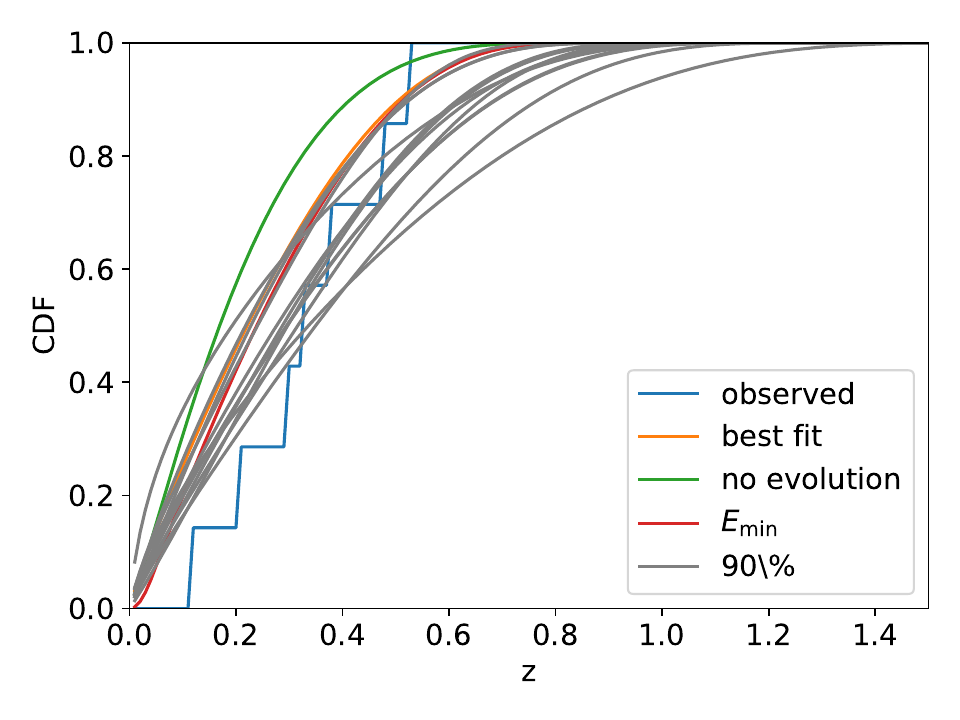}
    \caption{Observed and predicted cumulative distribution functions for the DM from all considered surveys (above) and $z$ from the ASKAP ICS survey (below).  Predictions show the best-fit over the entire parameter space, when constrained to no source evolution, and when allowing $E_{\rm min}$ to vary.}
    \label{fig:ks_cdfs}
\end{figure}

\begin{figure}
    \centering
    \includegraphics[width=0.95\columnwidth]{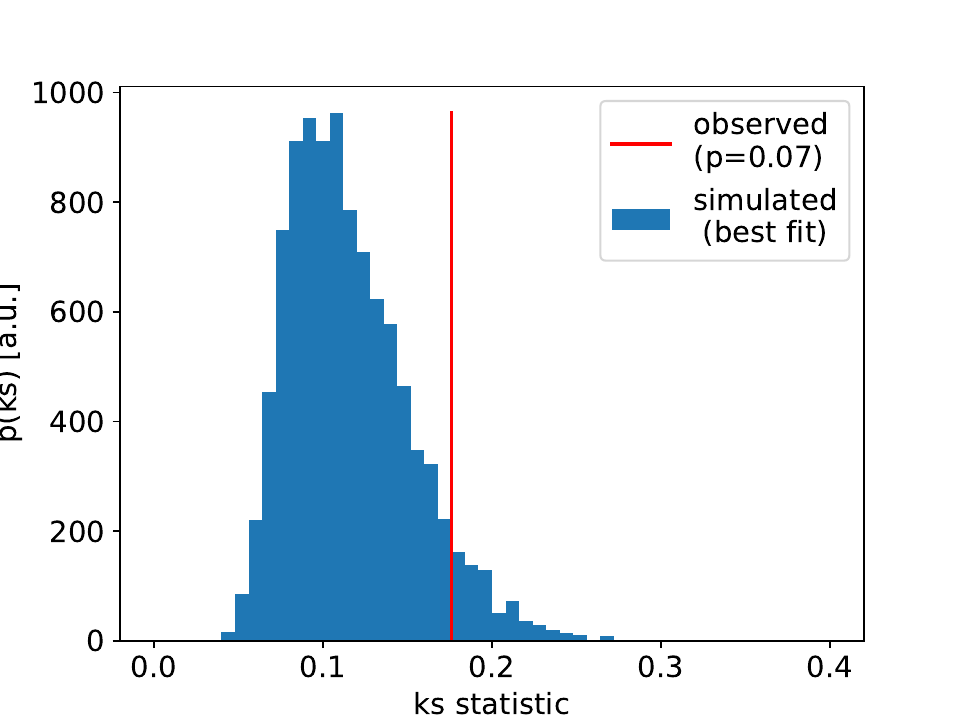}
    \includegraphics[width=0.95\columnwidth]{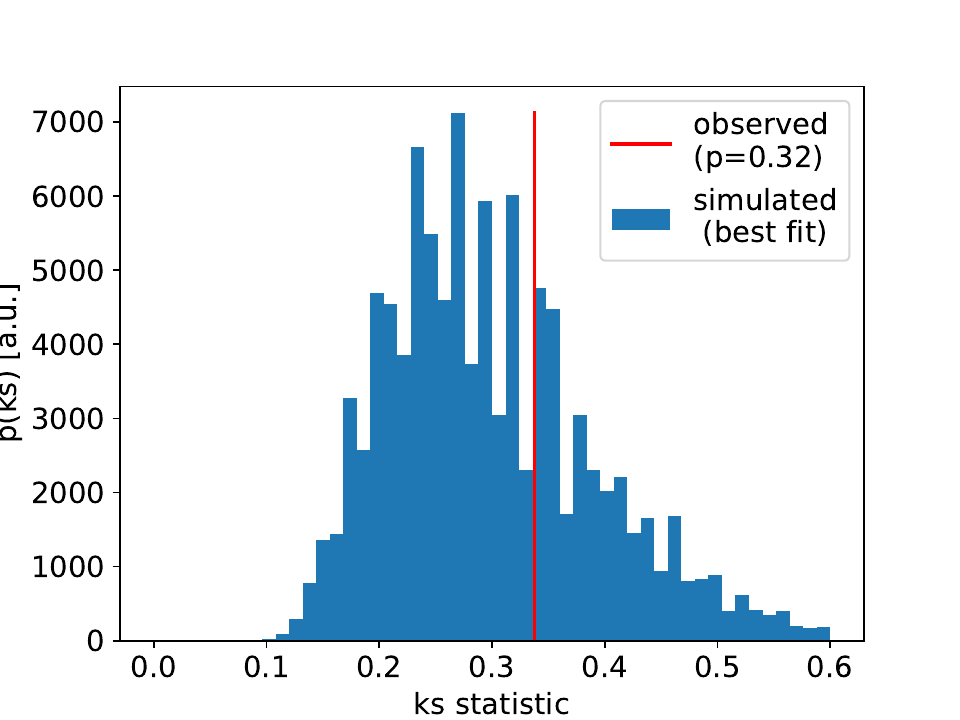}
    \caption{Observed value (red line) and simulated distribution (blue histogram) of the KS statistics for the DM (top) and redshift (bottom) distributions under the best-fit hypothesis. Shown are the corresponding p-values to reject this hypothesis.}
    \label{fig:ks_distribution}
\end{figure}

The observed deficit of low-DM and low-z FRBs may simply be a statistical fluctuation. To evaluate this, we perform Kolmogorov-Smirnov tests on the ASKAP ICS z distribution and total DM distribution \citep{kolmogorov,smirnov}. Predicted and measured cumulative distributions of both DM and z are shown in Figure~\ref{fig:ks_cdfs} --- the KS-statistic is the maximum absolute difference between the two curves. To evaluate the significance of this statistic, we take the best-fit curves as the truth, and randomly generate 10,000 samples from each. Histogramming the results produces the expected distributions of the KS statistic under the null hypothesis that the best-fit prediction is true. Comparing this distribution to the observed value of the KS statistic in Figure~\ref{fig:ks_distribution} shows that our observed values of DM and $z$ are consistent with predictions, with larger values of the KS statistics observed in 7\% and 42\% of cases for the DM and z distributions respectively. Performing a similar analysis using the $n=0$ distribution gives p-values of 8\% and 6\% for the DM and z distributions respectively --- evidently, the DM distributions of the ASKAP/FE and Parkes samples show a much better fit for $n>0$ than $n=0$. We therefore conclude that the apparent deficit of FRBs at low DM and redshift compared with predictions is consistent within statistical fluctuations of expectations. Nonetheless, we proceed with further analysis since the presence or otherwise of a minimum energy, or effects due to a large fraction of the population being repeaters, is of great interest.

\subsection{Evidence for a minimum energy $E_{\rm min}$}
\label{sec:emin}

In our standard modelling, we have set the minimum FRB energy $E_{\rm min}$ to an extremely low value of $10^{30}$\,erg --- well below the characteristic energies of observed FRBs, effectively making it zero. This is because values of $\gamma > -1.5$ render FRB observations primarily sensitive to $E_{\rm max}$ \citep{Macquart2018b}, while bursts from FRB~121102 have been observed at much lower energies than are likely to be probed by Parkes and ASKAP observations \citep[e.g.][]{Law2017}.

However, a clear possible explanation for the apparent deficit of FRBs at low DM-z is a minimum FRB energy. Such has recently been observed for the repeating FRB~121102 by FAST \citep{FAST_2021_energy_dist}, with the burst rate suppressed below $4.8 \, 10^{37}$\,erg. Without such a cut-off, as telescopes probe ever lower values of $E$ in the nearby Universe, the predicted number of FRB observations per comoving volume will increase without limit when $\gamma \le -1$. If $-1 \ge \gamma > -1.5$, the reducing volume will somewhat compensate, and the total rate will remain finite. This gives rise to the sharp increase in the best-fit expected redshift distributions near $z=0$ in Figure~\ref{fig:all_z_1D}, although such a peak may not be present within 90\% error margins.

\begin{figure}
    \centering
    \includegraphics[width=\columnwidth]{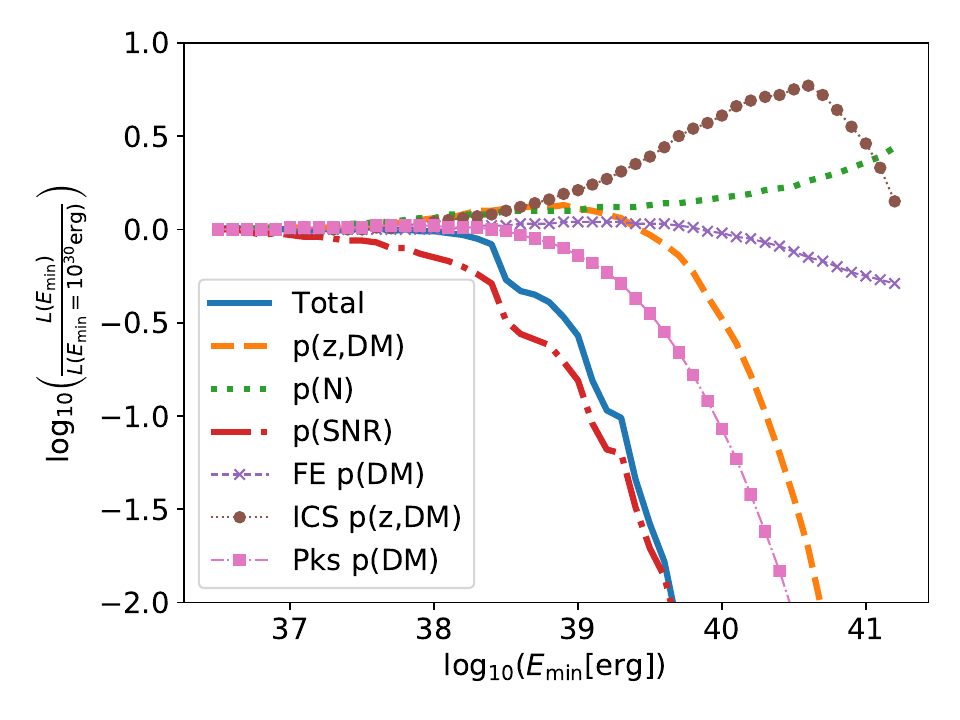}
    \caption{Evolution of likelihoods as a function of minimum energy $E_{\rm min}$, relative to the value at $E_{\rm min}=10^{30}$\,erg. Shown is the total likelihood; the likelihood over all surveys, split into contributions from the z--DM distribution p(z,DM), the number of events p(N), and the signal-to-noise probability p(SNR). The p(z,DM) contribution is further split into components from the ASKAP/FE, ASKAP/ICS, and Parkes/Multibeam surveys.}
    \label{fig:emin_likelihoods}
\end{figure}

To investigate the effect of $E_{\rm min}$, we fix the best-fit parameters, and vary $E_{\rm min}$. The evolution of the likelihoods is shown in Figure~\ref{fig:emin_likelihoods}. Interestingly, the total likelihood decreases with increasing $E_{\rm min}$, with $E_{\rm min}=10^{39.0}$\,erg the 90\% C.L.\ upper limit. Why? As expected from Figures~\ref{fig:all_dm_1D} and \ref{fig:all_z_1D}, the $p(z,DM)$ contribution increases with $E_{\rm min}$, peaking near $10^{39}$\,erg. This peak is a combination of the ASKAP/ICS observations strongly favouring a large $E_{\rm min} \approx 10^{40.5}$\,erg, and the Parkes multibeam observations strongly favouring $E_{\rm min}<10^{39}$ erg, with the ASKAP/FE observations being relatively neutral until $E_{\rm min} > 10^{40}$\,erg.

The reason why the Parkes observations are strongly against high values of $E_{\rm min}$ is that sensitive telescopes with small fields of view are highly unlikely to observe low-DM bursts, unless the small volume of the near Universe corresponding to such a low DM is populated by many (necessarily low-energy) FRBs. The lowest-DM burst detected by Parkes during the included surveys, FRB\,110214, had a DM of $168.5$\,pc\,cm$^{-3}$ \citep{Petroff_110214_2019}. Setting $E_{\rm min}=10^{39.0}$ implies reduced FRB rates for redshifts closer than $z=0.332$, assuming a limiting fluence of $0.3$\,Jy\,ms, with those bursts that are detected likely to have a SNR significantly greater than threshold. A redshift of $z=0.332$ implies a most likely dispersion measure of approximately 322\,pc\,cm$^{-3}$, being composed of a cosmological contribution of DM$_{\rm cosmic}=142$\,pc\,cm$^{-3}$, local contribution of 82\,pc\,cm$^{-3}$, and our model best-fit value of $\mu_{\rm host}/(1+z)=97.6$\,pc\,cm$^{-3}$. The observation of FRBs by Parkes with DMs below this value therefore disfavour a significant minimum energy $E_{\rm min}$.

\begin{figure} 
    \centering
    \includegraphics[width=\columnwidth]{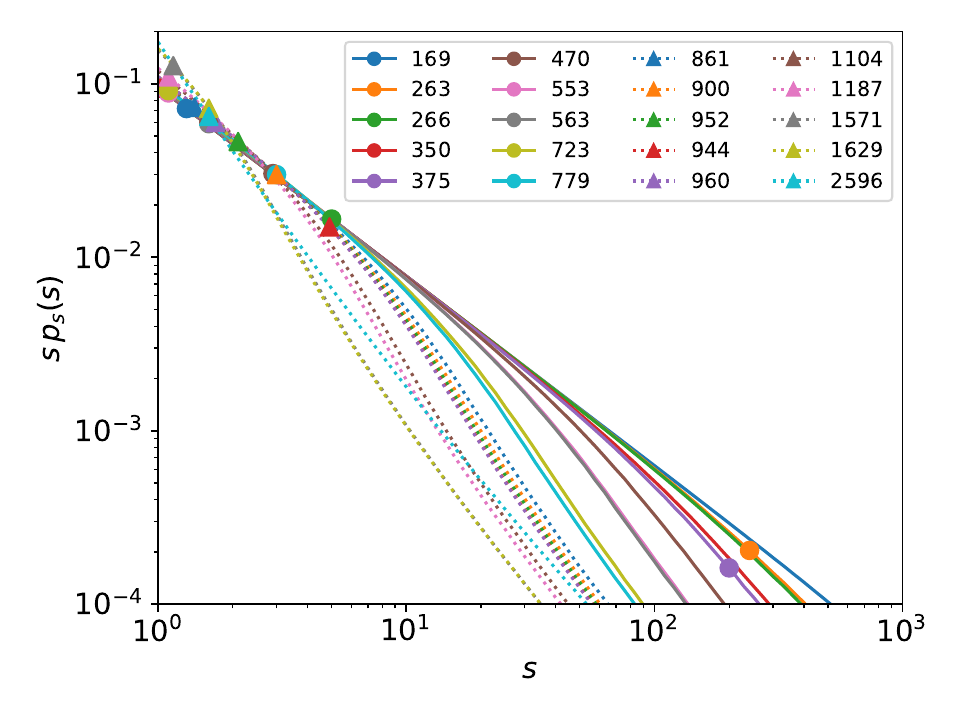}
    \includegraphics[width=\columnwidth]{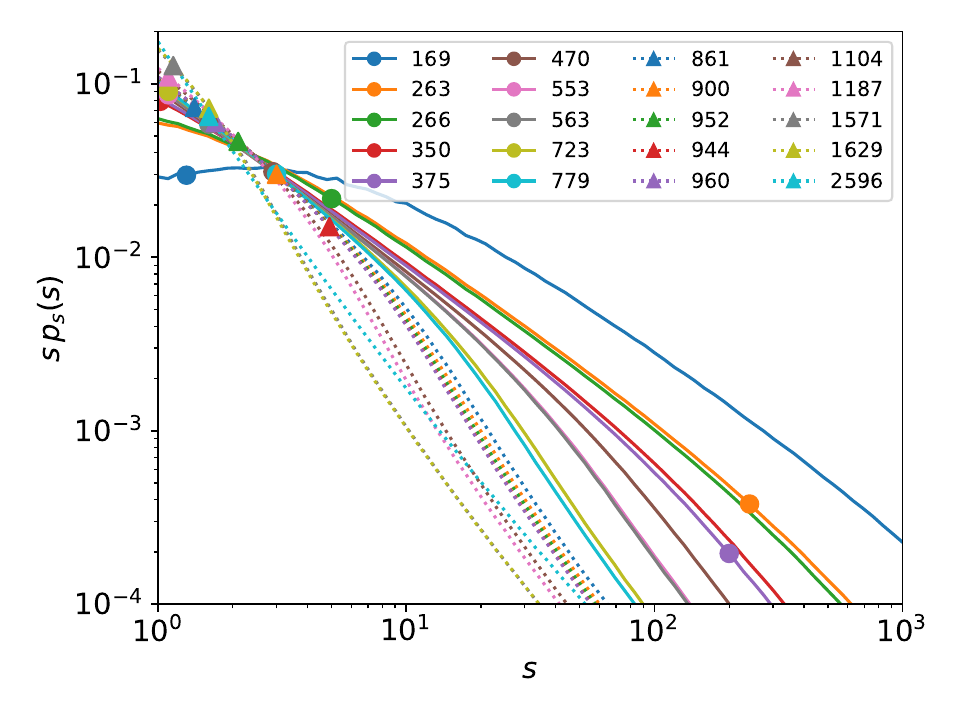}
    \caption{Likelihoods of observing FRBs with a given $s$, $p_s(s)$, as a function of their dispersion measure, DM, for the Parkes multibeam observations, weighted by $s$ for clarity. This is calculated for best-fit parameters using $E_{\rm min}=10^{30}$\,erg (top), and $E_{\rm min}=10^{39.0}$\,erg (bottom).}
    \label{fig:pks_psnr}
\end{figure}

This is illustrated in Figure~\ref{fig:pks_psnr}, where we show model predictions of $p_s$, and the observed values of $s$, for each Parkes-detected FRB. This is done for the best-fit model, and when using $E_{\rm min}=10^{39.0}$\,erg. While the $E_{\rm min}$ model predicts that the high fluence of FRB~180309 with a DM of $\sim$263\,pc\,cm$^{-3}$ and $s=41$ is slightly more likely, it predicts that FRB~110214, with a DM of $\sim$169\,pc\,cm$^{-3}$, is much less likely to have its observed value of $s=1.44$.

In this work, we have not included $E_{\rm min}$ as a global minimisation parameter due to computational constraints. Might there be some other combination of parameters for which a significant $E_{\rm min}$ is found? To investigate this, we have repeated the $E_{\rm min}$ optimisation for all parameter sets in Table~\ref{tab:systematic_sets}. Only for the parameter set minimising $\gamma$ do we find a significant value of $E_{\rm min}$ to be preferred, with a best-fit value of $10^{38.2}$\,erg, and 90\% upper limit of $10^{39.1}$\,erg. This makes sense, since otherwise a steep energy function would over-predict the number of near-Universe FRBs. The resulting gain in likelihood acts to weaken our confidence in the lower limits on $\gamma$, e.g.\ the 90\% lower limit shifts to 60\% confidence.

Our findings clearly do not indicate that there is no minimum FRB energy. Within the confines of our power-law model, the data used appear insensitive to $E_{\rm min} < 10^{37}$\,erg, and we can only rule out $E_{\rm min} > 10^{39.0}$\,erg at 90\% C.L.\ --- but it does run counter to the findings of \citet{Luo2020}. These authors find a most likely minimum luminosity of $10^{42}$\,erg\,s$^{-1}$, which is approximately $10^{39}$\,erg assuming a standard 1\,ms burst, although they conclude rather that this finding is due to the limit of their sample.

\subsection{Influence of repetition}
\label{sec:explanation_repetiton}

In this study, we have ignored repeating FRBs on the basis that none of the FRBs detected by ASKAP and Parkes have been observed more than once --- even though some are known to repeat \citep{Kumar2019,CHIME2019AtelRepeat,Kumar2020,CHIME_catalog1_2021}, the surveys were not sensitive enough to probe this. However, the method of Section~\ref{sec:methodology} covers the entirety of z--DM space, regardless of whether or not an FRB has been detected at that point. Therefore, for any hypothesised repeating FRB, there will always be a sufficiently nearby volume of the Universe where any survey would be expected to detect more than one burst if the FRB landed in its field of view. If such a repeating FRB happens to be located in that volume, the observed burst rate will be greater than expected --- but if one does not, it will be less.

This effect is analysed in the context of 
Canadian Hydrogen Intensity Mapping Experiment (CHIME) observations by \citet{Gardenier2020FRBpoppy}, who note that the DM distribution of repeating FRBs should be lower than that of repeaters observed only once. While this has not yet been observed \citep{CHIME2019a,CHIME2019b,CHIME2019c,CHIME2020a}, this may be due to a very broad distribution of intrinsic repetition rates \citep{James2020b}.

As discussed by \citet{James2019b}, the ASKAP/FE observations have deeply probed some regions of sky with over 50 antenna-days spent on individual fields. This has allowed strong limits to be placed on FRB repetitions --- but also made the observations more susceptible to whether or not a strong repeater at $z < 0.2$ was present in these fields. We observe that the low-DM deficit is greatest in ASKAP/FE observations, and lowest for Parkes observations, consistent with this prediction.

Until more is known about the population of repeating FRBs, we cannot further quantify this effect, except to add that this effect is guaranteed to be present to some degree (some FRBs definitely do repeat), with its importance increasing if all FRBs are explained by strong repeaters, and lessening as a lower fraction of all FRBs are due to repeaters, and those repeaters are weak.

\subsection{Minimum search DM}
\label{sec:min_search_dm}

Many FRB surveys use a minimum searched DM, either to exclude Galactic FRB candidates, or due to local RFI with intrinsic DM of $0$ that nonetheless contaminates searches. For example, \citet{Thornton2013} reject bursts with DM$<100$\,pc\,cm$^{-3}$, although in our model nearby FRBs could have a DM of as little as DM$ \approx 80$\,pc\,cm$^{-3}$. This effect will sometimes therefore exclude FRBs in the very nearby Universe, potentially resulting in the observed deficit.

At initial completion of this work, only two bursts --- 171020 \citep{Shannonetal2018} and 180729.J0558+56 \citep{CHIME2019a} --- had been reported near this limit, however during the referee period, \citet{CHIME_M81_2021} reported an FRB with DM 87.82\,pc\,cm$^{-3}$. That CHIME have only reported one such FRB from several years' worth of observations however implies that the number of these events is low --- as are the number of FRBs predicted to be in this range by our model. The observed deficit extends well above 100\,pc\,cm$^{-3}$, so we consider this an unlikely cause.

Could FRB search algorithms be less efficient at low DM? Tests on Galactic pulsars have shown the FREDDA algorithm used in ASKAP FRB observations to be equally efficient ($\sim90$\% of bursts detected) at 67.99~\,pc\,cm$^{-3}$ as at 478.8\,pc\,cm$^{-3}$ \citep{James2019cSensitivity}, and follow-up (unpublished) studies have shown that Parkes/MB observations have similarly high levels of efficiency. Analysis of historical Parkes data by \citet{Zhang2020} has found one new FRB at 350\,pc\,cm$^{-3}$ (which was too recent to include in this analysis), but none at lower DM. We therefore do not consider a DM-dependent search efficiency to be a likely cause of the observed deficit.

\subsection{Summary: redshift and dispersion measure distributions}

Having established the robustness of these results, we summarise the predictions for the redshift distributions from Figure~\ref{fig:all_z_1D}. In all scenarios, the redshift distribution of the ASKAP/FE detections lies in the range $z_{\rm max}<0.8$, and most fits find $z_{\rm max} < 0.6$. Over all parameter sets, between 27\% and 48\% of ASKAP/FE bursts should originate from within $z<0.1$, confirming that these bursts are ideal follow-up targets due to their likely proximity. This suggests that the limits set on the repeatability of individual FRBs by \citet{James2020a} are significantly stronger than published, since those authors conservatively assume maximal redshifts. It also lends additional
weight to the possible association of FRB~171020 with a galaxy at 40\,Mpc by \citet{Mahony2018}.

The predicted $z$-distribution of Parkes bursts is significantly broader than that of ASKAP/FE observations, although 3--20\% (over all parameter sets) of observed FRBs are predicted to being within $z<0.1$.

A key test of our prediction of a large number of near-Universe FRBs will be future ASKAP/ICS detections. So-far, all bursts detected by ASKAP's ICS mode have been in a limited range of both DM and $z$, which seem to be from the central part of all predicted distributions. Our best-fit model predicts that 23\% of ASKAP/ICS localisations should lie within $z<0.1$, with a range over all sets of 14--33\%. To date, none have been observed --- even amongst the unpublished ones not included here.

\subsection{Source counts distribution}
\label{sec:source_counts}

\begin{figure}
    \centering
    \includegraphics[width=\columnwidth]{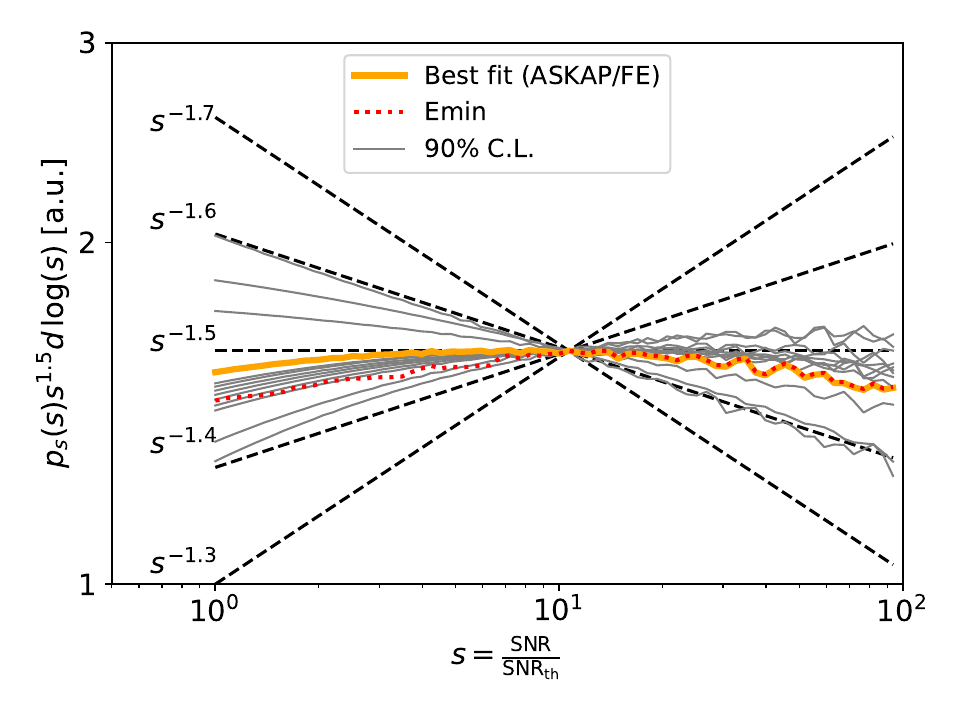}
    \includegraphics[width=\columnwidth]{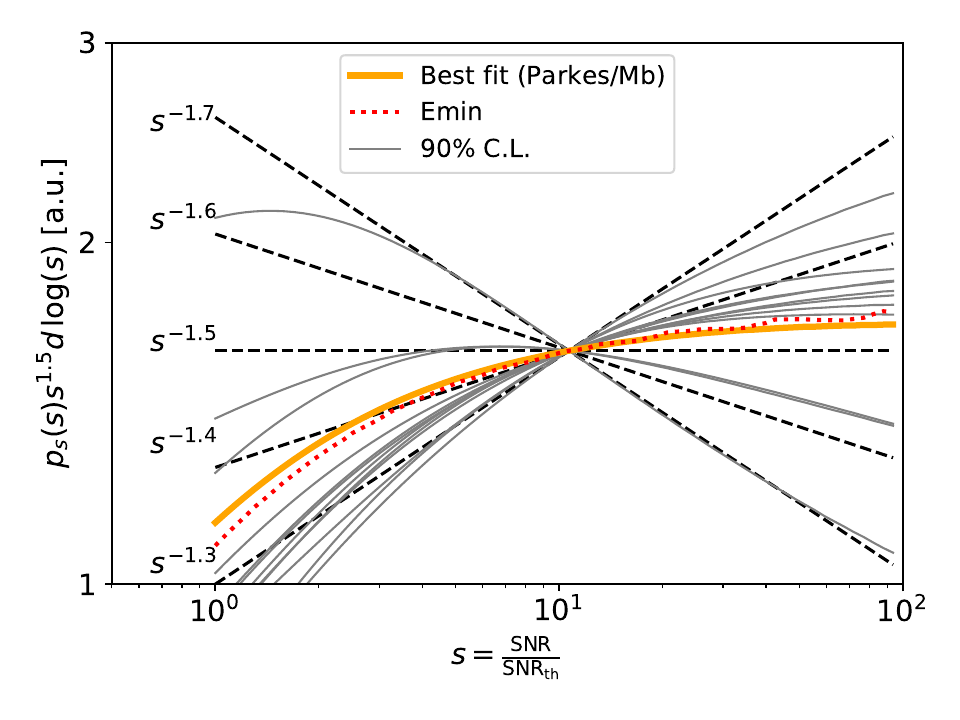}
    \caption{Predicted source counts distribution, $p(s)$, multiplied by $s^{1.5}$ for clarity, for top: ASKAP/lat50, and bottom: Parkes/Mb, according to the best-fit parameter set (black solid), when setting $E_{\rm min}=10^{39}$\,erg (red dashed), and when varying parameters within their 90\% C.L.\ (thin grey). Also shown are lines of constant power-law slopes to guide the eye. The fluctuations above $s=10$ are due to finite gridding in $z$ and DM.}
    \label{fig:source_counts}
\end{figure}

The slope of the source counts (``logN--logS'') distribution was one of the first FRB observables to be analysed. Adapted from its original use in the study of radio galaxies characterised by their flux S \citep{Ryle68}, applied to FRBs, the source counts distribution is the number N of observed FRBs as a function of fluence threshold $F_{\rm th}$. In an infinite Euclidean Universe, the distribution is expected to have a form
\begin{eqnarray}
N(F_{\rm th}) & = & C \left(\frac{F_{\rm th}}{F_0}\right)^a,
\end{eqnarray}
with $a=-1.5$, and $C$ and $F_0$ normalising constants. Studies using a variety of methods, with different treatments --- or neglect of --- observational biases in $F_{\rm th}$, and using data from telescopes with different detection thresholds, found inconsistent values of $a$ in the range of $-0.8 \ge a \ge -2.6$ \citep{Vedantham2016,Oppermannetal2018,Lawrence2017_poisson,Macquart2018a,Bhandarietal2018}.
\citet{Jamesetal2019a}, reviving methods applied to radio galaxy studies by \citet{Crawford1970}, argue that $s$ --- defined in Eq.~\ref{eq:SNRratio} --- is a bias-independent measure of the source-counts slope, and find $a=-1.18 \pm 0.24$ (68\% C.L.) for Parkes FRBs, and $a=-2.2 \pm 0.47$ for ASKAP/FE FRBs, equating to a $2.6\sigma$ tension. This was qualitatively consistent with \citet{Macquart2018b}, who argue that at high values of $F_{\rm th}$, the slope should be Euclidean ($a=-1.5$), while the parameters of the FRB population will lead to a flattening at lower thresholds.

Our model can be used to derive the expected distribution of $s$ using Eq.~\ref{eq:psu}, by integrating over all values of DM and $z$, then converting this differential distribution to a cumulative distribution. The results for ASKAP/FE and Parkes/Mb observations are given in Figure~\ref{fig:source_counts}.

We immediately see that in all scenarios, ASKAP/FE observations, with a higher base threshold of $26$\,Jy\,ms, are expected to follow a Euclidean ($a \sim -1.5$) distribution, while near-threshold ($1 \le s \le 10$) Parkes/Mb observations exhibit a flatter source-counts index near $a=-1.3$. Different scenarios predict source-counts indices in the range $-1.3 \le a \le -1.7 $ above $s=10$ for Parkes/Mb observations. The existence or otherwise of a minimum energy does not significantly affect the distribution.

Our results in all scenarios are consistent with the findings of \citet{Jamesetal2019a}, with the greatest tension --- about $1.5\sigma$ --- being between the source-counts slope that those authors find for ASKAP of $-2.2 \pm 0.47$ (68\% C.L.), and the approximate range of $-1.4$ to $-1.6$ found here. In particular, we also find that the source-counts index for Parkes/Mb is expected to be flatter than for ASKAP/lat50, and furthermore, the apparent `deficit' of FRBs with low values of $s$ is potentially attributable to the true behaviour of the FRB population, rather than statistical fluctuations or a measurement bias. This also suggests that the lower values of $a$ found by previous authors ---  \citet{Vedantham2016,Oppermannetal2018,Lawrence2017_poisson} --- incorporating data from telescopes more sensitive than Parkes may have been correct.

%% file: Results3_pzgdm.tex
\section{The $z$--DM distribution}
\label{sec:results3_zdm}

\begin{figure} 
    \centering
    \includegraphics[width=\columnwidth]{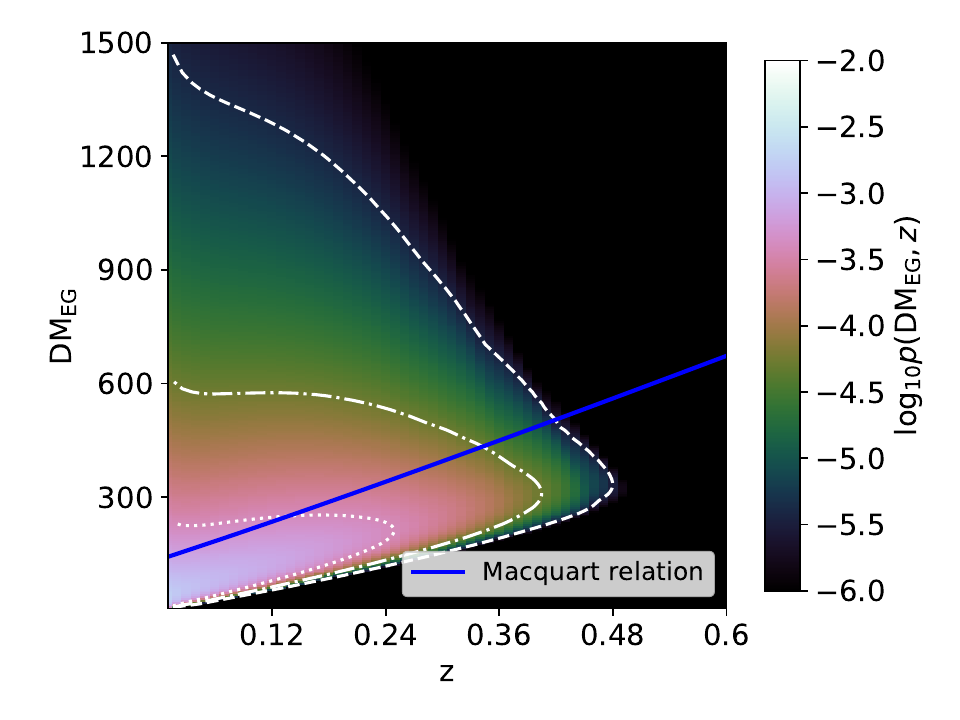}
   \includegraphics[width=\columnwidth]{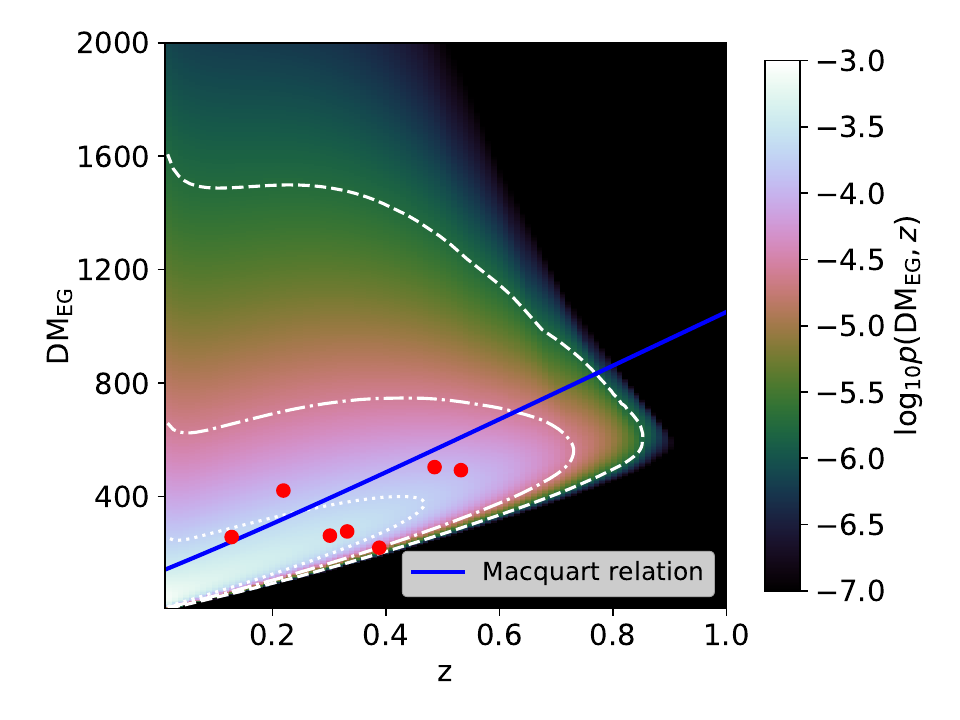}
   \includegraphics[width=\columnwidth]{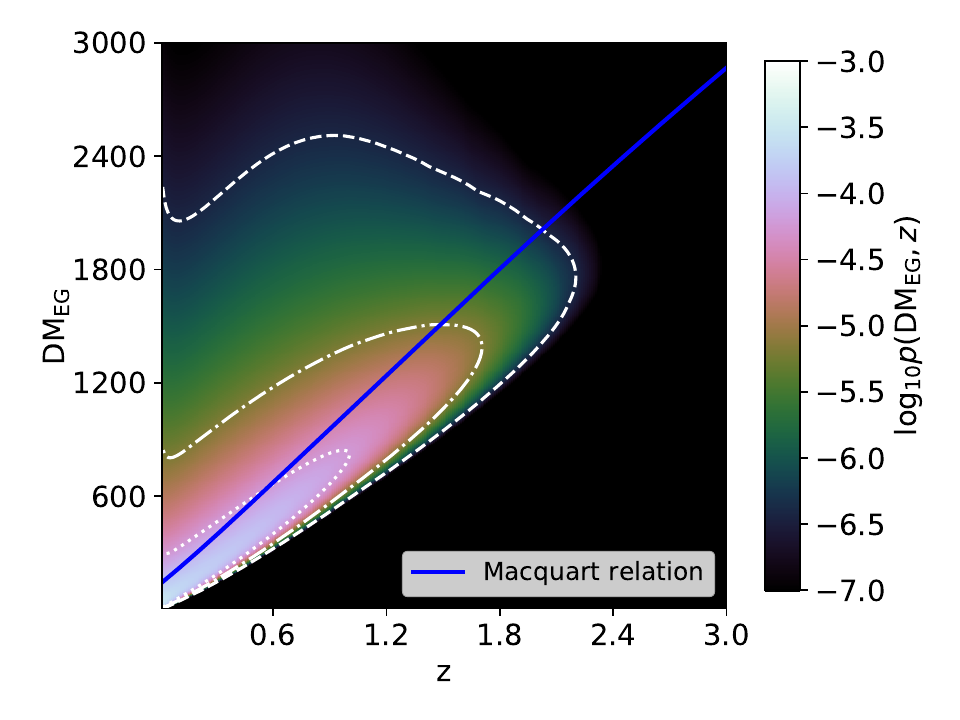}
    \caption{Normalised probability distribution $p(z,{\rm DM}_{\rm EG})$ of the redshift $z$ and extragalactic dispersion measure ${\rm DM}_{\rm EG}$ of FRBs detected in ASKAP Fly's Eye (top), ICS (middle) and Parkes multibeam (bottom) surveys. Contours indicate the intervals in which 50\% (dotted), 90\% (dot-dash), and 99\% (dashed) of all FRBs would be detected. Red dots indicate the values of DM$_{\rm EG}$ and host galaxy redshift $z$ for ASKAP ICS localized FRBs.}
    \label{fig:zdms}
\end{figure}

We argue in this work that the best representation of the FRB population observable by a telescope is a two-dimensional function of extragalactic (cosmological plus host) dispersion measure, and red shift. Necessarily, the observable fraction of this distribution is a function of survey parameters, and also local DM contribution, which reduces sensitivity as it increases at lower Galactic latitudes. Our best-fit z--DM distributions for the three surveys considered are plotted in Figure~\ref{fig:zdms}.

We describe the general features of these plots. Most FRBs are expected to lie near to, or below, the Macquart relation, being the 1--1 correspondence of FRB redshfit with DM. This relation, with a slope of approximately 100\,pc\,cm$^{-3}$ per 0.1 in redshift, continues up to some maximum detectable distance, being approximately $z$=0.5, 0.85, and 2.2 for ASKAP/FE, ASKAP/ICS, and Parkes/Mb respectively. However, the majority of FRBs will not be found near the maximum redshift, simply because there are very many more FRBs with lower energy, and much more of the sky covered at lower sensitivity. This is most evident with Parkes/Mb observations, where the most likely half of all FRB observations will arise on the Macquart relation with $z<1$.

Off the Macquart relation, there is a significant fraction of FRBs expected to be found with higher than expected DMs, due to a combination of their host and cosmological contributions. Only our ASKAP/ICS (localised) FRB sample has provable examples, being FRB~190714 (504.7\,pc\,cm$^{-3}$ from $z=0.209$) and FRB~190608 (339.5\,pc\,cm$^{-3}$ from $z=0.1178$), although a third (FRB~191001, with 506.92\,pc\,cm$^{-3}$ from $z=0.23$) was excluded due to being detected at a lower frequency. This effect is less pronounced for more-sensitive surveys, since excess host contributions are dominated by cosmological ones. We emphasise that much of the structure above the Macquart relation --- i.e.\ in the high-DM region --- is poorly constrained, since our adopted log-normal distributions may not reflect reality.

At very high DMs, only near-Universe FRBs are observable, since a burst must be observed with very high fluence to overcome the detection bias against high DM.
The upper bound of the 99\% region (dashed lines) slopes backward, against the Macquart relation, since more low-energy FRBs with large excess DM are predicted than high-energy FRBs lying on the Macquart relation. We do not expect our quantitative estimates in this region to be accurate until it is directly probed with localised FRBs. Since the cause of this effect is a well-understood observational bias however, it will clearly be present to some degree.

\subsection{The Macquart Relation}
\label{sec:MacquartRelation}

The `Macquart Relation' is the general one-to-one-ness of the relationship between redshift and DM of FRBs. It is predicted from the distribution of baryonic matter in the Universe \citep{Inoue2004}, and first evidence for it was given in \citet{Shannonetal2018}, by comparing the DM distributions of ASKAP/FE and Parkes/Mb populations, where the higher sensitivity of Parkes allowed it to probe more-distant FRBs with higher DMs. The relation was first observed directly by \cite{Macquart2020}, who showed that the redshifts of localized FRBs were consistent with the baryonic content of the Universe.

For the purpose of comparing survey results, we propose a useful distinction: the `weak' Macquart relation (which might more accurately be titled the `Shannon' relation), where telescopes with higher sensitivity on average observe more-distant FRBs with higher DM; and the `strong' version (or true Macquart relation), where the DMs of FRBs \emph{within} a survey are a good proxy for their redshift. Several authors use a 1--1 z--DM relation in performing estimates of the FRB population from the DMs of un-localised FRBs \citep{Shannonetal2018,Deng2019,Lu2019,James2020a}.

\begin{figure} 
    \centering
    \includegraphics[width=\columnwidth]{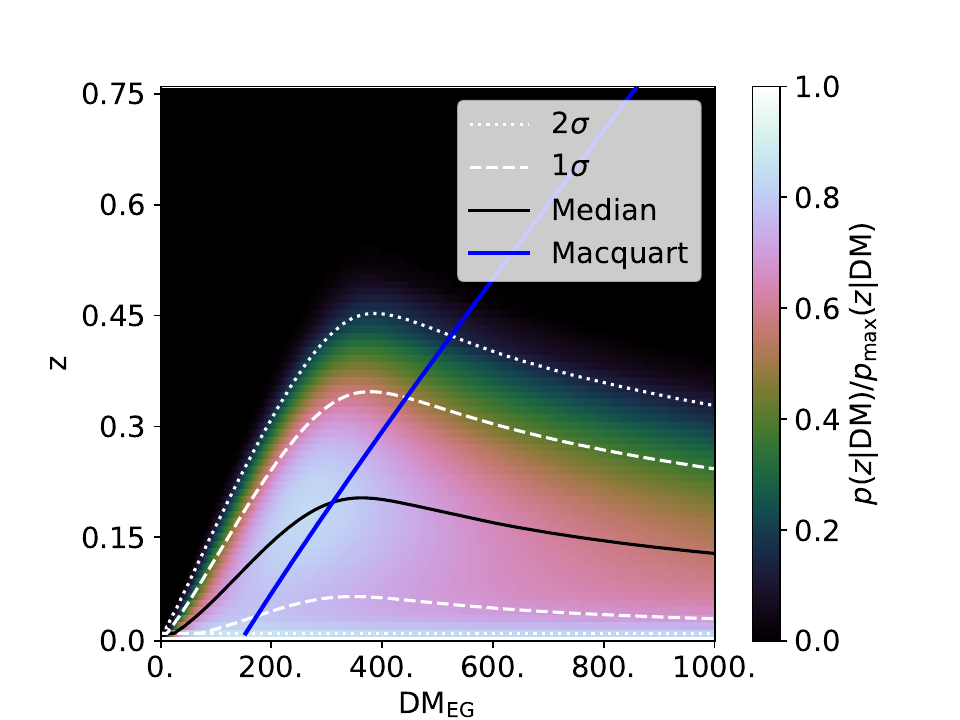}
   \includegraphics[width=\columnwidth]{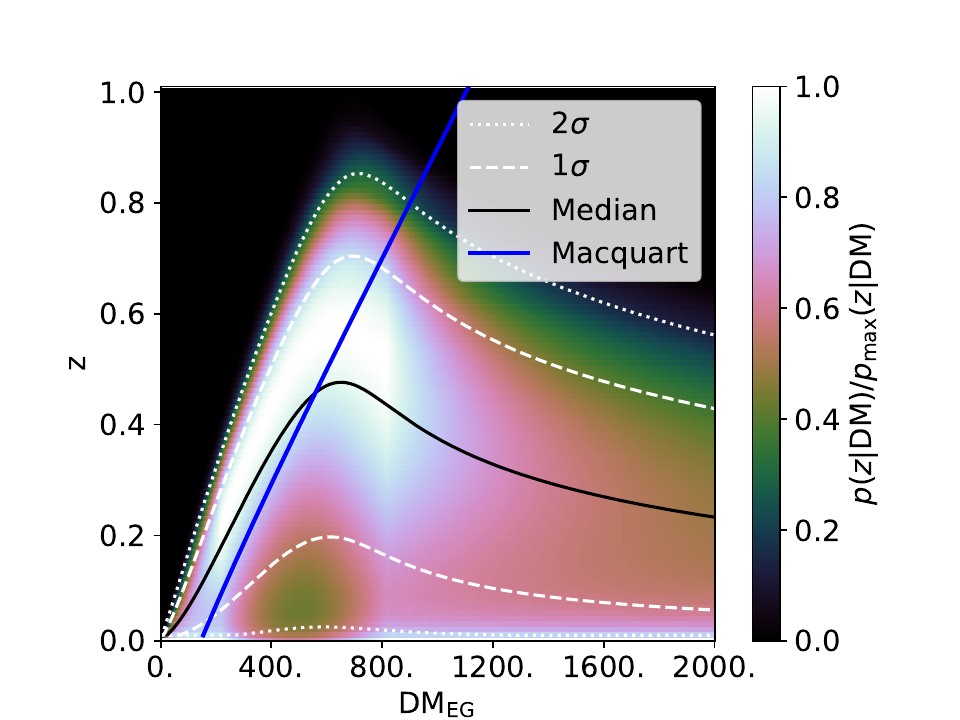}
   \includegraphics[width=\columnwidth]{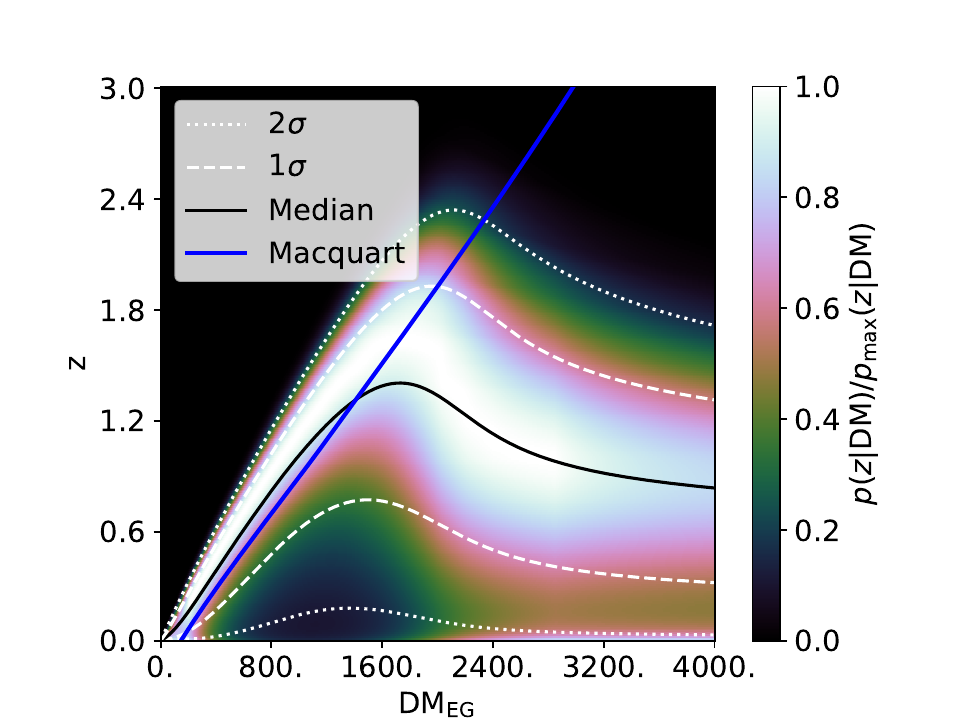}
    \caption{Probability $p(z|\DM)$ of an FRB originating from redshift $z$ given it has been observed with the specified DM in the ASKAP/FE (top), ASKAP/ICS (middle), and Parkes/Mb (bottom) surveys.}
    \label{fig:macquart_relation}
\end{figure}

Clearly, the weak version of the relation is well-established, and an obvious consequence of the cosmological nature of FRBs. Here, we test the strong version, which can be readily tested by calculating
\begin{eqnarray}
p(z|\DM) & = & \frac{p(z,\DM)}{p(\DM)}
\end{eqnarray}
for a given survey. This is shown for each survey in Figure~\ref{fig:macquart_relation}.

In each survey, the Macquart relation applies up to a maximum redshift $z_{\rm max}$, beyond which it reverses. The reversal can be intuitively understood by realising that at the maximum redshift at which an FRB can be detected, FRBs with any excess of DM can not be detected, due to DM smearing reducing sensitivity. Therefore, the only way to detect FRBs with a DM lying above that expected from a burst originating at $z_{\rm max}$ is to have the burst originate at a nearer redshift. As noted in Section~\ref{sec:ingredients2} and Fig.~\ref{fig:basic_F0}, the increased number of FRBs in the local Universe is related to the cumulative luminosity index $\gamma=-1.09$, with more negative values leading to more nearby high-DM events.

The reversal of the Macquart relation has several practical consequences. Firstly: for surveys with a large sample of FRBs, the burst with the greatest DM will \emph{not} be the most distant. An excellent example of this phenomenon is FRB 170428 \citep[ASKAP;][]{Shannonetal2018}, which is most likely to originate below $z=0.3$, rather than the value of $z\sim1$ expected from the Macquart relation. FRB~160102, observed by Parkes, is another likely candidate. The implication is that works using a 1--1 DM--z relation will vastly over-estimate the maximum FRB energy, since they will necessarily attribute a large distance and therefore high energy to the highest DM event, which may in fact be quite local.

A key test of this reversal would be the detection of an FRB with DM$\gtrsim 1000$\,pc\,cm$^{-3}$ by ASKAP in ICS mode, and its subsequent localisation to a redshift $z \lesssim 0.6$. Again, we note that this reversal of the Macquart relation will always be present to some extent, since it is fundamentally due to observational effects which are known and understood --- it is merely the extent of this effect, and the DM for any given survey above which it occurs, that is currently uncertain,

Finally, we note that the existence of FRBs with very high DMs has raised the possibility of using FRBs to probe for the signature of Helium reionisation \citep{DengCosmology2014,CalebHelium2019,Linder2020}. While this is by no means ruled out, it emphasises that doing so will require FRBs to be localised, since simple measures of FRB properties as a function of DM will yield a very large scatter in redshifts, and hence reduced statistical power.

%% file: DiscussionConclusion.tex
\section{Conclusion}
\label{sec:conclusion}

We have developed a precise model of FRB observations, including observational biases due to the full telescope beamshape, degradation in efficiency due to DM, and intrinsic burst width.  None of these effects are fundamentally new; many others should take credit for highlighting their importance, and \citet{Luo2020} should be attributed with a first analysis using these techniques. Here, we have improved upon this method by using an unbiased data sample, adding new observations of localized FRBs, studying the effects of source evolution, including the likelihood of the observed signal strength, and improving the beam model. We show that ignoring, or incorrectly modelling, these factors leads to significant biases in the expected redshifts of observable FRBs.

We have also highlighted how uncertainties in the spectral index $\alpha$, and indeed even in the \emph{interpretation} of $\alpha$ as either a true FRB spectral index or a frequency-dependent rate, is a systematic uncertainty in the modelling, and resolving this question should be a focus of future research. We suggest treating both interpretations of $\alpha$ as being equally plausible for the time being.

We have used our approach to model FRB observations with ASKAP in both fly's eye and incoherent sum mode, and Parkes multibeam observations. We have carefully selected our data to ensure it is not biased due to under-reporting of observation time, or due to large local DM contributions reducing sensitivity. Crucially, we have included a sample of localized FRBs from ASKAP for which the redshift of the host galaxies is measured.

The $z$, DM, and SNR distributions of FRBs predicted by our best-fit population estimates, presented in \citet{James2021Lett}, are tested against observations. We find no evidence for, and some evidence against, a lower bound to the FRB energy distribution, although we only exclude $E_{\rm min} \ge 10^{39.0}$\,erg (90\% C.L.). No such minimum energy is expected for the magnetar-origin hypothesis, which links the observed extragalactic FRB population to radio bursts from magnetar flares in our own Galaxy.

Our model also allows us to make inferences on the redshifts of the un-localized samples of FRBs detected by ASKAP and Parkes. We find these to be somewhat lower than expected from the Macquart relation, and indeed that the highest-DM events are likely not the most distant, due to observational effects causing an inversion of the Macquart relation, and the relatively steep best-fit value of the FRB energy distribution. The ability of this model to place priors on the expected redshift of FRBs given their measured DMs will also aid FRB localisation efforts, especially for those bursts --- such as FRB~171020 --- which have uncertain, but promising, host associations.

For the first time, we have incorporated the measured signal-to-noise ratio $s$ into FRB population modelling, allowing the use of this observable to constrain population parameters, and to predict the source-counts (``logN-logS'') distribution. In all scenarios, we find a steepening of this distribution from Parkes to ASKAP, consistent with the predictions of \citet{Macquart2018b} and the observations of \citet{Jamesetal2019a}.

Anomalies between observations and our model include a lack of ASKAP incoherent sum (localised) FRBs within $z<0.1$, an over (under) prediction of FRBs from Parkes multibeam (ASKAP fly's eye) observations, and an over-prediction of low-DM bursts. None of these have a high statistical significance --- however, we identify several potential explanations for these discrepancies. These are a spectral break in the FRB luminosity function, and the influence of the repeating FRB population. We recommend future works attempt to improve this modelling, and also improve understanding of the spectral beahviour of FRBs.

Future observations of localised FRBs with very large excess DMs, and/or from the local Universe, would verify predictions from this work. FRB surveys with very sensitive radio telescopes such as the 
Five-hundred-meter Aperture Spherical Telescope (FAST), or repeating previous Parkes and ASKAP surveys at different frequencies, would help to further constrain FRB population models. In particular, the application of this model to the large sample of bursts observed by CHIME would be particularly useful, although it would then need to be adapted to include repeating FRBs. We also aim to investigate improved numerical/computational methods to speed up calculations to allow the inclusion of further data.

%% file: appendix.tex
\section{Neglected effects}
\label{sec:neglected_effects}

Here we discuss effects that affect search sensitivity which have been neglected in this work.

\subsection{Fine sensitivity effects}
\label{sec:fine_effects}

The precise sensitivity of an incoherent FRB search to an FRB depends not only on the `coarse' effects of effective pulse width, but also the exact alignment of a burst with the time--frequency binning of the data compared to the FRB arrival time, and the DMs which are searched compared to the exact DM of the FRB. This effect is analysed by \citet{Keanefluence} for early FRB search methods, finding fluctuations of up to $\pm$15\%, although in current FRB search methods fluctuations are at the level of a few percent. This was also found to be the case for an internal investigation into the CRAFT FRB searches with ASKAP for \citet{Shannonetal2018}, with fluctuations of $\pm$3\%. Such effects are thus ignored in this work, and likely should be in all future works.

The presence of radio-frequency interference (RFI) during observations can result in a loss of effective bandwidth --- and hence sensitivity --- via vetoed frequency channels; loss of effective observing time if the RFI results in FRBs being completely unobservable; and reduced sensitivity to FRBs in a certain parameter space (particularly low-DM or high-width FRBs). The sensitivity of ASKAP to such effects has been studied using pulsar calibration observations by \citet{James2019cSensitivity}, finding a typical 10\% loss of effective observation time and $\sim$10\% fluctuation in sensitivity.

Further details of search pipelines can effect burst sensitivity. The zero-DM subtraction method --- i.e.\ subtracting the mean detected power prior to dedispersion --- will reduce sensitivity to low-DM events and very bright bursts, although typically only if the total dispersion sweep across the detection bandwidth is low. For instance, we have found a few percent bias in the estimated SNR for ASKAP searches when the expected SNR is greater than 100. An analysis of the {\sc HEIMDALL} software --- used in most Parkes FRB searches --- using real-time injected FRBs by \citet{Gupta2021} finds a 20\% reduction in SNR compared to expectations. This mostly affected bursts with widths above 20\,ms however, and we find in Section~\ref{sec:ingredients3} our estimates of the DM--z distribution to be insensitive to the population of very wide bursts. For ASKAP, the expected SNR of pulsar bursts was found by \cite{Shannonetal2018} to be within a few percent of that found by the Fast Real-time Engine for Dedispersing Amplitudes (FREDDA) algorithm used by the CRAFT collaboration. Nonetheless, for future precision cosmological studies, the response of FRB detection systems should be quantified in more detail.

We also note that in this work we use SNR as the true SNR of the FRB, when by definition this has a $1 \sigma$ error of unity. \citet{Murdoch1973} find that for SNR above six, the measured SNR can be approximated as the true SNR for statistical purposes, and so we have here ignored this effect, rather than marginalising over it. Possible non-linearities between FRB fluence and SNR become important at high SNR, e.g.\ as found for FRB~180309, which saturated the Parkes digitiser system and may have had a true SNR as high as 2616 \citep{Oslowski2019}. However, these effects can be corrected-for offline; care must be taken however to use the corrected values for model evaluation. Small effects are also possible at lower SNR, e.g.\ as found for high-width FRBs at UTMOST \citep{Gupta2021}.

Our beamshape model for Parkes uses the mean observation frequency, while for ASKAP it is uniformly weighted over all frequencies. A preference for e.g.\ low-frequency bursts would result in a slight increase in the survey effective area for both instruments, with only the difference in increase being relevant when estimating population parameters other than the absolute FRB rate. We consider this a third-order effect for the frequencies and bandwidths considered here.

Finally, we have not discussed errors in $F_0$. The absolute scale of the mean ASKAP threshold was calculated by \citet{James2019cSensitivity} by referencing observations to Parkes, and using Hydra A as an absolute flux calibrator. Thus we expect errors in flux calibration for Parkes and ASKAP to be linked, and cancel to first order. However, individual antenna sensitivity for ASKAP was found to vary by $\sim \pm 5\%$, and this is similar to the level of uncertainty found when performing calibration observations for \citet{Bannister2019}. Perhaps the largest source of uncertainty in $F_0$ comes from the quoted threshold of 0.5\,Jy\,ms for Parkes, which has an inherent rounding uncertainty of $\pm10\%$, and thus a corresponding rate uncertainty of $\pm15\%$ for a cumulative source counts index of $-1.5$ (see Section~\ref{sec:source_counts}).

\subsection{Pointing}
\label{sec:integral_galactic}

Here we treat a given FRB survey as having a constant local DM component, DM$_{\rm local}=$DM$_{\rm MW}+$DM$_{\rm host}$. This is because a greater local DM reduces sensitivity to extragalatic FRBs due to DM smearing.

When this effect becomes significant, this requires extending the integral in Eq.~\ref{eq:Ni} to
\begin{eqnarray}
<N_i> & = & \sum_k T_{i,k} \int dz \Phi(z) \frac{d V(z)}{d \Omega dz} \int d\DM_{\rm EG} p(\DM_{\rm EG} |z) \nonumber \\
&& \int dB \Omega(B) \int dw p(w)  p(E>E_{\rm th}) \label{eq:integral_galactic} \\
E_{\rm th} & \sim & E_{\rm th}(B,w,z,\DM = \DM_{\rm EG}+\DM_{\rm local,k})
\end{eqnarray}
where $T_{ik}$ is the total time spent observing at $\DM_{\rm local}=\DM_{\rm local,k}$.

\subsection{Scattering}
\label{sec:scattering_appendix}

In Section~\ref{sec:ingredients3}, we combine the intrinsic burst width $w_{\rm int}$ and scatter-broadening width $w_{\rm scat}$ into the incident width $w_{\rm inc}$. However, while $w_{\rm int}$ can reasonably be assumed to be independent of other FRB properties, it is plausible that $w_{\rm inc}$ will be correlated with both DM and/or $z$, and it will certainly be frequency-dependent. This is due to interstellar scattering, which is implicitly included through modelling of observed FRB widths, but is not explicitly accounted for as per e.g.\ \citet{Calebetal2016}. We largely avoid this question in this work by choosing surveys of similar frequency --- however, we consider evidence for such a correlation here.



Most FRB searches have time resolutions in the range of 100\,$\mu$s--1\,ms, and resolving scattering tails from the intrinsic burst structure is difficult: \citet{Harry2020} are able to do this for only six of the 33 FRBs observed at the typically 1.26\,ms real-time resolution of ASKAP FRB searches. The broader time-frequency structure exhibited by many repeating FRBs \citep{CHIME2019c,CHIME2020a,CHIME_catalog1_2021} however can often be resolved at these resolutions \citep[e.g.][]{Hessels2019}, while observations down to 10\,$\mu$s have revealed a 1\,GHz scattering time of only 24\,$\mu$s in FRB~121102, which is obtained indirectly from the measured scintillation bandwidth \citep{Michilli2018_121102}.

The most reliable way to resolve these two contributions is to use time--frequency data at the Nyquist resolution. ASKAP \citep{Cho2020,Day2020} and UTMOST \citep{Farah2018,Farah2019} have analysed such data for six FRBs each. Bursts were found to have strong sub-burst structures down to 50\,$\upmu$s. Scattering was conclusively measured for a total of four ASKAP FRBs, being in the range 40\,$\mu$s--3.3\,ms at 1.27\,GHz, while for 181112, it was at most 20\,$\mu$s. UTMOST FRBs had 835\,MHz scattering times of 4\,$\mu$s--30\,ms, with one upper limit at 0.2\,ms.

Clearly, there is a broad distribution of FRB intrinsic widths and scattering times. There are general effects that this distribution can have on the DM-$z$ distribution of FRBs.


The most complicated potential interaction of scattering and the intrinsic width is one that is dependent on the exact position of the FRB in DM--z space.

If indeed FRBs do arise from two source populations with different cosmological evolutions or different host galaxy properties, or else-wise one population of objects with properties that age on cosmological timescales, the intrinsic width distribution may have some DM--$z$ dependence. This possibility should not be ignored. However, the most likely redshift dependence arises from the scattering term.

Theoretical studies have examined expectations for scattering of FRBs during cosmic propagation, with effects attributed to the intergalactic medium (IGM), intra-cluster medium (ICM), and the halos and interstellar medium (ISM) of intersected galaxies.

The general form of the redshift dependence is analysed by \citet{MacquartKoay2013}, finding that scattering due to the IGM will increase as $(1+z)^2$ to $z \lesssim 1$, and as $(1+z)^{0.2 - 0.5}$ for $z \gtrsim 1$. This is in contrast to the contribution from hosts, which scales as $(1+z)^{-3}$.  Results on the absolute magnitude of the scattering depend on the assumed minimum and maximum length scales of the turbulence. \citet{MacquartKoay2013} argue that for realistic turbulence parameters, the total amount of scattering from the IGM and ICM is expected to be low, at $\lesssim 1$ and $\lesssim 5$\,ms at 300\,MHz, respectively. \citet{zhu2018} simulate FRBs propagating in a clumpy IGM. Examining the dependence of the mean scattering time $\bar{\tau}$ on DM, they find $\bar{\tau} \sim {\rm DM}^{1-2}$ for voids, clusters, and filaments over a range of simulation parameters. They also conclude that unreasonably high turbulence scales would be required to achieve scattering values comparable to that observed in FRBs.

In some cases, FRBs localisations have been sufficient to identify intersections of the line-of-sight with Galactic halos, placing upper limits on the degree of scattering caused by such intersections \citep{Prochaska2019,Apertif2020}.

\cite{Harry2020} find no evidence however for a DM-dependence of scattering. Both measurements and expectation therefore suggest that the observed distribution of scattering arises from the host galaxies. We do not however attempt to model this in this work, since the $(1+z)^2$ reduction in width at high redshift will be in any case insignificant against the DM smearing effect. A full model of scattering will however become important when including observations over a wide frequency range.

\subsection{Influence of specific parameters}

This work uses measured FRB detection numbers, dispersion measures, redshifts, and strengths from three different FRB searches. It is useful to probe the influence of each of these on our overall result. This also places an absolute limit on the aforementioned effects discussed in Appendix \ref{sec:neglected_effects} --- no matter how untrustworthy we consider the number of FRBs detected, the information contained in detection numbers will still be better than no information at all.

\begin{figure*} 
    \centering
    \includegraphics[width=0.7\textwidth]{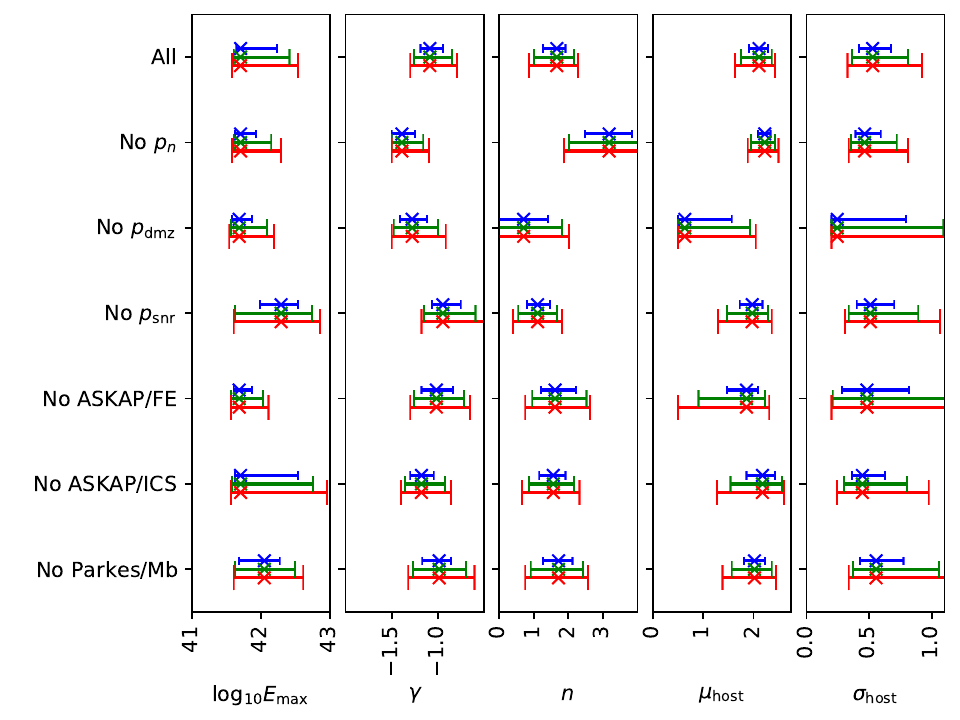}
    \includegraphics[width=0.7\textwidth]{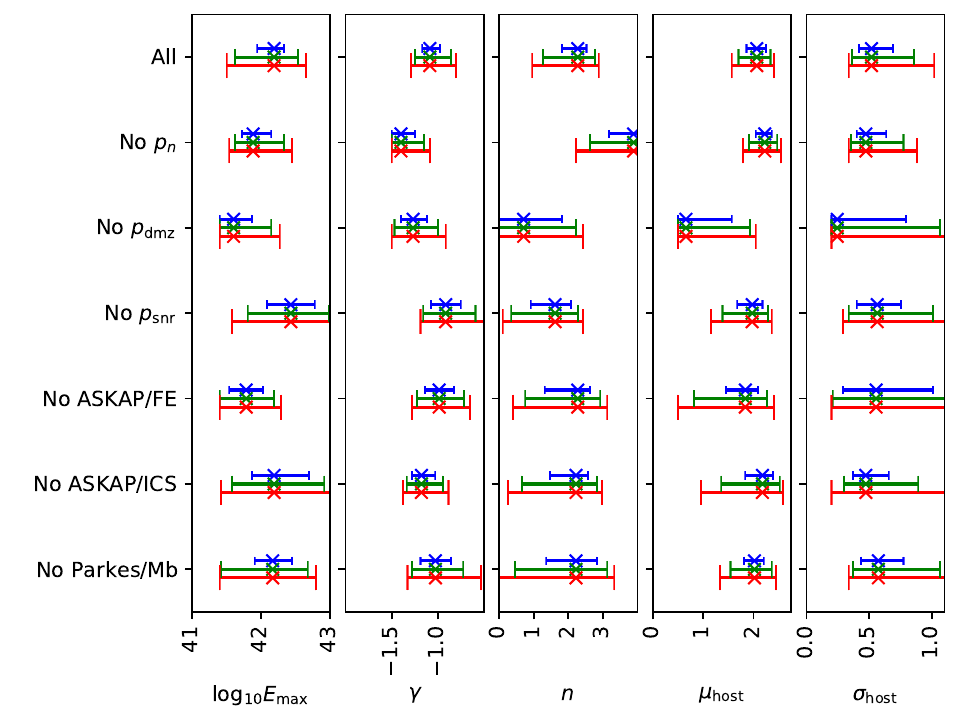}
    \caption{Resulting 68\%, 90\%, and 95\% confidence limits (respectively: blue, upper; green, middle; red, lower) on individual parameters under the spectral index interpretation of $\alpha$, using in descending order: all available information, removing information on the number of detected FRBs, their DMs and redshifts, their measured signal-to-noise ratios, and data from ASKAP/FE, ASKAP/ICS, and Parkes/MB, respectively. Top: using a prior on $\alpha$; bottom: no prior on $\alpha$.}
    \label{fig:ignored}
\end{figure*}

To do so, we show in Figure~\ref{fig:ignored} our confidence limits for each parameter, both with (top) and without (bottom) our prior on $\alpha$, when removing the likelihood corresponding to $p_{\rm N}$ (eq.\ref{eq:ni_given_n}), $p_{\rm DMZ}$ (eq.\ref{eq:pzdm}), $p_s$ (eq.\ref{eq:psu}), and when removing ASKAP/FE, ASKAP/ICE, and Parkes/Mb observations. We do this under the spectral index interpretation of $\alpha$ --- results are similar under the rate interpretation. 

Unsurprisingly, the most dramatic effect occurs when removing information on measured DM and $z$. In this limit, we do not find lower bounds on $E_{\rm max}$ in the range $\log_{10} E_{\rm max} {\rm [erg]} > 40$, and limits on other parameters are very broad. This is not a very interesting, informative, or unexpected result, and we ignore the ``No $p_{\rm DMZ}$'' in the following discussion. However, the fact that some constraints are nonetheless derived is evidence for the source-counts distribution of FRBs containing useful information \citep{Macquart2018b}.

For all other parameters, we obtain a variety of modified limits. Clearly, the ASKAP/ICS sample is most constraining to $E_{\rm max}$, while without the ASKAP/FE sample, the constraints are narrower. This is natural, since the ASKAP/FE sample probes the most intrinsically luminous bursts, without which a fit for a narrow population distribution is possible.

The slope of the cumulative luminosity function $\gamma$ is constrained to by flat by $p_n$, without which a steeper (lower) value would be obtained, while both the Parkes/Mb sample, and $p_{\rm snr}$, prevent flat (higher) values.
This is a very interesting result. As noted in our companion paper, our best fit under-predicts the number of ASKAP/FE FRBs (13.9 vs.\ 20 in 1274.6 days),, and over-predicts those for Parkes/Mb (17.3 vs 12 in 164.4 days). One would expect that removing either Parkes/Mb or $p_n$ would have a similar effect by removing this tension. However, removing Parkes/Mb also removes the Parkes DM distribution. This then suggests that the Parkes DM distribution is indicative of a steep source-counts spectrum, whereas the number of FRBs detected by Parkes is indicative of a flat spectrum. We do not have a full solution to this quandry, although it is clearly related to the discussion in Section~\ref{sec:results2}. We do not elaborate further, since this discrepancy is of only marginal statistical significance.

Perhaps our most significant result --- strong evidence for source evolution with redshift, i.e.\ $n>0$ --- is most strongly disfavoured by $p_n$, such that when this is removed, we prefer a strongly evolving population. Removing any of the three FRB surveys, or $p_{\rm snr}$, simply results in mildly less evidence for $n>0$ --- there is no single result strongly favouring $n>0$. We thus conclude that our conclusion on $n$ robust, especially considering uncertainties discussed in Appendix~\ref{sec:fine_effects}.

Our host galaxy parameters, $\mu_{\rm host}$ and $\sigma_{\rm host}$ --- recall these are in log DM space --- are most affected by the highest- and lowest-sensitivity surveys, ASKAP/FE and Parkes/Mb.  Without these observations, very low values of $\mu_{\rm host}$ (and hence very large values of $\sigma_{\rm host}$) become possible. This validates the use of DM distributions only when studying these parameters. Neither $p_{\rm snr}$ nor $p_n$ have a large effect.

Are the changes when ignoring measurements indicative of statistical robustness? The most statistically significant change in best-fit values occurs when ignoring $p_n$. This, predictably, is related to the low number of bursts observed by Parkes/Mb relative to ASKAP/FE compared to expectations discussed above, which alone argues against source evolution. The best-fit value when ignoring $p_n$ predicts that Parkes/Mb surveys should have observed fives times as many FRBs as ASKAP/Fe, as opposed to the 60\% (12/20) observed. We do not consider it possible that any of the fine sensitivity effects discussed in this Appendix could be responsible for a factor of five reduction in observable bursts at Parkes. Furthermore, ignoring $p_n$ favours a very steep luminosity function ($\gamma = -1.4$), which will tend to counteract the effect of the best-fit strong source evolution ($n=3.2$). In isolation, these values are inconsistent with those found when including $p_n$ at approximately the 99\% level, although including trial factors due to looking at the effects on five parameters when excluding six subsets of data reduces this significance to the 10\% level (since parameters are correlated, it is difficult to quote precise statistical significances). We thus interpret the large changes in best-fit values of $\gamma$ and $n$ when excluding $p_n$ as being due to the fit trying to account for anomalies in the z--DM distributions previously discussed in Section~\ref{sec:observed_predicted_z_dm}.

\subsection{Influence of uncertainty in Parkes sensitivity}

To provide a more realistic estimate of potential systematic effects, we perform two tests. Firstly, we test the effect of a systematic error in FRB detection threshold, as might occur if the reduced SNR found for UTMOST \citep{Gupta2021} also applied for Parkes. We take a more extreme value, and artificially increase the Parkes/Mb detection threshold $F_0$ by 60\% to 0.8\,Jy\,ms. Secondly, we assume all Parkes measurements below a SNR of 16 to be untrustworthy, as suggested by \citet{Macquart2018a} and \citet{Jamesetal2019a}, and discussed in Section~\ref{sec:source_counts}. As well as also raising the detection threshold to 0.8\,Jy\,ms, this removes seven FRBs --- six during the normalisable observation time --- from the analysis. Interestingly, removing six of twelve normalisable FRBs is exactly what is predicted from unbiased observations of a Euclidean source-counts distribution, where $N_{\rm FRB} \propto {\rm SNR}_{\rm th}^{-1.5}$, since $(16/10)^{-1.5}=0.49$. This suggests that any observational bias was present only in the less systematic FRB searches with Parkes.
Re-running the fit optimisation under the `rate' assumption on $\alpha$ (because numerical evaluation is quicker), the best-fit values of the fit parameters are shown in Table~\ref{tab:parkes_sys}.

\begin{table*} 
\renewcommand{\arraystretch}{1.3}
    \centering
    \begin{tabular}{c|c c c|c c c}
     & \multicolumn{3}{|c|}{Uniform prior on $\alpha$} & \multicolumn{3}{|c|}{Gaussian prior on $\alpha$}\\
     Parameter & Best Fit & $F_{\rm th}^{\rm Pks}=0.8$\,Jy\,ms & SNR$_{\rm th}^{\rm Pks}=16$ & Best Fit & $F_{\rm th}^{\rm Pks}=0.8$\,Jy\,ms & SNR$_{\rm th}^{\rm Pks}=16$  \\
     \hline
$\log_{10} E_{\rm max}$ & $42.19_{-0.24}^{+0.14}$ & 41.44$_{-0.06}^{+0.28}$ & 41.40$_{-0.02}^{+0.08}$ &    $41.70_{-0.06}^{+0.53}$ & 41.40$_{-0.02}^{+0.30}$ & 41.40$_{-0.02}^{+0.14}$ \\
$\gamma$ & -1.09$_{-0.08}^{+0.11}$ & -1.11$_{-0.17}^{+0.14}$ & -0.92$_{-0.16}^{+0.16}$ &  -1.09$_{-0.10}^{+0.14}$  & -1.11$_{-0.14}^{+0.13}$ & -0.95$_{-0.14}^{+0.20}$ \\
$n$ & 2.27$_{-0.45}^{+0.25}$  & 2.00$_{-1.64}^{+0.36}$ & 0.06$_{-0.30}^{+1.52}$ & 1.67$_{-0.40}^{+0.25}$  & 1.58$_{-0.30}^{+0.36}$ & 1.15$_{-0.36}^{+0.48}$ \\
$\mu_{\rm host}$ & 2.07$_{-0.20}^{+0.18}$ & 2.20$_{-0.18}^{+0.16}$ & 2.23$_{-0.16}^{+0.11}$ &   2.11$_{-0.20}^{+0.18}$ & 2.20$_{-0.16}^{+0.16}$ & 2.20$_{-0.18}^{+0.14}$ \\
$\sigma_{\rm host}$ & 0.52$_{-0.10}^{+0.17}$ & 0.48$_{-0.09}^{+0.14}$ & 0.46$_{-0.07}^{+0.17}$ &   0.53$_{-0.11}^{+0.15}$  & 0.48$_{-0.09}^{+0.14}$ & 0.51$_{-0.11}^{+0.15}$   \\
    \end{tabular}
    \caption{Comparison of best-fit parameter values to those calculated assuming an increase in the Parkes FRB detection threshold, $F_{\rm th}^{\rm Pks}$, to 0.8\,Jy\,ms; and when also excluding Parkes FRBs with SNR$<16$ from the analysis. Shown are results calculated under the `rate' approximation to $\alpha$, both with and without a prior on $\alpha = -1.5 \pm 0.3$. Errors correspond to 68\% confidence intervals calculated using Wilks' theorem.}
    \label{tab:parkes_sys}
\end{table*}

 Increasing the Parkes threshold to $0.8$\,Jy\,ms has almost negligible impact on the parameter estimates. We attribute this to the change in threshold being small compared to the difference in sensitivity between Parkes and ASKAP observations. This highlights the utility of fitting data from telescopes with very different thresholds. Setting SNR$_{\rm th}=16$ however has a greater impact: a flatter luminosity function is preferred, as is a reduced $n$. With a uniform prior on $\alpha$, the source evolution ($n\sim0$) is preferred, although with a Gaussian prior on $\alpha$, evolution consistent with the star-formation rate ($n\sim1$) is favoured.
 
 In general, varying these two systematics produces effects of smaller size than the 68\% parameter confidence intervals. Except in the case that the Parkes suffered from missing FRBs below SNR$_{\rm th}=16$, and the measurements of \citet{Macquart2019a} are somehow flawed, would the our main conclusion on the evolving nature of the FRB population be affected.
 
 Note that confidence intervals will be correlated between columns in Table~\ref{tab:parkes_sys}, and are shown to evaluate the importance of these systematics against random variation. 
 
 In concluding this section, we remind readers that the $F_{\rm th}^{\rm Pks}=0.8$\,Jy\,ms case is larger than we consider plausible, while the SNR$_{\rm th}^{\rm Pks}=16$ case discards a significant fraction of data from the most sensitive FRB sample; thus we expect the true systematic errors in our analysis to be less than the variation evident in Table~\ref{tab:parkes_sys}. Nonetheless, since $p_n$ is an important constraint on FRB parameters, improved modelling may be needed of the Parkes experiment to avoid uncertainties limiting the accuracy of future, otherwise more-precise calculations.
 
 \subsection{Validity of statistical assumptions}
 \label{sec:mc}
 
 In this work, We have by default used a frequentist approach to setting confidence intervals. This is more typical of the particle physics community, where the validity of models are evaluated using maximum likelihood methods. Here, we have  taken this approach, and invoked Wilks' theorem in setting confidence intervals.
 
 A more robust approach would be to use Monte Carlo methods to set these intervals. This would involve, for each combination of parameters:
 
 \begin{enumerate}
     \item For each parameter set $p_{\rm set}$, generating `pseudo-experiments' by sampling FRBs from the simulated distributions in z, DM, s space for each of the ASKAP/FE, ASKAP/ICS, and Parkes/Mb experiments;
     \item evaluating the likelihood of each FRB sample using the parameter set it was generated with, and the parameter set $p_{\rm max}$ maximising the likelihood for the actual, observed FRBs;
     \item include $p_{\rm set}$ in a C\% confidence interval if at least C\% of all Monte Carlo FRB samples from $p_{\rm set}$ had a likelihood difference between $p_{\rm max}$ and $p_{\rm set}$ less than or equal to the observed value.
 \end{enumerate}

\begin{figure}
    \centering
    \includegraphics[width=0.49\textwidth]{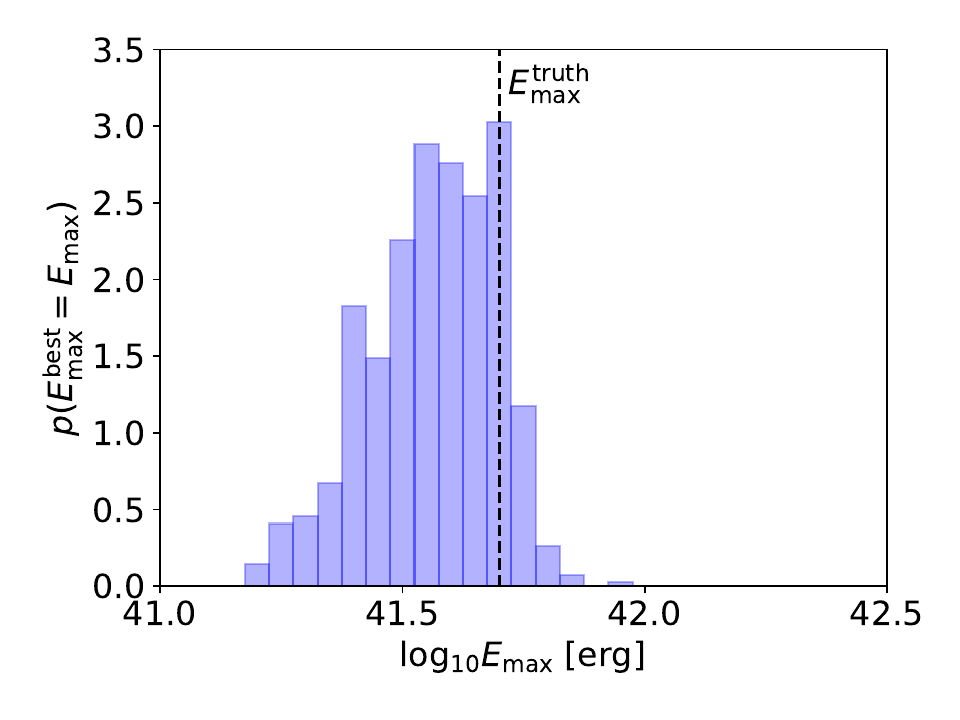}
    \includegraphics[width=0.49\textwidth]{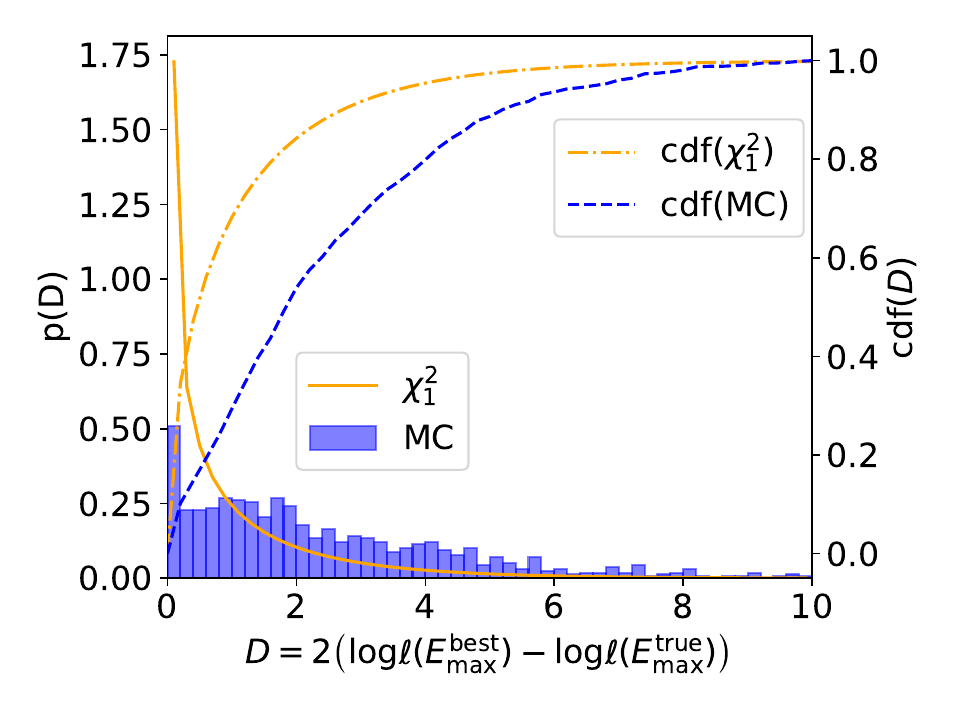}
    \caption{Upper: histogram of the values of $E_{\rm max}$ which maximise the likelihood, $E_{\rm max}^{\rm best}$, compared to the simulated true $E_{\rm max}^{\rm true}$, when varying $E_{\rm max}$ only, and using a Monte Carlo sample of FRBs as per Appendix~\ref{sec:mc}. Lower: differential and cumulative distributions of the test statistic $D$ compared to the predicted $\chi^2_1$ distribution from Wilks' theorem.}
    \label{fig:mc}
\end{figure}

Such a procedure would, however, be too computationally complex for this multi-dimensional space. Rather, we test the validity of Wilks' theorem for this particular data sample in one dimension, by varying $E_{\rm max}$ only. We choose $E_{\rm max}$ because the sharp drop in the likelihood distribution below the observed energy of localised FRBs makes this the least well-behaved parameter. To further simplify the problem, we only generate pseudo-experiments at one simulated truth, being the set $p_{\rm max}$ reported for the best-fit with a Gaussian prior on $\alpha$ given in Table~\ref{tab:parkes_sys} (and in our companion paper). We do not vary the number of observed FRBs $N_{\rm FRB}$, and simply re-generate the properties $z$, DM, and $s$ for each FRB. We then evaluate these while varying $E_{\rm max}$, and measure the test statistic
\begin{eqnarray}
D & = & 2 \log ( \ell(E_{\rm max}^{\rm best})/\ell(E_{\rm max}^{\rm true})).
\end{eqnarray}
The resulting distribution of best-fitting $E_{\rm max}$ values, $E_{\rm max}^{\rm best}$, and the distribution of $D$, are given in Figure~\ref{fig:mc}, upper and lower respectively, using $10^4$ pseudo-experiments.

From Figure~\ref{fig:mc}(upper), $E_{\rm max}^{\rm best}$ is clearly biased towards low values. This is expected, since the fit naturally prefers values of $E_{\rm max}$ only marginally above the highest observed FRB energy. The MC never generates FRBs with energy above $E_{\rm max}^{\rm true}$, but can generate samples with FRBs with lower energy. Thus the distribution of $E_{\rm max}^{\rm best}$ is similarly shaped to the likelihood function $\ell(E_{\rm max})$ when evaluated on actual data, but skewed in the opposite direction.

According to Wilks' theorem, $D$ should follow a $\chi^2_1$ distribution. It is evident from Figure~\ref{fig:mc}(right) that while the shapes are similar, the true distribution of $D$ is significantly skewed to the right compared to the expectation. The cumulative distributions illustrate that the confidence intervals calculated using Wilks' theorem may therefore suffer from undercoverage.

We note that we have tested Wilks' theorem for the most extreme case. Our confidence intervals, when marginalised over other parameters, should be more trustworthy than indicated by Figure~\ref{fig:mc}. Nonetheless, we proceed to calculate Bayesian confidence intervals to assess the sensitivity of our conclusions to the statistical method used.

\subsection{Bayesian confidence intervals}
\label{sec:bayesian}

A Bayesian posterior probability distribution on a single parameter $\theta_i$ in a parameter set $\vec{\theta}$ can be calculated by integrating the likelihood function $\ell$, weighted by appropriate priors, over all dimensions except $i$:
\begin{eqnarray}
p(\theta_i) & = & \int \ell( \vec{\theta}) p(\vec{\theta}) d\vec{\theta}_{\ne i}.
\end{eqnarray}
We choose as our priors uniform distributions in $\mu_{\rm host}$ and $\sigma_{\rm host}$ (which are already defined in log--space), $\gamma$, and $n$, and log-uniform priors in $E_{\rm max}$, in the ranges shown in Table~\ref{tab:bayes_ranges}. As previously, we take both uniform and Gaussian priors on $\alpha$.

\begin{table}
    \centering
    \begin{tabular}{c|c c}
    Parameter & Min & Max \\
    \hline
$\log_{10} E_{\rm max}$ & 41 & 43.4 \\
$\alpha$ & -2.50 & 1.0 \\
$\gamma$ & -0.5 & -1.5 \\
$n$ & -2 & 5 \\
$\mu_{\rm host}$ & 0.5 & 2.75 \\
$\sigma_{\rm host}$ & 0.2 & 1.1 \\
    \end{tabular}
    \caption{Ranges for the uniform priors on parameters used in the calculation of Bayesian confidence intervals. In all cases other than $\alpha$, the likelihood is negligible outside the interval.}
    \label{tab:bayes_ranges}
\end{table}

\begin{figure*}
    \centering
    \includegraphics[width=0.32\textwidth]{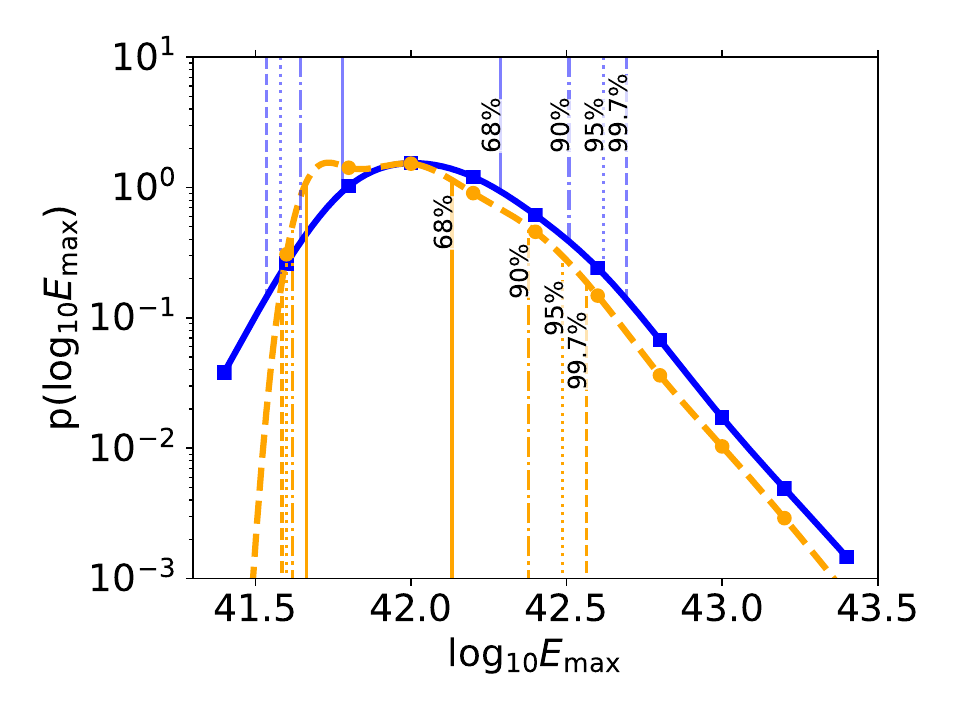}
     \includegraphics[width=0.32\textwidth]{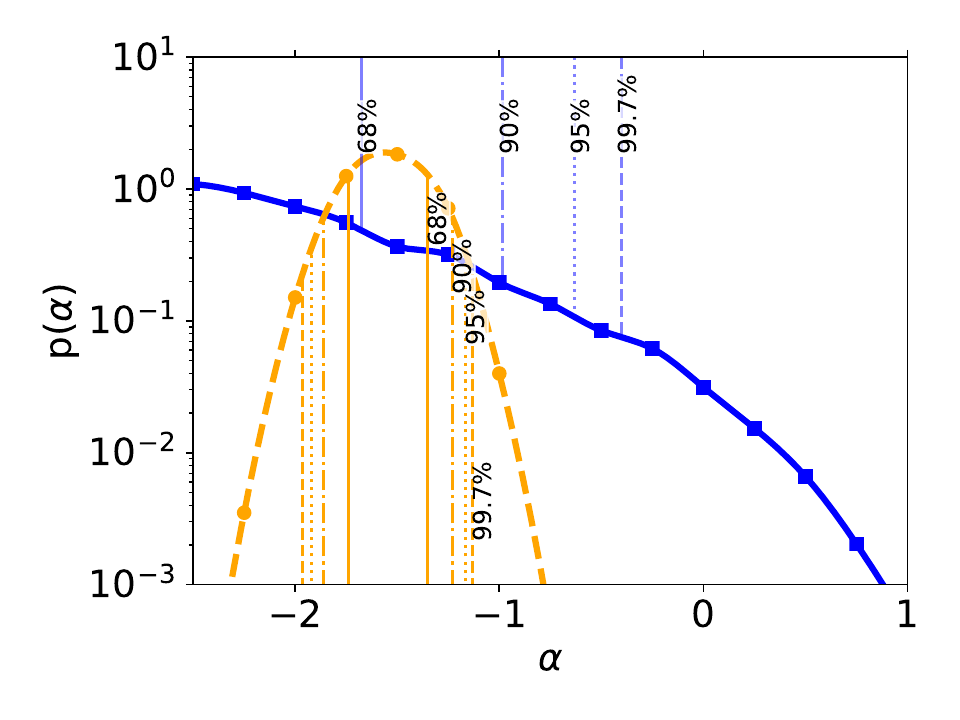}
      \includegraphics[width=0.32\textwidth]{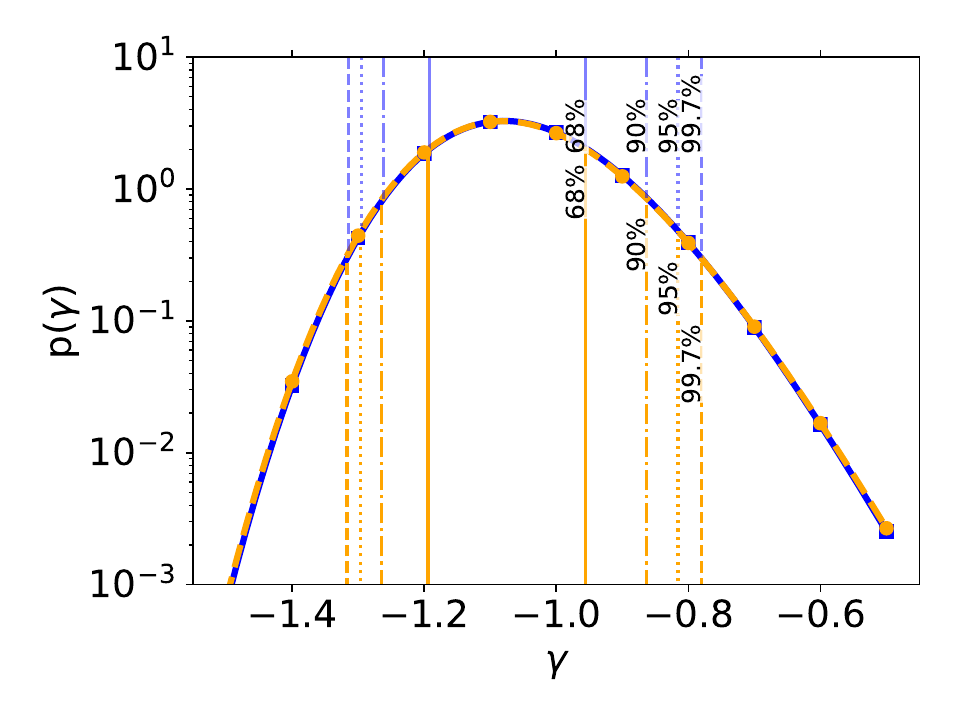} \\
       \includegraphics[width=0.32\textwidth]{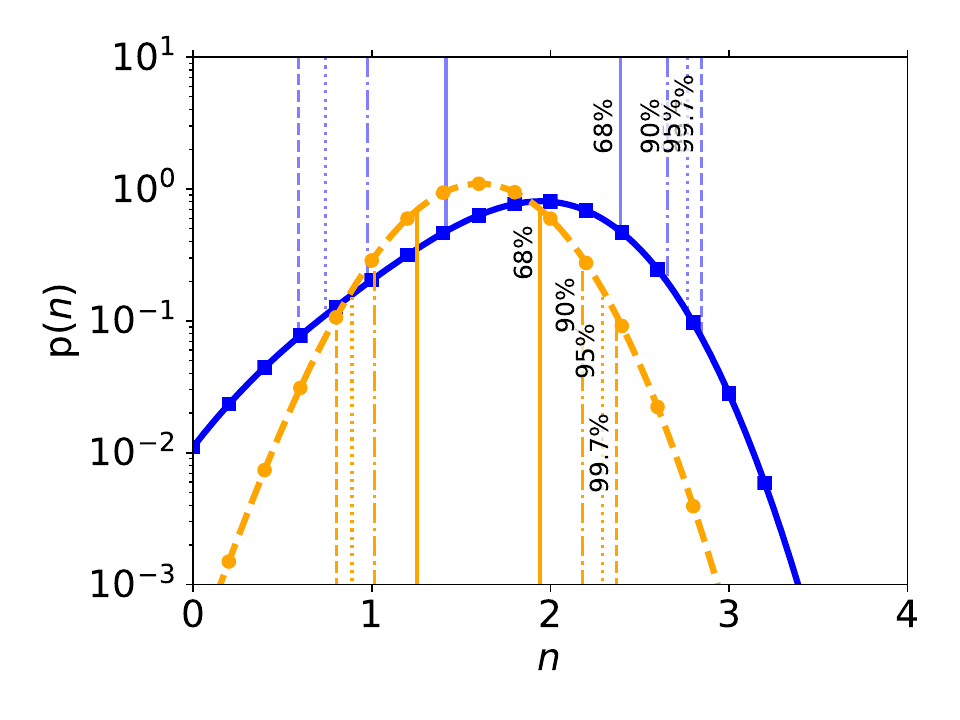}
        \includegraphics[width=0.32\textwidth]{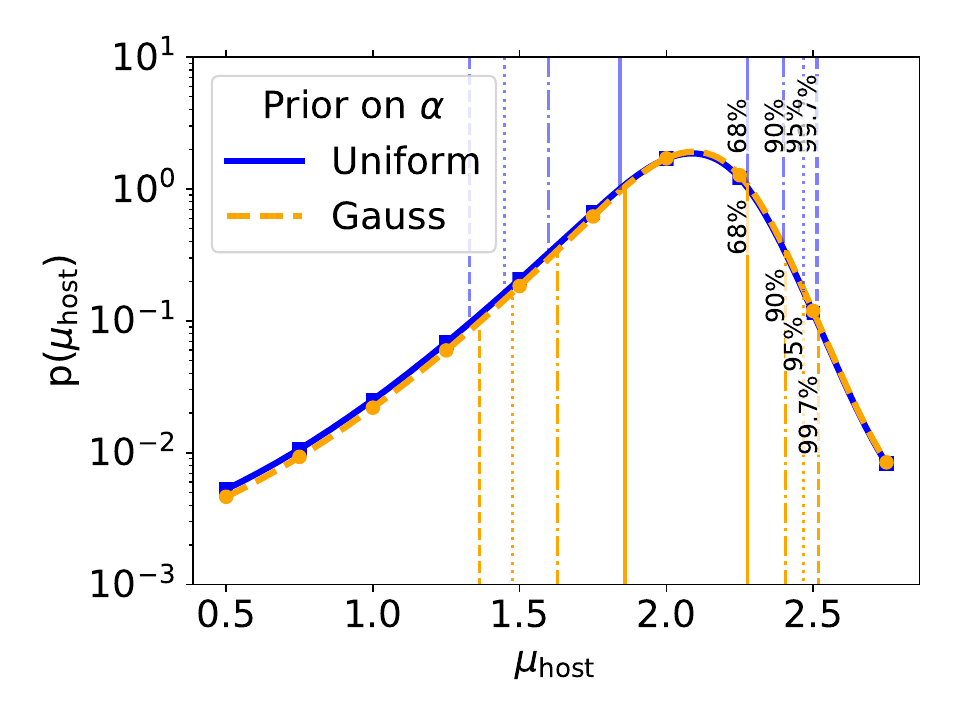}
         \includegraphics[width=0.32\textwidth]{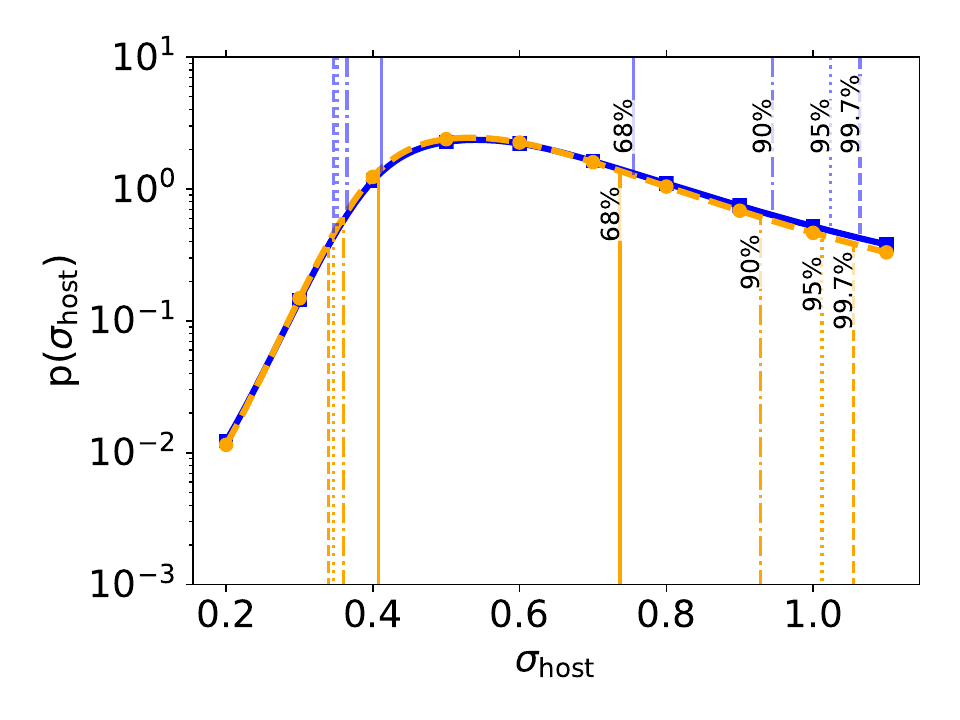}
    \caption{Bayesian confidence intervals on parameters $E_{\rm max}, \alpha, \gamma, n, \mu_{\rm host},\sigma_{\rm host}$ when marginalised over the other five, under the spectral index interpretation of $\alpha$, using a Gaussian (orange, lower) and uniform (blue, upper) prior on the spectral index $\alpha$. Calculation results are given by points, with lines drawn using cubic spline smoothing. Vertical lines are single-parameter intervals at the labelled degree of confidence calculated.}
    \label{fig:bayesian_1d}
\end{figure*}

\begin{figure*}
    \centering
    \includegraphics[width=0.32\textwidth]{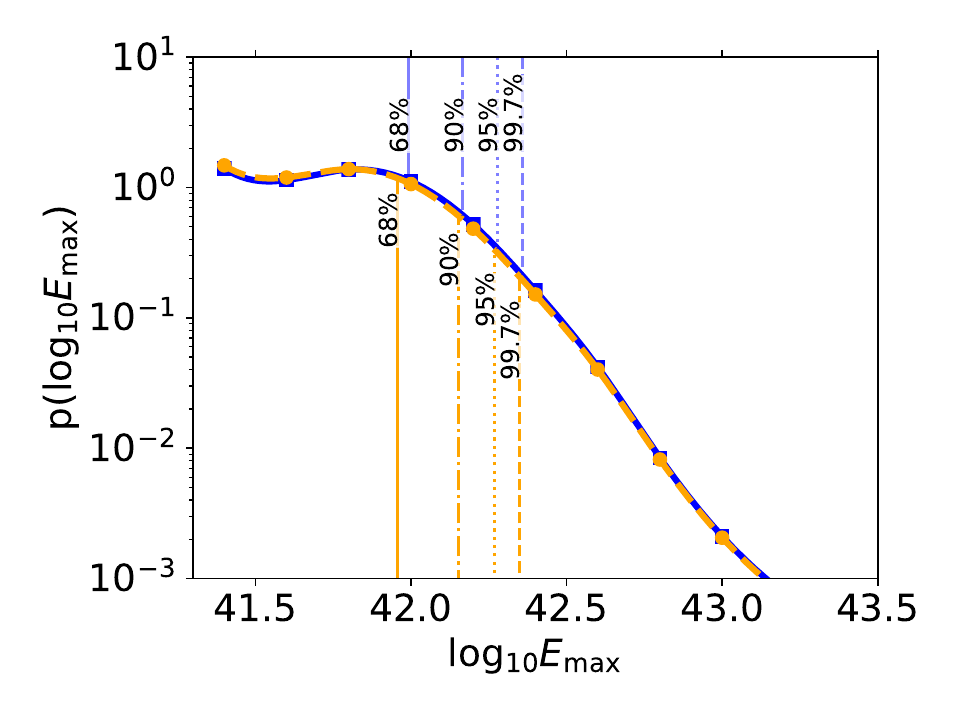}
     \includegraphics[width=0.32\textwidth]{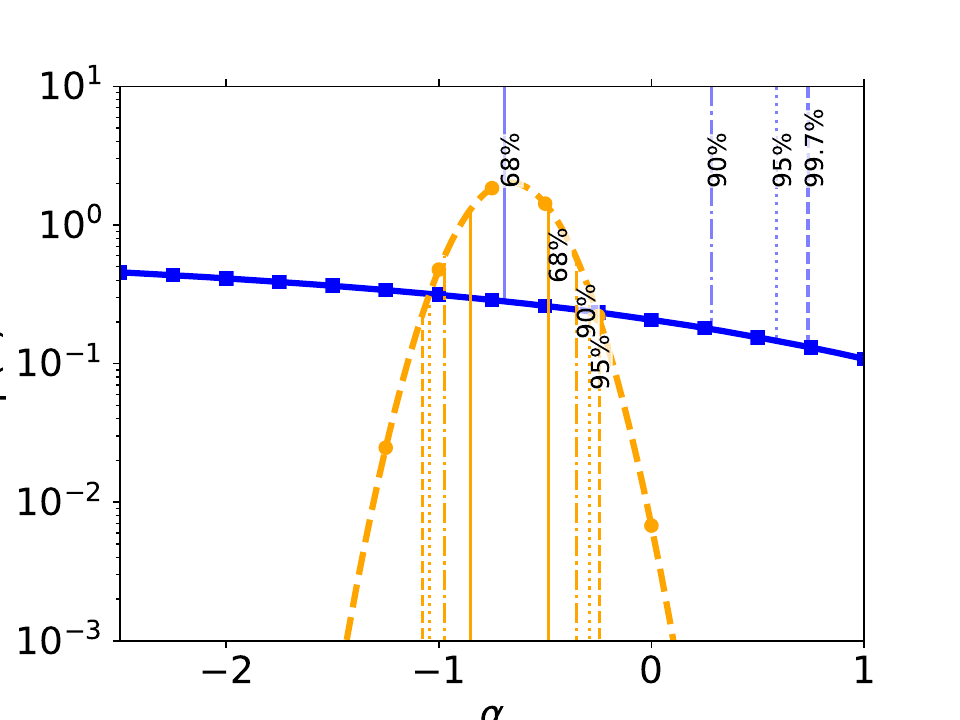}
      \includegraphics[width=0.32\textwidth]{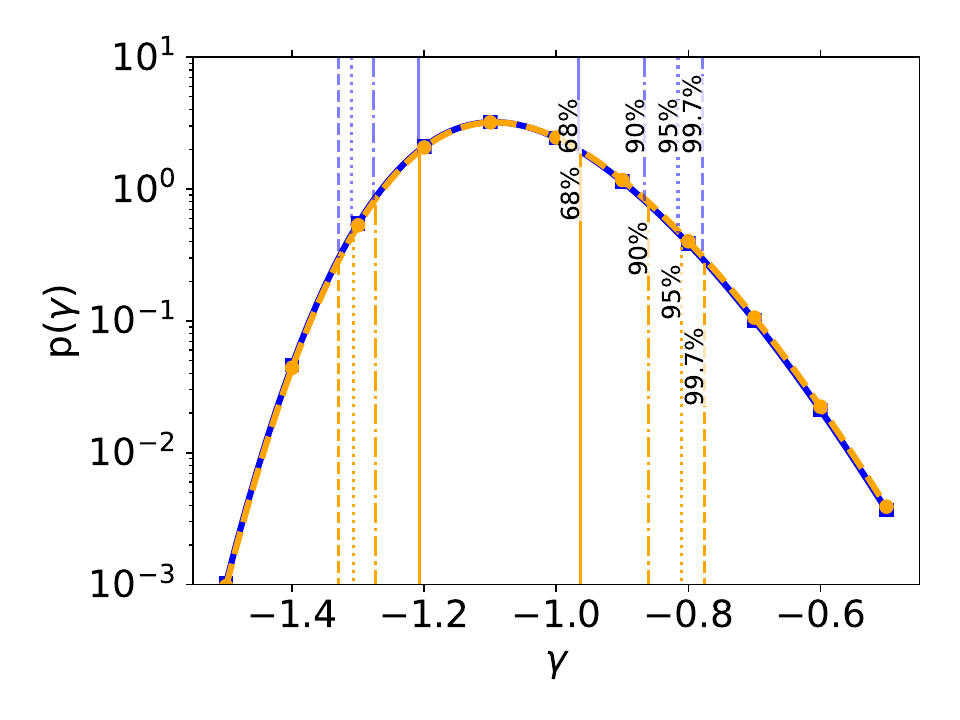} \\
       \includegraphics[width=0.32\textwidth]{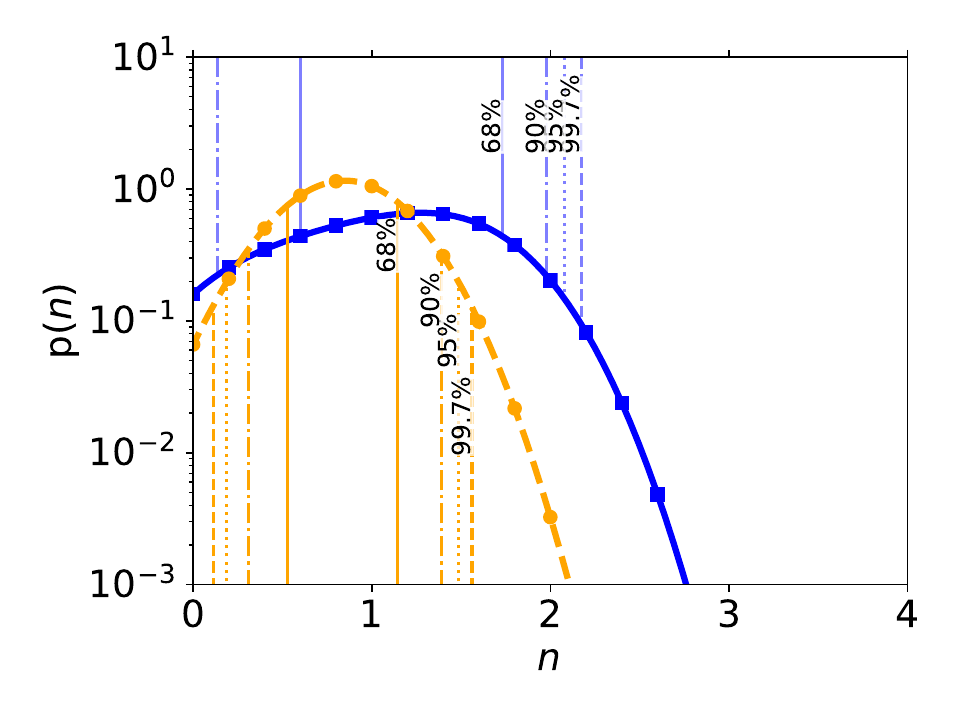}
        \includegraphics[width=0.32\textwidth]{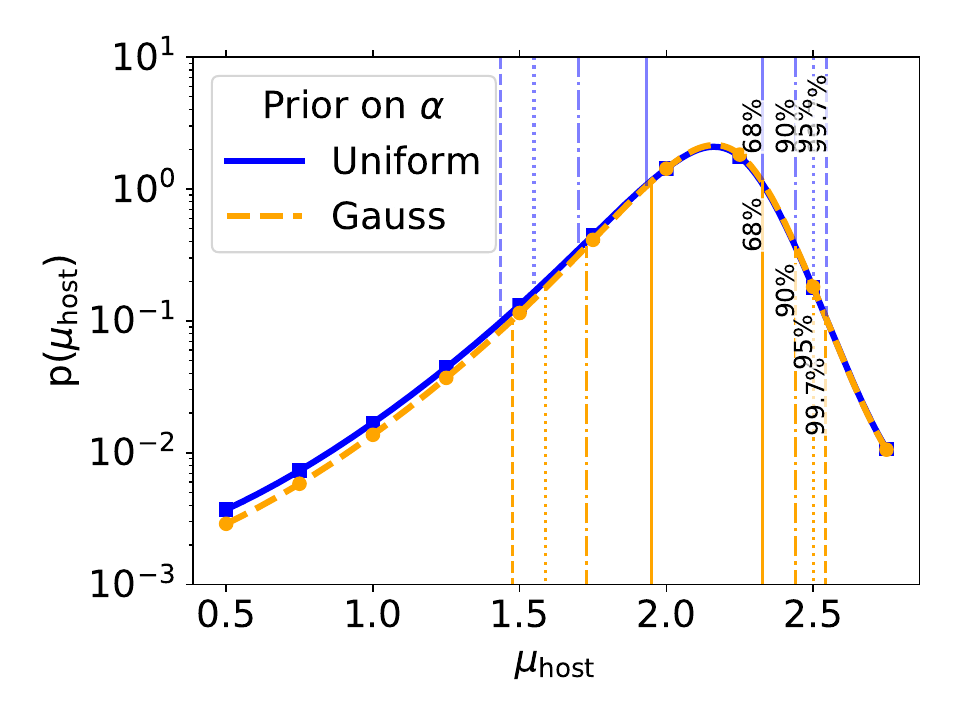}
         \includegraphics[width=0.32\textwidth]{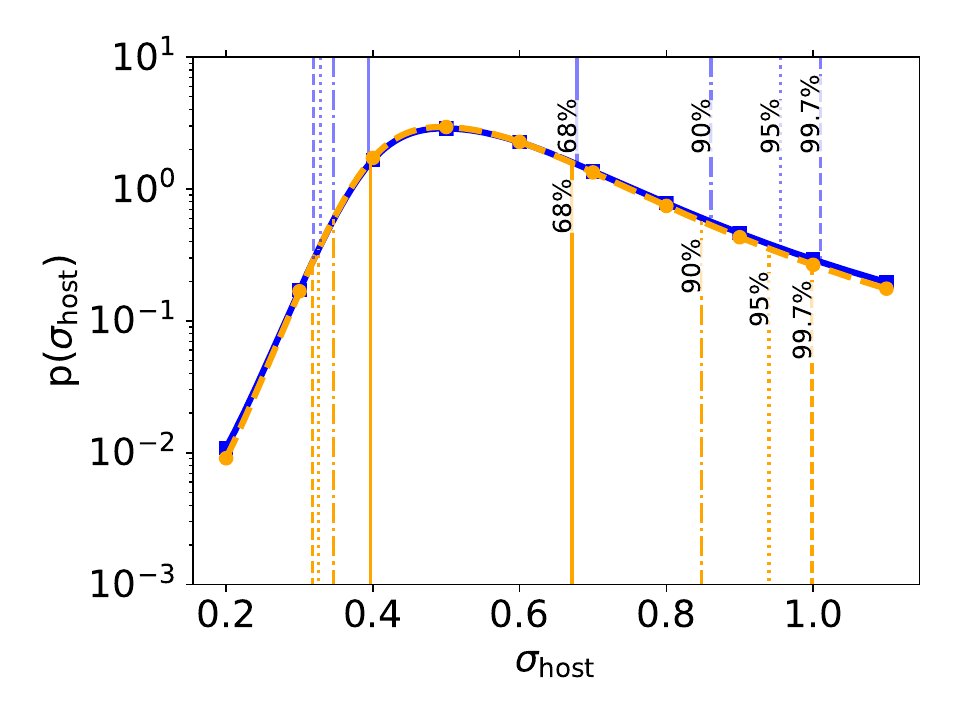}
    \caption{As per Figure~\ref{fig:bayesian_1d}, except calculated using the rate interpretation of $\alpha$.}
    \label{fig:bayesian_rate_1d}
\end{figure*}

The result using the spectral (rate) interpretation of $\alpha$ is plotted in Figure~\ref{fig:bayesian_1d} (Figure~\ref{fig:bayesian_rate_1d}).
Confidence intervals are constructed by including regions of parameter space with the greatest $p(\theta_i)$ until the desired level of confidence is reached. These intervals are also listed in Table~\ref{tab:bayes_ranges}. We observe that Bayesian confidence intervals are generally more constraining than those calculated using Wilks' theorem in the frequentist approach. Thus while it is possible that the latter intervals do suffer from undercoverage, they are nonetheless more conservative than the intervals which would be generated using the more standard Bayesian approach, as used by e.g.\ \citet{Luo2020}.

\begin{table*} 
\renewcommand{\arraystretch}{1.3}
    \centering
    \begin{tabular}{c|c c c c c|c c c c c}
     & \multicolumn{5}{|c|}{Uniform prior on $\alpha$} & \multicolumn{5}{|c|}{Gaussian prior on $\alpha$}\\
     Parameter & Best Fit & 68\% C.L.  & 90\% C.L.  & 95\% C.L. & 99.7\% C.L. & Best Fit & 68\% C.L.  & 90\% C.L.  & 95\% C.L. & 99.7\% C.L. \\
     \hline
$\log_{10} E_{\rm max}$ & 42.01 & $_{-0.23}^{+0.28}$  & $_{-0.36}^{+0.50}$  & $_{-0.43}^{+0.61}$  & $_{-0.47}^{+0.69}$  &  41.74 & $_{-0.07}^{+0.40}$  & $_{-0.12}^{+0.64}$  & $_{-0.14}^{+0.75}$  & $_{-0.15}^{+0.83}$ \\
$\gamma$ & -1.08 & $_{-0.11}^{+0.12}$  & $_{-0.18}^{+0.22}$  & $_{-0.22}^{+0.26}$  & $_{-0.24}^{+0.30}$  &  -1.08 & $_{-0.11}^{+0.13}$  & $_{-0.18}^{+0.22}$  & $_{-0.22}^{+0.27}$  & $_{-0.24}^{+0.30}$ \\
$n$ & 1.94 & $_{-0.53}^{+0.45}$  & $_{-0.96}^{+0.71}$  & $_{-1.20}^{+0.83}$  & $_{-1.35}^{+0.90}$  &  1.60 & $_{-0.35}^{+0.34}$  & $_{-0.59}^{+0.58}$  & $_{-0.71}^{+0.69}$  & $_{-0.80}^{+0.76}$ \\
$\mu_{\rm host}$ & 2.08 & $_{-0.24}^{+0.19}$  & $_{-0.48}^{+0.32}$  & $_{-0.64}^{+0.38}$  & $_{-0.76}^{+0.43}$  &  2.10 & $_{-0.24}^{+0.18}$  & $_{-0.47}^{+0.31}$  & $_{-0.62}^{+0.37}$  & $_{-0.73}^{+0.42}$ \\
$\sigma_{\rm host}$ & 0.54 & $_{-0.13}^{+0.21}$  & $_{-0.18}^{+0.40}$  & $_{-0.19}^{+0.48}$  & $_{-0.19}^{+0.52}$  &  0.53 & $_{-0.13}^{+0.20}$  & $_{-0.17}^{+0.39}$  & $_{-0.19}^{+0.48}$  & $_{-0.19}^{+0.52}$ \\
    \end{tabular}
    \caption{Bayesian confidence limits on single parameters, with both a uniform prior and Gaussian prior on $\alpha$, calculated assuming the spectral interpretation of $\alpha$ (see Section~\ref{sec:ingredients2}). The best-fit value is quoted such that the posterior probability is maximised.}
    \label{tab:bayes_spectral}
\end{table*}

\begin{table*} 
\renewcommand{\arraystretch}{1.3}
    \centering
    \begin{tabular}{c|c c c c c|c c c c c}
     & \multicolumn{5}{|c|}{Uniform prior on $\alpha$} & \multicolumn{5}{|c|}{Gaussian prior on $\alpha$}\\
     Parameter & Best Fit & 68\% C.L.  & 90\% C.L.  & 95\% C.L. & 99.7\% C.L. & Best Fit & 68\% C.L.  & 90\% C.L.  & 95\% C.L. & 99.7\% C.L. \\
     \hline
$\log_{10} E_{\rm max}$ & 41.40 & $_{-0.20}^{+0.59}$  & $_{-0.20}^{+0.77}$  & $_{-0.20}^{+0.88}$  & $_{-0.20}^{+0.96}$  &  41.40 & $_{-0.20}^{+0.56}$  & $_{-0.20}^{+0.75}$  & $_{-0.20}^{+0.87}$  & $_{-0.20}^{+0.95}$ \\
$\gamma$ & -1.09 & $_{-0.12}^{+0.13}$  & $_{-0.18}^{+0.23}$  & $_{-0.22}^{+0.28}$  & $_{-0.24}^{+0.32}$  &  -1.09 & $_{-0.11}^{+0.13}$  & $_{-0.18}^{+0.23}$  & $_{-0.21}^{+0.28}$  & $_{-0.24}^{+0.32}$ \\
$n$ & 1.27 & $_{-0.66}^{+0.47}$  & $_{-1.13}^{+0.71}$  & $_{-1.33}^{+0.81}$  & $_{-1.42}^{+0.91}$  &  0.85 & $_{-0.32}^{+0.29}$  & $_{-0.54}^{+0.54}$  & $_{-0.66}^{+0.64}$  & $_{-0.74}^{+0.71}$ \\
$\mu_{\rm host}$ & 2.16 & $_{-0.23}^{+0.16}$  & $_{-0.46}^{+0.28}$  & $_{-0.61}^{+0.34}$  & $_{-0.73}^{+0.38}$  &  2.16 & $_{-0.21}^{+0.16}$  & $_{-0.43}^{+0.28}$  & $_{-0.58}^{+0.34}$  & $_{-0.69}^{+0.38}$ \\
$\sigma_{\rm host}$ & 0.50 & $_{-0.10}^{+0.18}$  & $_{-0.15}^{+0.36}$  & $_{-0.17}^{+0.46}$  & $_{-0.18}^{+0.51}$  &  0.49 & $_{-0.10}^{+0.18}$  & $_{-0.15}^{+0.35}$  & $_{-0.17}^{+0.45}$  & $_{-0.18}^{+0.51}$ \\
    \end{tabular}
    \caption{As per Table~\ref{tab:bayes_spectral}, assuming the rate interpretation of $\alpha$ (see Section~\ref{sec:ingredients2}). A lower limit on $\log_{10} E_{\rm max}$\,[erg] of $41.2$ has been assumed.}
    \label{tab:bayes_rate}
\end{table*}